\documentclass[12pt]{article}

\usepackage[title]{appendix}
\usepackage{lscape}
\usepackage{amsfonts}
\usepackage{amssymb}
\usepackage{amsmath}
\usepackage{epstopdf}
\usepackage{float}
\usepackage{natbib}
\usepackage[margin=.75in]{geometry}
\usepackage[hang]{caption}
\usepackage{color}
\usepackage{graphicx}
\usepackage{psfrag}
\usepackage{rotating}
\usepackage{tikz}
\usetikzlibrary{arrows}
\usepackage{bbm}
\usepackage[multiple]{footmisc}

\makeatletter
\newcommand*\rel@kern[1]{\kern#1\dimexpr\macc@kerna}
\newcommand*\widebar[1]{%
  \begingroup
  \def\mathaccent##1##2{%
    \rel@kern{0.8}%
    \overline{\rel@kern{-0.8}\macc@nucleus\rel@kern{0.2}}%
    \rel@kern{-0.2}%
  }%
  \macc@depth\@ne
  \let\math@bgroup\@empty \let\math@egroup\macc@set@skewchar
  \mathsurround\z@ \frozen@everymath{\mathgroup\macc@group\relax}%
  \macc@set@skewchar\relax
  \let\mathaccentV\macc@nested@a
  \macc@nested@a\relax111{#1}%
  \endgroup
}
\makeatother

\renewcommand{\baselinestretch}{1.2}
\allowdisplaybreaks

\DeclareMathOperator{\argmin}{arg\,min}

\usepackage{booktabs}
\usepackage{subcaption}

\usepackage[utf8]{inputenc}
\usepackage[acronym]{glossaries}

\makenoidxglossaries

\newglossaryentry{Activation function}
{
	name=Activation function,
	description={A function that takes the weighted sum of inputs (from the previous layer), adds a bias term, and then generates a nonlinear output, which serves as the input for the next layer}
}

\newglossaryentry{Adam}
{
	name=Adam,
	description={Adam or "adaptive moments" is an adaptive learning rate optimization algorithm. It can be seen, as a combination of the RMSprop and momentum algorithm with a few distinctions}
}

\newglossaryentry{Algorithm}
{
	name=Algorithm,
	description={Is a finite sequence of well-defined functions or series of instructions used to solves a class of problems}
}

\newglossaryentry{Boosting}
{
	name=Boosting,
	description={A machine learning technique that iteratively combines a set of weak learners to a strong learner}
}

\newglossaryentry{Epoch}
{
	name=Epoch,
	description={The number of times the algorithm is trained over the entire dataset}
}

\newglossaryentry{Feature}
{
	name=Feature,
	description={Feature, explanatory variable, independent variable, and input variables are different terms for the same thing}
}

\newglossaryentry{Hidden layer}
{
	name=Hidden layer,
	description={A synthetic layer in a neural network. The layer is between the input layer (input features) and the output layer (the prediction)}
}

\newglossaryentry{Hyperparameters}
{
	name=Hyperparameters,
	description={Hyperparameters are higher-level parameters of a model such as how fast it can learn (learning rate) or the complexity of a model}
}

\newglossaryentry{Input layer}
{
	name=Input layer,
	description={The first layer in the neural network, which receives the input features}
}

\newglossaryentry{Learning Rate}
{
	name=Learning rate,
	description={The learning rate is the step size of each iteration}
}

\newglossaryentry{Momentum}
{
	name=Momentum,
	description={A gradient descent based algorithm which helps accelerate the convergence}
}

\newglossaryentry{Node Regression tree}
{
	name=Nodes - Regression tree,
	description={\textit{Root nodes}: Entry points to a collection of data. \textit{Inner nodes}: A set of binary questions where each child node is available for every possible answer. \textit{Leaf nodes}: Respond value if reached}
}

\newglossaryentry{Nodes Neural network}
{
	name=Nodes - Neural network,
	description={A node or neuron in a neural network takes multiple input values and transforms them into an output value. The transformation is conducted through a nonlinear activation function}
}

\newglossaryentry{Output layer}
{
	name=Output layer,
	description={The final layer of a neural network. The layer containing the predictions}
}

\newglossaryentry{Regularization}
{
	name=Regularization,
	description={Regularization is a technique used for handling the overfitting problem. By adding a complexity term to the loss function, the sample variance of the model will decrease}
}

\newglossaryentry{RMSprop}
{
	name=RMSprop,
	description={RMSprop (for Root Mean Square propagation) is an adaptive learning rate algorithm. The learning rate is adapted for each of the parameters. RMSprop divides the learning rate by an exponentially decaying average of squared gradients}
}

\newglossaryentry{Shrinkage}
{
	name=Shrinkage,
	description={Reducing the effect of sample variation by shrinking the coefficient towards zero}
}

\newglossaryentry{Sparsity}
{
	name=Sparsity,
	description={Sparsity occurs when a large number of elements are set to zero in a vector or matrix}
}

\newglossaryentry{Test Set}
{
	name=Test set,
	description={A subset of data, at the end of the dataset, to assess the out-of-sample predictive power of the model}
}

\newglossaryentry{Training Set}
{
	name=Training set,
	description={A subset of data used to fitting/training the parameters in the model}
}

\newglossaryentry{Validation Set}
{
	name= Validation set,
	description={A subset of data used to provide an unbiased evaluation of the current model specification. Often used for tuning the hyperparameters}
}

\newglossaryentry{Subset selection}
{
	name= Subset selection,
	description={Selecting a subset of relevant variables}
}

\newglossaryentry{Weak learner}
{
	name= Weak learner,
	description={A weak learner in a machine learning model that consistently beats a random guess}
} 

\begin{document}

\title{A machine learning approach to volatility forecasting\thanks{Christensen was partly funded by the Independent Research Fund Denmark (DFF 1028--00030B). This work was also supported by CREATES. We appreciate feedback from the audience in our session at the 13th International Conference on Computational and Financial Econometrics (CFE 2019) in London, UK, and in seminars at Aarhus University, Denmark, and Scuola Normale Superiore di Pisa, Italy. We are also indebted to Torben G. Andersen and Tim Bollerslev for their extensive comments on an earlier draft of this article. At the Journal of Financial Econometrics, we thank co-editor Bryan Kelly, the associate editor, and two anonymous referees for their input during the revision process. We are grateful to Francis X. Diebold for highlighting our paper on his ``No Hesitations'' blog. E-mail correspondence to \texttt{kim@econ.au.dk}.}}
	
\author{Kim Christensen\thanks{CREATES, Department of Economics and Business Economics, Aarhus University, Fuglesangs All\'{e} 4, 8210 Aarhus V, Denmark.}
\and Mathias Siggaard\footnotemark[2]
\and Bezirgen Veliyev\footnotemark[2] \vspace*{0.5cm}}

\date{May, 2022}

\maketitle

\begin{abstract}
We inspect how accurate machine learning (ML) is at forecasting realized variance of the Dow Jones Industrial Average index constituents. We compare several ML algorithms, including regularization, regression trees, and neural networks, to multiple Heterogeneous AutoRegressive (HAR) models. ML is implemented with minimal hyperparameter tuning. In spite of this, ML is competitive and beats the HAR lineage, even when the only predictors are the daily, weekly, and monthly lags of realized variance. The forecast gains are more pronounced at longer horizons. We attribute this to higher persistence in the ML models, which helps to approximate the long-memory of realized variance. ML also excels at locating incremental information about future volatility from additional predictors. Lastly, we propose a ML measure of variable importance based on accumulated local effects. This shows that while there is agreement about the most important predictors, there is disagreement on their ranking, helping to reconcile our results.

\vspace*{0.5cm}


\end{abstract}

\vfill

\thispagestyle{empty}

\pagebreak

\section{Introduction} \label{section:introduction}

\setcounter{page}{1} \renewcommand{\baselinestretch}{1.5}

The unprecedented levels of volatility in financial markets during the last decades has led to increased awareness about the importance of explaining key drivers behind such movements. On the one hand, the structure of financial markets is complex, highly nonlinear and has a low signal-to-noise ratio, which makes it nearly impossible to predict short-term asset returns \citep*[see, e.g.,][]{gu-kelly-xiu:20a, chen-pelger-zhu:24a}. On the other hand, various stylized facts (e.g., the slow decay of autocorrelation in absolute returns) implies that volatility is to a large extent predictable.

With its crucial role in asset pricing, portfolio allocation, and risk management, volatility forecasting has attracted great attention. The origins of this literature go back to \citet*{engle:82a, bollerslev:86a, taylor:82a}, who developed discrete-time GARCH and stochastic volatility models for autoregressive conditional heteroskedasticity. As argued by \citet*{corsi:09a}, however, standard volatility models are not able to reproduce several salient features of financial markets. He proposes a Heterogeneous Auto-Regressive (HAR) model, which combines nonparametric realized variance measured at different frequencies with a parametric autoregressive model, in order to capture heterogeneous types of market participants and approximate long-memory of volatility. This has been an incredibly successful model, which is by now ``the benchmark,'' to a large extent occupying the former role of the GARCH(1,1) \citep*[e.g.][]{hansen-lunde:05b}. However, because the HAR model is so parsimonious, the complicated structure of the underlying data renders it inadequate in some directions. Therefore, a number of extensions of the baseline HAR have been proposed \citep*[e.g.,][]{andersen-bollerslev-diebold:07a, corsi-reno:12a, patton-sheppard:15a, bollerslev-patton-quaedvlieg:16a}. These are, however, still based on past returns as conditioning information, which is surprising, since a large amount of research has documented an intimate relationship between news and volatility \citep*[e.g.,][]{schwert:89a, engle-ghysels-sohn:13a, bollerslev-li-xue:18a}.

The reason that no or only few additional covariates are typically included in volatility prediction is partly explained by the observation that past volatility is a very powerful predictor of its future self. That is, after all, how this literature evolved. However, it is also motivated by the fact that traditional models, often estimated with linear regression, break down when the explanatory variables are strongly correlated, exhibit low signal-to-noise ratios, or if the underlying structure is nonlinear. As explained above, this is a trademark of financial markets, so it is unsurprising that relatively few studies conduct volatility forecasting in a data-rich environment. The recent access to large datasets therefore highlights a need for investigating more powerful tools that take these problems into account.

As emphasized by \citet*{varian:14a}, machine learning (ML) techniques overcome these deficiencies. Up to now, however, only a few studies combine ML and volatility forecasting using a single approach, e.g., \citet*{audrino-knaus:16a, audrino-sigrist-ballinari:20a, caporin-poli:17a} investigate lasso, \citet*{luong-dokuchaev:18a} use random forests, \citet*{mittnik-robinzonov-spindler:15a} apply component-wise gradient boosting\glsadd{Boosting}, \citet*{bucci:20a, donaldson-kamstra:97a, hillebrand-medeiros:10a, fernandes-medeiros-scharth:14a, rahimikia-poon:20a} explore neural networks. In related work, \citet*{carr-wu-zhang:20a} study the application of ML to low-frequency options data in order to forecast 30-day realized variance of the S\&P 500 index.

In contrast, we conduct a comprehensive analysis of the out-of-sample performance of multiple ML techniques for volatility forecasting, i.e. linear regularized models\glsadd{Regularization} (ridge, lasso, elastic net), tree-based algorithms (bagging, random forest, gradient boosting), and neural networks, when a reasonably large number of additional controls appear in the information set. In such a setting, regularization is helpful to deal with the high-dimensionality of the problem and get less noisy parameter estimates. Furthermore, tree-based regression allows for estimation without imposing any distributional or functional form on the model, while neural networks can regularize and capture nonlinearities. We compare these rather different ML approaches to the HAR lineage. We aim not only to do a comparison of HAR models and ML methods but also to provide evidence of how and why ML improves the accuracy of forecasting volatility. We adopt a conservative approach to implement ML in that we do minimal hyperparameter tuning to assess how good the techniques are when implemented ``as is''.

The findings are five-fold. Firstly, we show superior out-of-sample forecasting with off-the-shelf implementations of ML compared to HAR. A \citet*{diebold-mariano:95a} test for equal predictive accuracy shows that these gains are often viewed as statistically significant. Secondly, we document that substantial incremental information about future volatility can be extracted with ML from additional volatility predictors. In contrast, the HAR models yield only minor improvements or outright deteriorate in forecast accuracy, unless regularization is applied to account for overfitting. Thirdly, we compute variable importance via accumulated local effects (ALE), which shows a general agreement about the set of the most dominant predictors, but disagreement on their ranking. The ALE also reveal a nonlinear interaction between covariates. Fourthly, we show that the forecast gains are stronger at longer horizons, which we attribute to ML techniques better capturing an underlying long-memory behavior in the realized variance. Fifthly, we conduct a Value-at-Risk application. We show that ML continues to excel at forecasting in this setting, although the differences are smaller and less statistically significant. Moreover, ML produces correct unconditional and conditional coverage.

The paper is organized as follows. Section \ref{Methodology} sets the theoretical foundation and introduces the various models applied in the paper. Section \ref{section:data-description} describes the high-frequency data used in the empirical analysis. Section \ref{Empirical Results} features an out-of-sample one-day-ahead forecast comparison and conducts a series of robustness checks. In Section \ref{section:week-month}, we examine the weekly and monthly horizon. In Section \ref{section:value-at-risk}, we perform an economic application by forecasting Value-at-Risk. We conclude in Section \ref{section:conclusion}. In the Appendix, we present supplemental information, including a glossary that can assist readers who are not acquainted with the jargon of ML.

\section{Methodology} \label{Methodology}

\subsection{The setting} \label{Realized Volatility and Bi-power Volatility}
We assume the log-price $X = \left(X_{t} \right)_{t \geq0}$ is supported by a filtered probability space $\big( \Omega, ( \mathcal{F}_{t})_{t \geq 0}, \mathcal{F}, \mathbb{P} \big)$. If the price is determined in an arbitrage-free frictionless market, then $X$ is a semimartingale process, see \citet*{delbaen-schachermayer:94a}. We assume $X_{t}$ can be represented as follows:
\begin{equation}
X_{t} = X_{0} + \int_{0}^{t} \mu_{s}ds + \int_{0}^{t} \sigma_{s}dW_{s} + \sum_{s=1}^{N_{t}}J_{s}, \quad t \geq 0, \label{eq: Price_Process}
\end{equation}
where $X_{0}$ is $\mathcal{F}_{0}$-measurable, $\mu = (\mu_{t})_{t \geq0}$ denotes the drift, $\sigma = ( \sigma)_{t \geq 0}$ is the stochastic volatility process, $W = (W_t)_{t \geq 0}$ is a standard Brownian motion, $N = (N_{t})_{t \geq 0}$ is a counting process, which represents the number of jumps in $X$, and $J = (J_{s})_{s=1, \dots,N_{t}}$ is a sequence of nonzero random variables of jump sizes with corresponding jump times $\tau = ( \tau_{s})_{s=1, \dots,N_{t}}$.
	
Our aim is to forecast the daily quadratic variation, which is $QV_{t} = \int_{t-1}^{t} \sigma_{s}^{2}ds + \sum_{t-1 \leq \tau_{s} \leq t} J_{s}^{2}$, for $t = 1, \ldots, T$, where $t$ is the day and $T$ is the total number of days in the sample. The quadratic variation is not directly observable, since in practice only discretely sampled measurements of $X$ are available. An estimator of QV is given by the realized variance:
\begin{equation}
RV_{t} = \sum_{j=1}^{n} | \Delta_{t-1,j}^{n} X|^{2},
\end{equation}
where $n$ is the number of intraday logarithmic returns, $\Delta_{t-1,j}^{n}X = X_{t-1+\frac{j}{n}} - X_{t-1 + \frac{j-1}{n}}$. Then, as shown by \citet*{andersen-bollerslev:98a, barndorff-nielsen-shephard:02a}, realized variance is a consistent estimator of the quadratic variation as the sampling frequency increases, i.e. $RV_{t} \overset{ \mathbb{P}}{ \longrightarrow} QV_{t}$ as $n \rightarrow \infty$.

Later on, we also employ a decomposition, where realized variance is broken into its positive and negative semivariance part \citep*[e.g.][]{barndorff-nielsen-kinnebrock-shephard:10a}:
\begin{equation}
RV_{t} \equiv RV_{t}^{+} + RV_{t}^{-},
\end{equation}
where
\begin{equation}
RV_{t}^{+} = \sum_{j \ : \ \Delta_{t-1,j}^{n} X > 0} | \Delta_{t-1,j}^{n} X|^{2} \quad \text{and} \quad RV_{t}^{-} = \sum_{j \ : \ \Delta_{t-1,j}^{n} X < 0} | \Delta_{t-1,j}^{n} X|^{2}.
\end{equation}

\subsection{The HAR model and its extensions} \label{section:har}

The parsimonious structure of the HAR model of \citet*{corsi:09a} makes it a default choice in the literature. Therefore, the HAR model is the benchmark throughout this study, ensuring comparability with earlier work.

The HAR is defined as:
\begin{equation} \label{equation:har}
RV_{t} = \beta_{0} + \beta_{1}RV_{t-1} + \beta_{2}RV_{t-1 \mid t-5} + \beta_{3}RV_{t-1 \mid t-22} + u_{t},
\end{equation}
where $RV_{t-1 \mid t-h}=\frac{1}{h} \sum_{i=1}^{h}RV_{t-i}$. The explanatory variables are the daily, weekly, and monthly averages of lagged realized variance.

After its inception, a large number of extensions of the basic HAR has been developed. To ensure our results are robust against a broader suite of recent HAR-type models, we include several of the more popular ones as ``horses'' in the race. In particular, to impose a nonlinear relationship between present and future volatility and attenuate the influence of outliers, \citet*{corsi:09a} also proposed the log-version of the HAR (LogHAR):
\begin{equation}
\log(RV_{t}) = \beta_{0} + \beta_{1}\log(RV_{t-1}) + \beta_{2}\log(RV_{t-1 \mid t-5}) + \beta_{3}\log(RV_{t-1 \mid t-22}) + u_{t}.
\end{equation}
To capture a leverage effect, \citet*{corsi-reno:12a} suggested to include past aggregated negative returns into the HAR model, which is dubbed LevHAR:
\begin{equation}
RV_{t} = \beta_{0} + \beta_{1}RV_{t-1} + \beta_{2}RV_{t-1 \mid t-5} + \beta_{3}RV_{t-1 \mid t-22} +\gamma_{1}r_{t-1}^{-} + \gamma_{2}r_{t-1\left|t-5\right .}^{-}  + \gamma_{3}r_{t-1\left|t-22\right.}^{-} + u_{t},
\end{equation}
where
\begin{equation}
r_{t-1 \mid t-h} = \frac{1}{h} \sum_{i=1}^{h} r_{t-i},
\end{equation}
and $r_{t-1 \mid t-h}^{-} = \min(0, r_{t-1 \mid t-h})$.

An alternative way to model an asymmetric impact of past positive and negative returns on future volatility is given by the semivariance HAR (SHAR) of \citet*{patton-sheppard:15a}:
\begin{equation}
RV_{t} = \beta_{0} + \beta_{1}^{-}RV_{t-1}^{-} + \beta_{1}^{+}RV_{t-1}^{+} + \beta_{2}RV_{t-1 \mid t-5} + \beta_{3}RV_{t-1 \mid t-22} + u_{t}.
\end{equation}
We also add the HARQ by \citet*{bollerslev-patton-quaedvlieg:16a}, which corrects an inherent error-in-variables problem in the HAR that arises because realized variance is a generated regressor. The HARQ is defined as:
\begin{equation} \label{equation:harq}
RV_{t} = \beta_{0} + \left( \beta_{1}+ \beta_{1Q} RQ_{t-1}^{1/2} \right) RV_{t-1} + \beta_{2}RV_{t-1 \mid t-5} + \beta_{3}RV_{t-1 \mid t-22} + u_{t},
\end{equation}
where
\begin{equation}
RQ_{t} = \frac{n}{3} \sum_{j = 1}^{n} | \Delta_{t-1,j}^{n} X|^{4},
\end{equation}
is the realized quarticity.

As readily seen, the formula for a general HAR model (except the LogHAR) is:
\begin{equation}
RV_{t} = \beta_{0} + \beta'Z_{t-1}+u_{t},
\end{equation}
where $Z_{t-1}$ is the information set, $\beta = ( \beta_{1}, \ldots, \beta_{J})'$ are the slope parameters, and $J$ is the number of explanatory variables. The HAR model has $Z_{t-1} = (RV_{t-1}, RV_{t-1 \mid t-5}, RV_{t-1 \mid t-22})$, whereas the HARQ has $Z_{t-1} = (RV_{t-1}, RQ_{t-1}^{1/2}RV_{t-1}, RV_{t-1 \mid t-5}, RV_{t-1 \mid t-22})$.

In general, $Z_{t-1}$ can include covariates that are not functions of past returns. We inspect such a ``big data'' setting. A complete list of extra explanatory variables included in this paper is available in Table \ref{table:covariate-list}. To ensure comparability with ML, we therefore adopt a distributed lag-type version of each HAR model to allow for a broader selection of variables to be included as conditioning information. We denote by HAR-X the extended version of the basic HAR. However, to keep notation down the other HAR models are still denoted with their original label even in the extended setting.

While the above list is comprehensive, it is necessarily incomplete. We do not include HAR models, where the quadratic variation is split into a continuous and jump part \citep*[e.g.][]{andersen-bollerslev-diebold:07a}.

In order to estimate the coefficients of the HAR models, we employ a least squares approach with objective function:
\begin{equation} \label{eq: loss}
(\hat{ \beta}_{0}, \hat{ \beta}) = \argmin_{ \beta_{0}, \beta} \mathcal{L}( \beta_{0}, \beta) = \argmin_{ \beta_{0}, \beta} \sum_{t \in \text{in-sample}} \big(RV_{t+1}- \beta_{0} - f(Z_{t}; \beta) \big)^{2},
\end{equation}
where $f(Z_{t}; \beta) = \beta'{Z}_{t}$ and in-sample is defined later.

\subsection{Regularization} \label{section:regularization}

When increasing the number of predictors in a setting with a low signal-to-noise ratio, the linear models at some point start to fit noise rather than relevant information, also known as overfitting. A common way to avoid overfitting and increase out-of-sample performance is to shrink\glsadd{Shrinkage} the regression coefficients by imposing a penalty term.

We define the penalized loss function:
\begin{equation}
\tilde{\mathcal{L}}(\beta_{0}, \beta;\theta)= \mathcal{L}(\beta_{0}, \beta)+ \phi( \beta; \theta),
\end{equation}
where $\phi( \beta; \theta)$ is a penalty term, and $\theta$ is a vector of hyperparameters\glsadd{Hyperparameters}, which are always determined in the validation set\glsadd{Validation Set}.\footnote{A thorough explanation of the validation procedure can be found in Appendix \ref{appendix:hyperparameter}.}

In this paper, we estimate three of the most widely used penalization methods, i.e. ridge regression (RR) by \citet*{hoerl-kennard:70a} and lasso (LA) by \citet*{tibshirani:96a}, where the penalty term is an $L^{2}$ and $L^{1}$ norm, and elastic net (EN) by \citet*{zou-hastie:05a}, which is a convex combination of ridge and lasso.

In ridge regression, the penalty term is given by:
\begin{equation}
\phi( \beta; \lambda) = \lambda \sum_{i=1}^{J} \beta_{i}^{2},
\end{equation}
where $\lambda \geq 0$ controls the amount of shrinkage. In the implementation, we choose a range for $\lambda$ that is effectively broad enough to include both the unregularized model and the constant model as potential solutions.

If dealing with a large feature space\glsadd{Feature}, shrinking is not always sufficient, and subset selection\glsadd{Subset selection} is preferable. Lasso addresses this with a penalty term of the form:
\begin{equation} \label{equation:lasso}
\phi(\beta; \lambda) = \lambda \sum_{i=1}^{J} | \beta_{i}|.
\end{equation}
Due to the geometry, lasso often forces coefficient estimates to zero and generates sparsity\glsadd{Sparsity} and subset selection.
The cost of this is that there is no closed-form solution to the problem, which makes it necessary to conduct numerical optimization.

The final penalization approach is the elastic net:
\begin{equation}
\phi(\beta; \lambda, \alpha) = \lambda \left( \alpha \sum_{i=1}^{J} \beta_{i}^{2} + (1- \alpha) \sum_{i=1}^{J} \left| \beta_{i} \right| \right),
\end{equation}
where $\alpha \in [0,1]$.\footnote{The elastic net nests lasso for $\alpha = 0$ and ridge for $\alpha = 1$.}

To gauge the effect of subset selection as a tool to pre-filter the data, we also include a post lasso HAR (P-LA), in which the penalization in \eqref{equation:lasso} is applied to the HAR model in a first-stage regression, and then only the variables with a non-zero slope coefficient estimate are included in an unrestricted second-stage HAR regression.

At last, following \citet*{zou:06a} we also estimate an adaptive lasso HAR model (A-LA). Here, the two-stage procedure of the post lasso are basically reversed. Firstly, an unrestricted HAR is estimated to get estimates of the slope parameters, say $\hat{ \beta}^{\text{1st}}$. Secondly, a lasso is run with a modified penalty term linked to the reciprocal of the elements in $\hat{ \beta}^{\text{1st}}$:
\begin{equation}
\phi( \beta; \lambda) = \lambda \sum_{i=1}^{J} \hat{w}_{i} | \beta_{i}|,
\end{equation}
where $\hat{w}_{i}= 1/ | \hat{\beta_{i}}^{\text{1st}}|$. Hence, a larger magnitude of the beta coefficient from the first stage leads to it being downweighted in the penalization during the second stage.\footnote{All variables are standardized prior to estimation to make the slope parameters scale invariant.}

\subsection{Tree-based regression} \label{section:tree-based}

A concern with linear models is that it is left to the researcher to impose the true association among predictors and the response variable. In contrast, a regression tree is a fully nonparametric approach, which not only allows for nonlinearity, but it also implicitly accounts for interaction effects between explanatory variables.

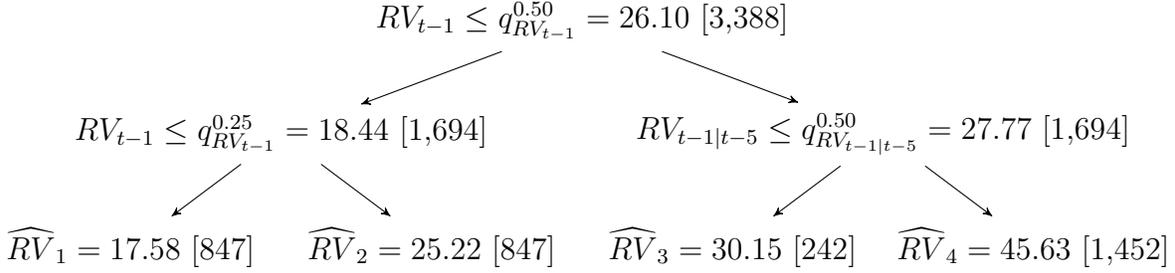
\begin{figure}[t!]
\begin{center}
\caption{Illustration of a regression tree. \label{figure:regression-tree}}	
\begin{tikzpicture}[->,>=stealth',level/.style = {level 1/.style={sibling distance = 8cm}, level 2/.style={sibling distance = 4cm}}, every node/.style = {align=center}]
\node{$RV_{t-1} \leq q_{RV_{t-1}}^{0.50} = 26.10$ [3,388]}
child{node{$RV_{t-1} \leq q_{RV_{t-1}}^{0.25} = 18.44$ [1,694]}
child{node{$\widehat{RV}_{1} = 17.58$ [847]}}
child{node{$\widehat{RV}_{2} = 25.22$ [847]}}
}
child{node{$RV_{t-1 \mid t-5} \leq q_{RV_{t-1 \mid t-5}}^{0.50} = 27.77$ [1,694]}
child{node{$\widehat{RV}_{3} = 30.15$ [242]}}
child{node{$\widehat{RV}_{4} = 45.63$ [1,452]}}
} 	
;
\end{tikzpicture}
\begin{scriptsize}
\parbox{\textwidth}{\emph{Note.} The figure shows an example of a possible regression tree constructed with the actual values of $RV_{t-1}$, $RV_{t-1 \mid t-5}$, and $RV_{t-1 \mid t-22}$ from the training and validation set of Apple (AAPL) high-frequency data in our empirical application. The variables are expressed as annualized standard deviation in percent for ease of interpretation. We arbitrarily set a stopping criteria to a maximum number of terminal nodes $M = 4$ and select the split points conveniently as the 1st and 2nd quartile ($q^{0.25}$ and $q^{0.50}$) of the associated variable for the purpose of the illustration. The numbers in square brackets are the total data points retained at each node, as we move through the tree, while $\widehat{RV}_{m}$, $m = 1, \dots, 4$, corresponds to the one-day-ahead prediction of $RV_{t}$ at each terminal node. Note the uneven distribution of data in the right-hand side of the tree is caused by using the overall median of $RV_{t-1 \mid t-5}$ as a split point.}
\end{scriptsize}
\end{center}
\end{figure}

A tree-based regression is, as the name suggests, based on a decision tree. A tree is grown by partitioning the domain (or feature space in ML parlance)\glsadd{Feature} of the explanatory variables $Z_{t}$, $\text{dom}(Z_{t}) \subseteq \mathbb{R}^{J}$, into smaller and smaller rectangular subspaces as we move through the tree (an illustrative example is provided in Figure \ref{figure:regression-tree}). This process is continued until a stopping criterion is reached. At the end of the tree, a constant prediction is assigned to all observations that fall in a given terminal node (also called a leaf).\glsadd{Node Regression tree}

Suppose that after constructing the tree there are $M$ terminal nodes. They correspond to a sequence of disjoint rectangles $R_{m}$, for $m = 1, \ldots, M$, such that $R_{m} \subseteq \text{dom}(Z_{t})$ and $\bigcup_{m=1}^{M} R_{m} = \text{dom}(Z_{t})$. The tree-based regression then predicts as follows:
\begin{equation}
\hat{f}(Z_{t}) = \sum_{m=1}^{M} \widehat{RV}_{m} 1_{ \{Z_{t} \in R_{m} \}},
\end{equation}
where $\widehat{RV}_{m}$ is a constant.

To ensure internal consistency, we adopt the minimum sum of squares criterion for selecting $\widehat{RV}_{m}$, i.e. $\widehat{RV}_{m} = \text{average}(RV_{t+1} \mid Z_{t} \in R_{m})$.

In essence, tree-based regression poses a combinatorial problem. There is no closed-form solution to find the best order in which to sort and split variables, and it may be too time-consuming to search over all possible choices. Hence, the need for a short-sighted, greedy algorithm arises. A popular approach to this problem is to start at the top node (called the root) and independently optimize nearby nodes in a stepwise fashion, known as the Classification And Regression Tree (CART). The CART, however, often yields inferior out-of-sample performance due to high variance from overfitting sample noise.

\citet*{breiman:96a} proposed bagging (bootstrap aggregation, BG) to refine the decision tree. In bagging, we grow several trees by resampling the original data. We subsequently predict as follows:
\begin{equation}
\hat{f}(Z_{t}) = \frac{1}{B} \sum_{b=1}^{B} \hat{f}^{b}(Z_{t}),
\end{equation}
where $\hat{f}^{b}(Z_{t})$ is the tree-based prediction from the $b$th bootstrap sample.

The serial correlation in volatility suggests a block bootstrap may be appropriate. However, in our experience an i.i.d. bootstrap does not impair the recursive segmentation of the input space compared to more refined resampling.\footnote{As a verification, we applied a block bootstrap with $B$ random samples from a given dataset $\mathcal{M}$, resulting in a sequence $\mathcal{M}^{b}$, for $b = 1, \ldots, B$, where the block length $l$ and the number of blocks in each bootstrap sample $k$ was chosen such that $T = kl$. $l$ was found in the validation set\glsadd{Validation Set}. We estimated a regression tree for each bootstrap and computed the prediction function as $\hat{f}(Z_{t}) = \frac{1}{B} \sum_{b=1}^{B} \hat{f}^{b}(Z_{t})$. There was no discernible change in the results compared to the i.i.d. bootstrap.}

The benefit of bagging compared to CART is related to the reduced correlation between trees. Therefore, an alternative approach building on bagging is random forest (RF), see \citet*{breiman:01a}. Here, attention is restricted to a random subset of input features before finding the best split point. In this way a random forest is able to de-correlate the trees further. In particular, $B$ trees $\big\{ f(Z^{b}_{t}, \theta^{b}) \big\}_{b=1}^{B}$ are grown, where $\theta^{b}$ summarizes the $b$th random forest tree in terms of split variables, split points, and values at the terminal nodes. The prediction is then:
\begin{equation}
\hat{f}(Z_{t}) = \frac{1}{B} \sum_{b=1}^{B} \hat{f}^{b}(Z_{t}, \theta^{b}).
\end{equation}
To reduce the complexity of the random forest and bagging, we set all tuning parameters equal to their default values from the original Fortran code by \citet*{breiman-cutler:04a}. This sidesteps validation and facilitates reproduction of our findings.

Another tree-based approach is gradient boosting (GB)\glsadd{Boosting} by \citet*{friedman:01a}.\footnote{He extends the original boosting algorithm from \citet*{schapire:90a} and \citet*{freund:95a}.} Gradient boosting produces a sequential model based on weak learners\glsadd{Weak learner}, where each of $B$ trees is grown on information from the previous one. Firstly, $\hat{f}^{0}(Z_{t})$ is initialized as a constant determined by $\hat{f}^{0}(Z_{t}) = \argmin_{ \beta_{0}} \sum_{t \in \text{in-sample}} (RV_{t} - \beta_{0})^{2}$. With the MSE loss function, this constant is the sample average. Secondly, the negative gradient of the loss function with respect to the prediction is calculated. With the MSE loss function, this corresponds to the residuals. A shallow tree is then fitted to the residuals, yielding a set of terminal nodes $R_{jb}$, $j = 1, \ldots, J_{b}$, where $j$ denotes the leaf and $b$ the tree. This is followed by choosing a gradient descent size as $\rho^{jb} = \argmin_{ \beta_{0}} \sum_{Z_{t} \in R_{jb}} (RV_{t} - \beta_{0} - \hat{f}^{b-1}(Z_{t}))^{2}$, and in the last step $\hat{f}^b(Z_{t})$ is updated iteratively as follows:
\begin{equation}
\hat{f}^{b}(Z_{t}) = \hat{f}^{b-1}(Z_{t}) + \nu \sum_{j=1}^{J_{b}} \rho^{jb}1_{ \{Z_{t} \in R_{jb} \}},
\end{equation}
for $b = 1, \ldots, B$, where $\nu$ is a learning rate\glsadd{Learning Rate} and $\hat{f}(Z_{t}) \equiv \hat{f}^{B}(Z_{t})$ the final prediction.

As explained by \citet*{deprado:18a}, gradient boosting attempts to address an underfitting problem, whereas random forest deals with overfitting (generally perceived to be a greater evil in finance). As a consequence, gradient boosting places substantially more weight on misclassification and is therefore susceptible to outliers. With the typical heavy tails in the volatility distribution, we expect this to lead to poor results for gradient boosting in our framework.

\subsection{Neural network} \label{section:neural-network}

To ensure a complete investigation of ML space, at last we apply the artificial neural network (NN). With their highly flexible and nonlinear structure, neural networks show astonishing performance in handling complex problems. However, with a broad set of hyperparameters\glsadd{Hyperparameters} the replicability of the neural network is frequently questioned. Therefore, we construct the neural network as streamlined as possible and set the majority of the hyperparameters\glsadd{Hyperparameters} to standard suggestions from the literature. This is presented in Appendix \ref{appendix:hyperparameter}.

We briefly outline the mechanics of a so-called feed-forward network, while a more comprehensive overview is available in, e.g., \citet*{goodfellow-bengio-courville:16a}. A neural network is constructed from an input layer\glsadd{Input layer}, hidden layers\glsadd{Hidden layer}, and an output layer\glsadd{Output layer}. Suppose the total number of such layers is $L$. In the first layer, the neural network receives an input $Z_{t}$. The data is transformed through one or more hidden layers using an activation function\glsadd{Activation function} $g$.

The general equation for the $l$-th layer is denoted as:
\begin{equation} \label{eq: NN_layer}
a_{t}^{ \theta_{l+1},b_{l+1}} = g_{l} \left( \sum_{j=1}^{N_{l}} \theta_{j}^{(l)}a_{t}^{ \theta_{l},b_{l}} + b^{(l)} \right), \quad 1 \leq l \leq L,
\end{equation}
where $g_{l}$ is the activation function, $\theta^{(l)}$ is the weight matrix, $b^{(l)}$ is the error serving as an activation threshold for the neurons\glsadd{Nodes Neural network} in the next layer, $N_{l}$ denotes the number of hidden neurons, and $a_{t}^{ \theta_{l+1},b_{l+1}}$ is the prediction.\footnote{By definition, the prediction from the input layer is $a_{t}^{ \theta_{2}, b_{2}} = g_{1} \Big( \sum_{j=1}^{N_{1}} \theta_{j}^{(1)}Z_{t} + b^{(1)} \Big)$.}

The output layer constructs a forecast:
\begin{equation}
\hat{f} \left(Z_{t} \right) = \left(g_{L} \circ \cdots \circ g_{1} \right)(Z_{t}). \label{eq: NN_2}
\end{equation}
As seen, a neural network applies a series of functional transformations to construct the forecast. In principle, a single hidden layer with a large enough number of neurons (and appropriate activation function) is sufficient to approximate any continuous function. This follows from the Universal Approximation Theorem of \citet*{cybenko:89a}. In practice, however, it is often convenient and computationally more efficient to add extra hidden layers than to arbitrarily increase the number of neurons in a layer. Hence, the optimal architecture of a neural network depends on the problem and is determined through tuning.

To allow inspection of the inner workings of a neural network, we construct four models with an architecture inspired by the geometric pyramid \citep*[see, e.g.,][]{masters:93a, gu-kelly-xiu:20a}. The first, denoted NN$_{1}$, has a shallow structure with a single hidden layer and 2 neurons. NN$_{2}$ is two-layered with 4 and 2 neurons, NN$_{3}$ has three layers with 8, 4, and 2 neurons, and NN$_{4}$ has four layers with 16, 8, 4, and 2 neurons.

We apply the Leaky Rectified Linear Unit (L-ReLU) by \citet*{maas-hannun-ng:13a} as activation function for all layers, as it is able to infer when the activation is zero via gradient-based methods, in contrast to the standard ReLU.\footnote{Other common activation functions include Sigmoid, TanH, or ReLU. Switching to one of these has no material impact on our empirical results.} The L-ReLU is defined as:
\begin{equation}
\text{L-ReLU}(x) = \begin{cases}
cx, & \text{if }x<0, \\
x, & \text{otherwise,}
\end{cases}
\end{equation}
\noindent
where $c \geq 0$.\footnote{In this paper, $x = \sum_{j=1}^{N_{l}} \theta_{j}^{(l)}a_{t}^{ \theta_{l},b_{l}} + b^{(l)}$. We follow the majority of the literature by setting $c = 0.01$.}

The weight parameters are chosen to minimize the squared error loss function.\footnote{Alternatively, a penalized $L^{2}$ objective function can be applied. We found no additional improvement when combining $L^{2}$ and drop-out.} In lack of a closed-form solution, we adopt the Adaptive Moment Estimation (Adam\glsadd{Adam}) by \citet*{kingma-ba:14a}, where the optimizer updates the weights $( \hat{ \theta}, \hat{b})$ as a combination of Momentum\glsadd{Momentum} and RMSprop\glsadd{RMSprop} (Root Mean Square propagation).\footnote{To limit the number of hyperparameters and reduce the complexity of the optimization, we select exponential decay rates in line with \citet*{kingma-ba:14a}.} We further predetermine the learning rate\glsadd{Learning Rate} at 0.001, see \citet*{kingma-ba:14a}.

Regularization of a neural network is crucial to avoid overfitting. There is a large literature proposing different regularization techniques. In this study, we employ four of the most commonly used, namely learning rate shrinkage, drop-out, early stopping, and ensembles. Ensembling is a model averaging technique, in which numerous neural networks are fitted, and the prediction is constructed from a subset of the best performing models from the validation set. The idea is to reduce the variation in the forecast caused by the random initial weights assigned to a neural network by repeating the experiment several times, much as an ensemble average. We investigate ensembles of 1---corresponding to no averaging---and 10 (out of 100), which we denote NN$_{1}^{1}$ and NN$_{1}^{10}$ for the single hidden layer neural network, and so forth. In Appendix \ref{appendix:hyperparameter} -- \ref{appendix:regularization}, we explain the regularization techniques and choice of hyperparameters in more detail.

\tikzset{%
	every neuron/.style={
		circle,
		draw,
		minimum size=0.5cm
	},
	neuron missing/.style={
		draw=none,
		scale=2,
		text height=0.2cm,
		execute at begin node=\color{black}$\vdots$
	},
}
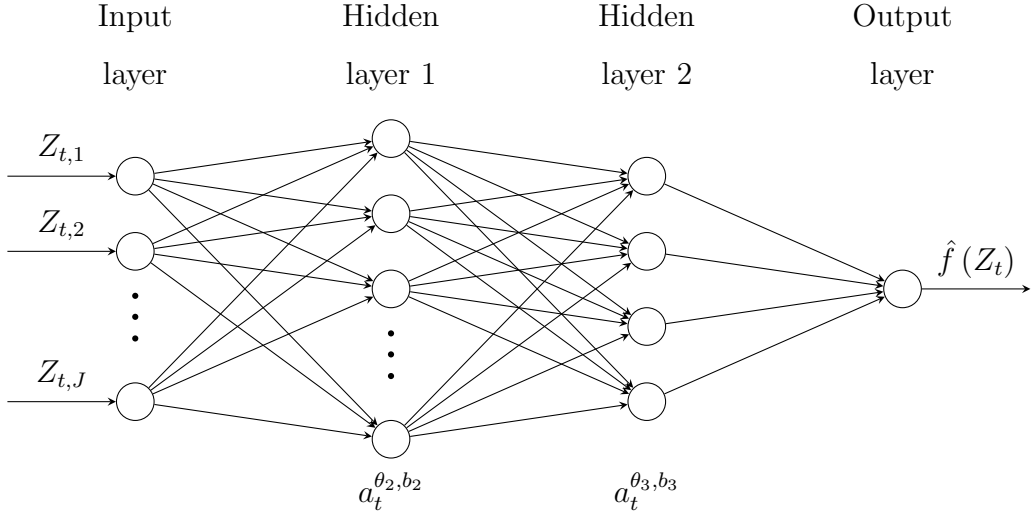
\begin{figure}[t!]
	\begin{center}
		\caption{Feed-forward neural network. \label{figure:reg_NN2}}	
		
		\begin{tikzpicture}[x=1.7cm, y=1cm, >=stealth]
		
		\foreach \m [count=\y] in {1,2,missing,3}
		\node [every neuron/.try, neuron \m/.try] (input-\m) at (0,1-\y) {};
		
		\foreach \m [count=\y] in {1,2,3,missing,4}
		\node [every neuron/.try, neuron \m/.try] (hidden1-\m) at (2,1.5-\y*1) {};
		
		\foreach \m [count=\y] in {1,2,3,4}
		\node [every neuron/.try, neuron \m/.try ] (hidden2-\m) at (4,1-\y) {};
		\foreach \m [count=\y] in {1}
		\node [every neuron/.try, neuron \m/.try ] (output-\m) at (6,-0.5-\y) {};
		
		\foreach \l [count=\i] in {1,2,J}
		\draw [<-] (input-\i) -- ++(-1,0)
		node [above, midway] {$Z_{t, \l}$};
		
		\foreach \i in {1,...,3}
		\foreach \j in {1,...,4}
		\draw [->] (input-\i) -- (hidden1-\j);
		\foreach \i in {1,...,4}
		\foreach \j in {1,...,4}
		\draw [->] (hidden1-\i) -- (hidden2-\j);
		
		\foreach \i in {1,...,4}
		\foreach \j in {1,...,1}
		\draw [->] (hidden2-\i) -- (output-\j);
		
		\foreach \l [count=\i] in {1}
		\draw [->] (output-\i) -- ++(1,0)
		node [above, midway] {$\hat{f} \left(Z_{t} \right)$};
		
		\foreach \l [count=\x from 0] in {Input}
		\node [align=center, above] at (\x,1) {\l \\ layer};
		\foreach \l [count=\x from 0] in {Hidden}
		\node [align=center, above] at (\x+2,1) {\l \\ layer 1};
		\foreach \l [count=\x from 0] in {Hidden}
		\node [align=center, above] at (\x+4,1) {\l \\ layer 2};
		\foreach \l [count=\x from 0] in {Output}
		\node [align=center, above] at (\x+6,1) {\l \\ layer};
		
		\foreach \l [count=\x from 0] in {$a_{t}^{ \theta_{2}, b_{2}}$}
		\node [align=center, below] at (\x+2,-3.8) {\l \\ };
		\foreach \l [count=\x from 0] in {$a_{t}^{ \theta_{3}, b_{3}}$}
		\node [align=center, below] at (\x+4,-3.8) {\l \\ };
	\end{tikzpicture}
\begin{scriptsize}
\parbox{\textwidth}{\emph{Note.} We illustrate a two-layered feed-forward neural network with an input layer, a pair of hidden layers, and the output layer. The input layer receives the raw features $Z_{t}$ with $J$ explanatory variables. Each hidden layer computes an affine transformation and applies an element-wise activation function\glsadd{Activation function}. The output layer returns the prediction. The arrows denote the direction of the information flow throughout the network during forward propagation.}
\end{scriptsize}
\end{center}
\end{figure}	

\subsection{Forecast comparison} \label{section:forecast-comparison}

As the mean squared error (MSE) laid the foundation for the optimal structure of the ML algorithms, a logical continuation is to employ it also as an out-of-sample evaluation measure:
\begin{equation} \label{eq: loss_MSE}
\mathcal{L} ( \hat{f}(Z_{t})) = \sum_{t \in \text{out-of-sample}} \big(RV_{t+1} - \hat{f}(Z_{t}) \big)^{2},
\end{equation}
where $\hat{f}(Z_{t})$ is the forecast.\footnote{The LogHAR produces a forecast $\hat{f}(Z_{t})$ of log-realized variance, but we require a forecast of realized variance $\exp \big( \hat{f}(Z_{t}) \big)$. This is a nonlinear transformation, so the forecast of realized variance is biased by Jensen's inequality, even if the forecast of log-realized variance is not. We bias-correct as follows: $E\left[ \exp \big( \hat{f}(Z_{t}) \big) \right] = \exp \left(E \big[ \hat{f}(Z_{t}) \big] + 0.5 \text{var} \big[ \hat{f}(Z_{t}) \big] \right)$, where $\text{var} \big( \hat{f}(Z_{t}) \big)$ is the variance of the residuals in the training and validation set. This is suitable when the distribution of log-realized variance is Gaussian, which is approximately true in practice, see, e.g., \citet*{andersen-bollerslev-diebold-ebens:01a, christensen-thyrsgaard-veliyev:19a}.}

If a model predicts volatility to be negative, we replace the forecast with the minimum in-sample realized variance $\min_{t \in \text{in-sample}} RV_{t}$. We also adopt the insanity filter from \citet*{bollerslev-patton-quaedvlieg:16a} for the HARQ.

To measure statistical significance, we compute a pairwise \citet*{diebold-mariano:95a} test for equal predictive accuracy with a one-sided alternative that the comparison model beats the benchmark model, as explained later. We also construct a Model Confidence Set (MCS) of \citet*{hansen-lunde-nason:11a}. It defines a collection of models containing the ``best'' one with a given level of confidence. Inferior forecasting models are removed via an elimination rule.

\section{Data description} \label{section:data-description}

The empirical investigation is based on high-frequency data from 29 of the 30 Dow Jones Industrial Average (DJIA) constituents prior to the recomposition on August 31, 2020. The complete list of ticker symbols is AAPL, AXP, BA, CAT, CSCO, CVX, DIS, DOW, GE, GS, HD, IBM, INTC, JNJ, JPM, KO, MCD, MMM, MRK, MSFT, NKE, PFE, PG, RTX, TRV, UNH, VZ, WMT, XOM.\footnote{We also ran the forecast comparison for the composite DJIA index and the SPY. The latter is a liquid ETF tracking the S\&P 500. The outcome of this analysis confirms the findings reported for individual equities, suggesting our conclusions hold more broadly on a market-wide level. Moreover, we attempted to forecast volatility of the DJIA index based on cross-sectional information of the volatilities from the individual member firms. However, we did not find any improvement in forecast accuracy, once we controlled for past DJIA index volatilities.} Visa (V) is excluded due to limited data availability, as its shares were only listed for public trading on NYSE on March 19, 2008. We construct a continuous stock price record for each company on a sample from 29 January 2001 to 31 December 2017, or $T = 4{,}257$ trading days. The sample includes several noteworthy events, such as the financial crisis, the European sovereign debt crisis, several rounds of negotiations about the limit of the U.S. debt ceiling, and the flash crashes of 6 May 2010 and 24 August 2015.

There has been several corporate actions changing the ticker symbol history for some of the current members of DJIA. We include high-frequency information from the main predecessor prior to such events. In particular, we employ 1) Chevron (CHV) prior to its merger with Texaco on October 9, 2001 to forge the company ChevronTexaco (CVX), 2) St. Paul Travelers Companies (STA) prior to its name change to Travelers Inc. (TRV) on February 27, 2007, 3) United Technologies (UTX) prior to its replacement by Raytheon Technologies (RTX) on April 6, 2020 (Raytheon is the fusion of United Technologies and Raytheon Company, which merged as of April 3, 2020), 4) Walgreens (WAG) prior to forming Walgreens Boots Alliance (WBA) by purchasing a majority stake in Alliance Boots on December 31, 2014, and 5) DowDuPont (DWDP) before the spin-off of Dow Chemical Company (DOW) on April 2, 2019. DowDupont itself was the result of a previous merger between Dow Chemical Company and DuPont (DD) on September 1, 2017, and here we take DuPont as predecessor.

The data is extracted from the NYSE Trade and Quote (TAQ) database. Prior to our analysis, the data were pre-processed and filtered for outliers using standard algorithms, see \citet*{barndorff-nielsen-hansen-lunde-shephard:09a, christensen-oomen-podolskij:14a}. Based on the cleaned transaction price record, we construct a five-minute log-return series and compute for each asset a time series of daily realized variance $RV_{t}$ with $n = 78$, for $t = 1, \dots, 4{,}257$. The goal is to forecast realized variance.

We split the data into a training\glsadd{Training Set}, validation\glsadd{Validation Set}, and test set\glsadd{Test Set}. The primary analysis is restricted to a training set of 70\% of the dataset (or 2,964 days), a validation set of 10\% (or 424 days), and a test set of 20\% (or 847 days).\footnote{A month worth of high-frequency data (or 22 days) are ``lost'' due to the lagging.}  The latter are reserved for out-of-sample evaluation. We conduct a robustness check, where the training set is modified to 1,000 and 2,000 days (with a validation set fixed at 200 days). The results, which are broadly in line with our reported findings, are available in Appendix \ref{appendix:test-set}.

In the non-regularized HAR models, we merge the training and validation set and employ time-varying parameters by constructing a rolling window forecast. This approach is also adopted for bagging and random forest to avoid any hyperparameter tuning\glsadd{Hyperparameters}, which results in off-the-shelf implementation of these methods. However, tuning is almost unavoidable for ridge, lasso, elastic net, and gradient boosting. In keeping with the spirit of our paper, a rolling scheme without concatenation of the training and validation set is conducted. An in-depth explanation of our tuning of hyperparameters is available in Appendix \ref{appendix:hyperparameter}. The vast majority of the hyperparameters are set equal to their default values proposed in the original research papers, often based on very different contexts. At last, we employ a fixed window estimation for the neural networks, where the weights are only found once in the initial validation sample and not rolled forward in the out-of-sample window. This implementation is strictly advantageous to the other competing models, which are allowed to adapt to new information. It may be possible to further improve the performance of the neural networks with a rolling window approach, however this requires a very large computer capacity, which is outside our budget.

To study the effect of tagging on additional explanatory variables, we construct two datasets $\mathcal{M}_{ \text{HAR}}$ and $\mathcal{M}_{ \text{ALL}}$. $\mathcal{M}_{ \text{HAR}}$ contains the daily, weekly, and monthly lag of realized variance (RVD, RVW, and RVM).\footnote{$\mathcal{M}_{ \text{HAR}}$ also includes variables that are employed in the various extensions of the HAR model. However, these variables are only passed to the concrete model (i.e., the square-root of realized quarticity appears as an interaction term with realized variance in HARQ, but not in the other models).} This ensures one-to-one comparison between the HAR models and ML algorithms. $\mathcal{M}_{ \text{ALL}}$ extends $\mathcal{M}_{ \text{HAR}}$ by including additional variables examined in previous research. We select a total of nine other key variables that are arguably important predictors of future volatility. In particular, we extract four firm characteristics: Model-free implied volatility (IV),\footnote{The data was downloaded from the OptionMetrics' IvyDB database, which estimates the implied volatility for each traded option contract. We followed the data cleaning suggested by \citet*{lei-wang-yan:20a} to deal with erroneous data points.}\footnote{To account for the correlation between implied volatility and realized variance, we also estimated an auxiliary model: $IV_{t} = \beta_{0} + \beta_{1}RV_{t} + u_{t}$ and employed the residuals from the fitted regression in place of implied volatility. The results were roughly unchanged.} an indicator for earnings announcements (EA), 1-week momentum (M1W), and dollar trading volume (\$VOL). We include five macroeconomic indicators: The CBOE volatility (VIX) index, the Hang Seng stock index daily squared log-return (HSI), the \citet*{aruoba-diebold-scotti:09a} business conditions (ADS) index, the US 3-month T-bill rate (US3M) and the economic policy uncertainty (EPU) index from \citet*{baker-bloom-nicholas:16a}.\footnote{We also constructed an intermediate dataset via subset selection\glsadd{Subset selection}. In particular, we fitted a separate least squares regression for each combination of the feature\glsadd{Feature} space of $\mathcal{M}_{ \text{ALL}}$. The best model was then selected with a BIC criterion. It contains the daily and weekly lag of realized variance, implied volatility, 1-week momentum, and the earnings announcement indicator, i.e. a total of five variables. The conclusions from the forecast comparison on this dataset is consistent with what we report below. Interestingly, if the evaluation criterion is changed from BIC to cross-validation error, the subset selection includes only the standard three HAR variables.}  We first-difference US3M to account for possible nonstationarity, and we replace \$VOL by first-differencing its log-transform. Moreover, in the LogHAR we also log-transform VIX and IV.

\begin{sidewaystable}
\setlength{\tabcolsep}{0.05cm}
\begin{center}
\caption{List and summary statistics of explanatory variables. \label{table:covariate-list}}
\smallskip
\begin{tabular}{rcrcrcrcrcrcrcrc}
\hline\hline
\multicolumn{1}{l}{No.} & Acronym  & \multicolumn{2}{c}{Mean} & \multicolumn{2}{c}{Median} &  \multicolumn{2}{c}{Maximum} &  \multicolumn{2}{c}{Minimum} & \multicolumn{2}{c}{Standard deviation} & \multicolumn{2}{c}{Skewness} &\multicolumn{2}{c}{Kurtosis} \\ \hline
1 & RVD &    20.67 &  {\scriptsize [14.76,27.34]}  & 17.26 &  {\scriptsize [12.26,23.14]} & 176.99 & {\scriptsize [116.58,313.61]} & 5.32 & {\scriptsize [3.51,6.77]} & 12.31 & {\scriptsize [8.10,19.90]}& 3.34 & {\scriptsize [1.99,5.94]}& 26.23 &  {\scriptsize [9.91,98.50]}\\
2 & RVW &   21.14 & {\scriptsize [15.15,28.09]}  & 17.77 &  {\scriptsize [12.74,24.34]} & 116.55 & {\scriptsize [80.33,220.78]} & 7.40 & {\scriptsize [5.38,9.61]} & 11.50 & {\scriptsize [7.47,19.16]}& 2.85 & {\scriptsize [1.65,4.51]}& 16.22  & {\scriptsize [6.94,34.29]}\\
3 & RVM &    21.52 & {\scriptsize [15.48,28.71]}  & 18.04 &  {\scriptsize [12.86,25.35]} & 88.62 & {\scriptsize [56.31,152.59]} & 9.26 & {\scriptsize [7.25,11.99]} & 10.89 & {\scriptsize [6.96,18.44]}& 2.55 & {\scriptsize [1.53,3.71]}& 12.29 & {\scriptsize [6.09,24.98]}\\
4 & IV &   26.27 & {\scriptsize [19.50,37.34]}  & 24.41 &  {\scriptsize [18.35,37.07]} & 67.16 & {\scriptsize [49.63,86.20]} & 12.42 & {\scriptsize [8.44,15.73]} & 8.17 & {\scriptsize [5.78,11.80]}& 1.26 & {\scriptsize [0.47,1.70]}& 5.08 & {\scriptsize [3.06,7.43]}\\
5 & EA &   & &  & &  & &  & & & & & & \\
6 & VIX &   19.63 & & 17.09 & &  80.86 & & 9.14 & & 8.94 & & 2.15 & & 9.89\\
7 & EPU &   101.04 & & 83.60 & &  719.07 & & 3.32 & & 68.66 & & 2.15 & & 11.52\\
8 & US3M &   -0.08&   & 0.00 & &  76.00 & & -81.00 & & 4.88 & & -1.46 & & 78.07 \\
9 & HSI &   0.00 &  & 0.00 & &  0.34 & & 0.00 & & 0.01 & & 15.76 & & 341.30\\
10 & M1W & 0.11 & {\scriptsize [-0.11,0.35]}  &  0.13 &  {\scriptsize [-0.07,0.35]} & 21.98 & {\scriptsize [12.19,51.33]} & -19.22 & {\scriptsize [-34.57,-11.62]} & 3.08 & {\scriptsize [1.96,4.15]}& 0.10 & {\scriptsize [-0.59,1.34]}& 8.30 & {\scriptsize [4.86,30.25]}\\
11 & \$VOL &   0.02 & {\scriptsize [-0.03,0.08]}  &  -1.21 &  {\scriptsize [-2.19,-0.12]} & 210.21 & {\scriptsize [150.82,327.68]} & -162.28 & {\scriptsize [-209.44,-124.94]} & 33.54 & {\scriptsize [26.38,39.68]}& 0.37 & {\scriptsize [0.19,0.85]}& 5.11 & {\scriptsize [4.31,8.47]}\\
12 & ADS &   -28.99  & & -15.37 & &  88.33 & & -379.75 & & 69.22 & & -2.38 & & 10.78 \\
\hline \hline
\end{tabular}
\smallskip
\begin{scriptsize}
\parbox{\textwidth}{\emph{Note.} EA is equal to one if the company makes an earnings announcement on a given day, zero otherwise. Hence, we omit the calculation of descriptive statistics for the EA variable. The square brackets contain interval-based measures of asset-specific variables. Here, the numbers are defined as the minimum and maximum value of that particular descriptive statistic when calculated individually over the cross-section of the included equities. The descriptive statistics for US3M and \$VOL are for the transformed variables. ADS is reported in percent.}
\end{scriptsize}
\end{center}
\end{sidewaystable}

A full list of explanatory variables and their acronyms is presented in Table \ref{table:covariate-list}, along with some standard descriptive statistics. Prior to estimation, we standardize the input data with the sample mean and sample variance from the training set\glsadd{Training Set} to render the estimation scale invariant and improve the numerical optimization procedure for ridge, lasso, elastic net, and the neural network.

\section{Empirical results} \label{Empirical Results}
	
\subsection{One-day-ahead forecasting} \label{Out-of-Sample Performance}

In Table \ref{table:har-1d.tex}, reading by column we report the out-of-sample MSE of each forecasting model relative to the benchmark in the selected row for the dataset $\mathcal{M}_{ \text{HAR}}$. The ratio is a cross-sectional average of the relative one-day-ahead realized variance forecast MSEs for each stock. Hence, the first row contains the out-of-sample MSE of every model relative to the basic HAR, averaged over stocks. In addition, for each stock and pairwise comparison we compute the Diebold-Mariano test. The number formatting shows whether the null hypothesis of equal predictive accuracy for a fixed pairwise comparison was rejected more than half of the times across stocks at various levels of significance. Note that in $\mathcal{M}_{ \text{HAR}}$ HAR-X is restricted access to lagged realized variance and is therefore identical to HAR.

\begin{sidewaystable}
\begin{scriptsize}
\setlength{ \tabcolsep}{0.09cm}
\begin{center}
\caption{One-day-ahead relative MSE and Diebold-Mariano test for dataset $\mathcal{M}_{\text{HAR}}$.
\label{table:har-1d.tex}}
\begin{tabular}{lcccccccccccccccccccccc}
\hline \hline
& HAR & HAR-X & LogHAR & LevHAR & SHAR & HARQ & RR & LA & EN & A-LA & P-LA & BG & RF & GB & NN$_{1}^{1}$ & NN$_{1}^{10}$ & NN$_{2}^{1}$ & NN$_{2}^{10}$ & NN$_{3}^{1}$ & NN$_{3}^{10}$ & NN$_{4}^{1}$ & NN$_{4}^{10}$ \\
\hline
HAR  & - & 1.000 & 0.995 & 1.073 & 1.009 & 1.059 & 1.000 & 1.003 & 0.999 & 1.007 & 1.005 & 1.147 & 1.020 & 1.054 & 0.980 & \textit{0.969} & \textit{0.966} & \textit{0.958} & \textit{0.955} & \textbf{\textit{0.954}} & 0.984 & 0.990 \\ 
HAR-X & 1.000  & - & 0.995 & 1.073 & 1.009 & 1.059 & 1.000 & 1.003 & 0.999 & 1.007 & 1.005 & 1.147 & 1.020 & 1.054 & 0.980 & \textit{0.969} & \textit{0.966} & \textit{0.958} & \textit{0.955} & \textbf{\textit{0.954}} & 0.984 & 0.990 \\ 
LogHAR & 1.008 & 1.008  & - & 1.076 & 1.017 & 1.063 & 1.007 & 1.010 & 1.006 & 1.016 & 1.013 & 1.151 & 1.025 & 1.059 & 0.987 & 0.976 & 0.972 & 0.964 & 0.962 & \textit{0.960} & 0.990 & 0.995 \\ 
LevHAR & 0.947 & 0.947 & 0.938  & - & 0.954 & 0.988 & 0.946 & 0.949 & 0.945 & 0.954 & 0.952 & 1.077 & 0.961 & 0.992 & 0.925 & 0.916 & 0.912 & \textit{0.905} & 0.902 & \textit{0.900} & 0.928 & 0.933 \\ 
SHAR & 0.994 & 0.994 & 0.989 & 1.066  & - & 1.055 & 0.994 & 0.997 & 0.993 & 1.002 & 0.999 & 1.140 & 1.014 & 1.048 & 0.975 & \textit{0.963} & 0.961 & 0.952 & \textit{0.950} & \textit{0.948} & 0.978 & 0.984 \\ 
HARQ & 0.968 & 0.968 & 0.959 & 1.025 & 0.978  & - & 0.966 & 0.970 & 0.965 & 0.975 & 0.973 & 1.104 & 0.982 & 1.015 & 0.944 & 0.935 & 0.932 & 0.924 & 0.921 & 0.920 & 0.948 & 0.953 \\ 
RR & 1.001 & 1.001 & 0.995 & 1.073 & 1.010 & 1.058  & - & 1.003 & 0.999 & 1.008 & 1.006 & 1.147 & 1.020 & 1.054 & 0.981 & \textit{0.969} & \textit{0.966} & \textbf{\textit{0.958}} & \textit{0.956} & \textbf{\textit{0.954}} & 0.984 & 0.990 \\ 
LA & \textbf{\textit{\underline{0.998}}} & \textbf{\textit{\underline{0.998}}} & 0.992 & 1.070 & 1.007 & 1.056 & 0.997  & - & 0.996 & 1.005 & \textbf{\textit{1.003}} & 1.143 & 1.017 & 1.051 & 0.978 & \textbf{\textit{0.966}} & \textit{0.964} & \textbf{\textit{0.955}} & \textbf{\textit{0.953}} & \textbf{\textit{0.951}} & 0.981 & 0.987 \\ 
EN & 1.002 & 1.002 & 0.996 & 1.073 & 1.011 & 1.059 & 1.001 & 1.004  & - & 1.009 & 1.007 & 1.147 & 1.020 & 1.055 & 0.981 & \textit{0.970} & \textit{0.967} & \textbf{\textit{0.959}} & \textbf{\textit{0.956}} & \textbf{\textit{0.955}} & 0.985 & 0.990 \\ 
A-LA & \textbf{\textit{0.993}} & \textbf{\textit{0.993}} & 0.988 & 1.065 & 1.002 & 1.052 & 0.992 & 0.995 & 0.992  & - & 0.998 & 1.139 & 1.012 & 1.047 & 0.973 & \textit{0.962} & \textit{0.959} & \textbf{\textit{0.951}} & \textbf{\textit{0.949}} & \textbf{\textit{0.947}} & 0.977 & 0.982 \\ 
P-LA & \textbf{\textit{0.995}} & \textbf{\textit{0.995}} & 0.990 & 1.067 & 1.003 & 1.054 & 0.995 & 0.998 & 0.994 & 1.003  & - & 1.141 & 1.015 & 1.049 & \textit{0.975} & \textit{0.964} & \textit{0.961} & \textbf{\textit{0.953}} & \textit{0.951} & \textbf{\textit{0.949}} & 0.979 & 0.985 \\ 
BG & \textit{0.891} & \textit{0.891} & 0.882 & 0.947 & 0.898 & 0.936 & \textit{0.889} & 0.892 & \textit{0.888} & 0.897 & 0.896  & - & \textit{0.901} & 0.930 & \textit{0.871} & \textbf{\textit{0.862}} & \textbf{\textit{0.859}} & \textbf{\textit{0.851}} & \textbf{\textit{0.849}} & \textbf{\textit{0.847}} & \textit{0.874} & \textit{0.879} \\ 
RF & 0.986 & 0.986 & 0.978 & 1.052 & 0.994 & 1.037 & 0.985 & 0.987 & 0.984 & 0.993 & 0.991 & 1.121  & - & 1.034 & \textit{0.964} & \textbf{\textit{0.954}} & \textit{0.951} & \textbf{\textit{0.943}} & \textit{0.940} & \textbf{\textit{0.938}} & 0.968 & 0.973 \\ 
GB & 0.958 & 0.958 & 0.949 & 1.020 & 0.966 & 1.008 & 0.957 & 0.960 & 0.956 & 0.965 & 0.963 & 1.088 & 0.972  & - & \textit{0.938} & \textbf{\textit{0.927}} & \textit{0.923} & \textit{0.916} & \textit{0.913} & \textit{0.912} & 0.940 & 0.945 \\ 
NN$_{1}^{1}$ & 1.022 & 1.022 & 1.016 & 1.093 & 1.032 & 1.075 & 1.021 & 1.025 & 1.021 & 1.030 & 1.028 & 1.170 & 1.040 & 1.076  & - & 0.989 & 0.986 & 0.978 & 0.976 & 0.974 & 1.004 & 1.010 \\ 
NN$_{1}^{10}$ & 1.033 & 1.033 & 1.027 & 1.106 & 1.042 & 1.091 & 1.032 & 1.036 & 1.032 & 1.041 & 1.039 & 1.184 & 1.053 & 1.087 & 1.012  & - & 0.997 & 0.989 & 0.986 & 0.984 & 1.015 & 1.021 \\ 
NN$_{2}^{1}$ & 1.037 & 1.037 & 1.030 & 1.109 & 1.046 & 1.094 & 1.036 & 1.039 & 1.035 & 1.045 & 1.042 & 1.187 & 1.056 & 1.091 & 1.015 & 1.004  & - & 0.992 & 0.989 & 0.987 & 1.018 & 1.024 \\ 
NN$_{2}^{10}$ & 1.045 & 1.045 & 1.038 & 1.118 & 1.054 & 1.103 & 1.044 & 1.047 & 1.043 & 1.053 & 1.051 & 1.197 & 1.064 & 1.100 & 1.023 & 1.012 & 1.008  & - & 0.997 & 0.996 & 1.027 & 1.032 \\ 
NN$_{3}^{1}$ & 1.048 & 1.048 & 1.042 & 1.122 & 1.058 & 1.105 & 1.047 & 1.051 & 1.047 & 1.056 & 1.054 & 1.200 & 1.067 & 1.103 & 1.027 & 1.015 & 1.012 & 1.003  & - & 0.999 & 1.030 & 1.036 \\ 
NN$_{3}^{10}$ & 1.050 & 1.050 & 1.043 & 1.123 & 1.059 & 1.107 & 1.049 & 1.052 & 1.048 & 1.058 & 1.056 & 1.202 & 1.069 & 1.105 & 1.028 & 1.016 & 1.013 & 1.005 & 1.002  & - & 1.031 & 1.037 \\ 
NN$_{4}^{1}$ & 1.019 & 1.019 & 1.012 & 1.089 & 1.029 & 1.074 & 1.018 & 1.021 & 1.018 & 1.027 & 1.025 & 1.167 & 1.037 & 1.072 & 0.998 & 0.987 & 0.983 & \textbf{\textit{0.975}} & \textit{0.972} & \textbf{\textit{\underline{0.970}}}  & - & 1.006 \\ 
NN$_{4}^{10}$ & 1.014 & 1.014 & 1.007 & 1.083 & 1.024 & 1.067 & 1.013 & 1.016 & 1.012 & 1.022 & 1.020 & 1.160 & 1.032 & 1.066 & 0.993 & \textit{0.981} & \textit{0.978} & \textbf{\textit{0.970}} & \textbf{\textit{0.967}} & \textbf{\textit{\underline{0.965}}} & 0.995  & - \\ 
\hline \hline
\end{tabular}
\smallskip
\begin{scriptsize}
\parbox{0.98\textwidth}{\emph{Note.} We report the out-of-sample realized variance forecast MSE of each model in the selected column relative to the benchmark in the selected row. Each number is a cross-sectional average of such pairwise relative MSEs for each stock. 
The formatting is as follows: \textit{number} (\textbf{\textit{number}}) [\underline{\textbf{\textit{number}}}] denotes whether the Diebold-Mariano test of equal predictive accuracy is rejected more than 50\% of the time at the 10\% (5\%) [1\%] level of significance across individual tests for each asset. 
The hypothesis being tested is $\text{H}_{0}: \text{MSE}_{i} = \text{MSE}_{j}$ against a one-sided alternative $\text{H}_{1}: \text{MSE}_{i} > \text{MSE}_{j}$, 
where model $i$ is the label of the selected row, whereas model $j$ is the label of the selected column.}
\end{scriptsize}
\end{center}
\end{scriptsize}
\end{sidewaystable}

Several interesting findings emerge. Firstly, the basic HAR is superior to its offspring. Compared to the extant literature, this is surprising. However, the results naturally depend on the split between training, validation, and test set. With a conventional shorter training set of 1,000 days, the extended HAR models outperform HAR by some distance, as reported in Appendix \ref{appendix:test-set}. A possible reconciliation of this is that with a shorter in-sample size, the financial crisis appears in the out-of-sample set, which has a huge impact on the relative performance of the models. For example, the quarticity correction in the HARQ is arguably more crucial in high volatility environments. Secondly, when the covariate list is restricted to $\mathcal{M}_{ \text{HAR}}$ the ML results are also mixed. This is not surprising. The lagged values of realized variance are strong determinants of future volatility, so there is not much room for improvement. In alignment with this interpretation, regularization is at par with HAR sporting an aggregated MSE ratio at or above one. So, as expected, there is no impending need to penalize the HAR model in $\mathcal{M}_{ \text{HAR}}$. The regression trees are also inferior to HAR, especially bagging, whereas the random forest is about equal in performance. In contrast, the cross-sectional relative MSEs for the neural networks are consistently below one by roughly five percent on average, and the Diebold-Mariano test is frequently rejected at the 10\% or 5\% significance level, when a neural network is evaluated against the HAR as a benchmark. This shows that for short-horizon forecasts of next-day volatility there are minor refinements associated with allowing for nonlinearities in the dynamic. What is perhaps the most striking outcome of Table \ref{table:har-1d.tex} is that even a shallow neural network beats the HAR. This can possibly be explained by the fact that a limited complexity of a neural network allows the benefits of early stopping and dropout to be emphasized.

\begin{sidewaystable}
\begin{scriptsize}
\setlength{ \tabcolsep}{0.09cm}
\begin{center}
\caption{One-day-ahead relative MSE and Diebold-Mariano test for dataset $\mathcal{M}_{\text{ALL}}$.
\label{table:all-1d.tex}}
\begin{tabular}{lcccccccccccccccccccccc}
\hline \hline
& HAR & HAR-X & LogHAR & LevHAR & SHAR & HARQ & RR & LA & EN & A-LA & P-LA & BG & RF & GB & NN$_{1}^{1}$ & NN$_{1}^{10}$ & NN$_{2}^{1}$ & NN$_{2}^{10}$ & NN$_{3}^{1}$ & NN$_{3}^{10}$ & NN$_{4}^{1}$ & NN$_{4}^{10}$ \\
\hline
HAR  & - & \textit{0.966} & \textbf{\textit{0.901}} & 1.003 & 1.080 & 1.289 & \textbf{\textit{0.919}} & \textbf{\textit{0.936}} & \textbf{\textit{0.916}} & \textit{0.957} & 0.987 & \textit{0.961} & \textbf{\textit{0.901}} & \textit{0.962} & \textbf{\textit{\underline{0.902}}} & \textbf{\textit{\underline{0.889}}} & \textbf{\textit{0.893}} & \textbf{\textit{0.885}} & \textbf{\textit{0.910}} & \textbf{\textit{0.898}} & \textit{0.929} & \textit{0.944} \\ 
HAR-X & 1.045  & - & \textit{0.938} & 1.034 & 1.128 & 1.298 & \textit{0.954} & 0.972 & 0.952 & 0.994 & 1.020 & 1.000 & 0.940 & 0.998 & 0.941 & \textit{0.928} & \textit{0.932} & \textit{0.923} & 0.944 & 0.935 & 0.966 & 0.980 \\ 
LogHAR & 1.113 & 1.070  & - & 1.109 & 1.201 & 1.413 & 1.019 & 1.039 & 1.017 & 1.062 & 1.093 & 1.067 & 1.001 & 1.066 & 1.002 & 0.988 & 0.992 & 0.983 & 1.009 & 0.997 & 1.031 & 1.047 \\ 
LevHAR & 1.018 & 0.971 & \textit{0.912}  & - & 1.096 & 1.246 & \textit{0.928} & 0.946 & 0.927 & 0.966 & 0.991 & 0.971 & 0.915 & 0.970 & \textit{0.916} & \textit{0.904} & \textit{0.907} & \textit{0.899} & 0.918 & 0.910 & 0.939 & 0.953 \\ 
SHAR & 0.942 & \textbf{\textit{0.909}} & \textbf{\textit{0.848}} & \textit{0.942}  & - & 1.219 & \textbf{\textit{\underline{0.865}}} & \textbf{\textit{\underline{0.881}}} & \textbf{\textit{\underline{0.863}}} & \textbf{\textit{0.900}} & \textbf{\textit{0.928}} & \textbf{\textit{0.903}} & \textbf{\textit{\underline{0.847}}} & \textbf{\textit{\underline{0.904}}} & \textbf{\textit{\underline{0.851}}} & \textbf{\textit{\underline{0.837}}} & \textbf{\textit{\underline{0.840}}} & \textbf{\textit{\underline{0.832}}} & \textbf{\textit{\underline{0.857}}} & \textbf{\textit{\underline{0.845}}} & \textbf{\textit{\underline{0.874}}} & \textbf{\textit{\underline{0.887}}} \\ 
HARQ & 0.882 & \textbf{\textit{\underline{0.832}}} & \textbf{\textit{\underline{0.788}}} & \textbf{\textit{0.852}} & 0.950  & - & \textbf{\textit{\underline{0.799}}} & \textbf{\textit{\underline{0.813}}} & \textbf{\textit{\underline{0.799}}} & \textbf{\textit{\underline{0.831}}} & \textbf{\textit{\underline{0.847}}} & \textbf{\textit{0.837}} & \textbf{\textit{0.792}} & \textbf{\textit{0.832}} & \textbf{\textit{\underline{0.796}}} & \textbf{\textit{\underline{0.785}}} & \textbf{\textit{\underline{0.788}}} & \textbf{\textit{\underline{0.781}}} & \textbf{\textit{\underline{0.791}}} & \textbf{\textit{\underline{0.787}}} & \textbf{\textit{\underline{0.809}}} & \textbf{\textit{0.818}} \\ 
RR & 1.093 & 1.050 & 0.983 & 1.087 & 1.181 & 1.375  & - & 1.018 & 0.998 & 1.042 & 1.070 & 1.049 & 0.984 & 1.046 & 0.985 & 0.971 & 0.975 & \textit{0.966} & 0.989 & 0.979 & 1.012 & 1.027 \\ 
LA & 1.076 & 1.033 & 0.967 & 1.070 & 1.161 & 1.352 & 0.984  & - & 0.981 & 1.024 & 1.052 & 1.033 & 0.968 & 1.028 & 0.970 & 0.955 & 0.959 & 0.950 & 0.972 & 0.963 & 0.996 & 1.010 \\ 
EN & 1.096 & 1.053 & 0.985 & 1.091 & 1.184 & 1.384 & 1.003 & 1.021  & - & 1.045 & 1.074 & 1.051 & 0.986 & 1.049 & 0.987 & 0.973 & 0.977 & 0.968 & 0.991 & 0.981 & 1.015 & 1.030 \\ 
A-LA & 1.051 & 1.009 & \textit{0.945} & 1.045 & 1.134 & 1.323 & \textit{0.962} & \textit{0.979} & \textbf{\textit{0.959}}  & - & 1.029 & 1.007 & \textit{0.946} & 1.006 & \textit{0.947} & \textit{0.933} & \textit{0.937} & \textbf{\textit{0.929}} & \textit{0.951} & \textit{0.941} & 0.973 & 0.987 \\ 
P-LA & 1.028 & 0.983 & \textbf{\textit{0.923}} & 1.017 & 1.109 & 1.275 & \textbf{\textit{0.938}} & \textbf{\textit{0.954}} & \textbf{\textit{0.936}} & 0.977  & - & 0.985 & \textit{0.925} & 0.980 & \textbf{\textit{0.927}} & \textbf{\textit{0.913}} & \textbf{\textit{0.916}} & \textbf{\textit{0.908}} & \textbf{\textit{0.926}} & \textbf{\textit{0.919}} & 0.950 & 0.963 \\ 
BG & 1.050 & 1.011 & 0.945 & 1.046 & 1.131 & 1.334 & 0.963 & 0.983 & 0.961 & 1.002 & 1.035  & - & \textit{0.944} & 1.008 & 0.946 & 0.933 & 0.937 & 0.929 & 0.954 & 0.942 & 0.972 & 0.988 \\ 
RF & 1.112 & 1.071 & 1.000 & 1.111 & 1.198 & 1.426 & 1.020 & 1.039 & 1.017 & 1.062 & 1.095 & 1.065  & - & 1.067 & 1.001 & 0.987 & 0.991 & 0.982 & 1.009 & 0.997 & 1.031 & 1.047 \\ 
GB & 1.049 & 1.007 & 0.942 & 1.042 & 1.132 & 1.310 & 0.959 & 0.976 & 0.957 & 1.000 & 1.026 & 1.006 & \textit{0.944}  & - & \textit{0.946} & \textbf{\textit{0.932}} & \textbf{\textit{0.936}} & \textbf{\textit{0.927}} & \textit{0.948} & \textit{0.939} & 0.970 & 0.984 \\ 
NN$_{1}^{1}$ & 1.113 & 1.074 & 1.001 & 1.114 & 1.204 & 1.436 & 1.022 & 1.042 & 1.019 & 1.064 & 1.099 & 1.068 & 1.002 & 1.070  & - & 0.987 & 0.992 & 0.983 & 1.012 & 0.998 & 1.032 & 1.050 \\ 
NN$_{1}^{10}$ & 1.128 & 1.088 & 1.015 & 1.129 & 1.218 & 1.457 & 1.035 & 1.055 & 1.032 & 1.078 & 1.112 & 1.083 & 1.015 & 1.084 & 1.014  & - & 1.005 & 0.996 & 1.024 & 1.011 & 1.046 & 1.063 \\ 
NN$_{2}^{1}$ & 1.123 & 1.083 & 1.010 & 1.123 & 1.211 & 1.446 & 1.030 & 1.049 & 1.027 & 1.073 & 1.106 & 1.078 & 1.010 & 1.078 & 1.011 & 0.996  & - & 0.991 & 1.018 & 1.006 & 1.041 & 1.057 \\ 
NN$_{2}^{10}$ & 1.133 & 1.093 & 1.020 & 1.134 & 1.222 & 1.459 & 1.039 & 1.058 & 1.036 & 1.082 & 1.115 & 1.088 & 1.020 & 1.088 & 1.020 & 1.005 & 1.009  & - & 1.028 & 1.015 & 1.051 & 1.067 \\ 
NN$_{3}^{1}$ & 1.109 & 1.064 & 0.996 & 1.103 & 1.197 & 1.396 & 1.013 & 1.031 & 1.011 & 1.056 & 1.083 & 1.064 & 0.998 & 1.060 & 0.999 & 0.984 & 0.987 & 0.978  & - & 0.991 & 1.026 & 1.040 \\ 
NN$_{3}^{10}$ & 1.117 & 1.075 & 1.005 & 1.114 & 1.205 & 1.424 & 1.023 & 1.042 & 1.021 & 1.066 & 1.096 & 1.072 & 1.005 & 1.070 & 1.006 & 0.991 & 0.995 & 0.986 & 1.011  & - & 1.035 & 1.050 \\ 
NN$_{4}^{1}$ & 1.081 & 1.039 & 0.972 & 1.076 & 1.167 & 1.365 & 0.990 & 1.008 & 0.987 & 1.031 & 1.061 & 1.035 & 0.973 & 1.034 & 0.974 & \textit{0.960} & \textbf{\textit{0.964}} & \textbf{\textit{\underline{0.955}}} & \textbf{\textit{0.979}} & \textbf{\textit{0.968}}  & - & 1.015 \\ 
NN$_{4}^{10}$ & 1.067 & 1.024 & 0.958 & 1.060 & 1.150 & 1.337 & 0.976 & 0.993 & 0.973 & 1.016 & 1.044 & 1.022 & 0.960 & 1.019 & \textit{0.961} & \textbf{\textit{\underline{0.947}}} & \textbf{\textit{\underline{0.951}}} & \textbf{\textit{\underline{0.942}}} & \textbf{\textit{\underline{0.964}}} & \textbf{\textit{\underline{0.955}}} & 0.986  & - \\ 
\hline \hline
\end{tabular}
\smallskip
\begin{scriptsize}
\parbox{0.98\textwidth}{\emph{Note.} We report the out-of-sample realized variance forecast MSE of each model in the selected column relative to the benchmark in the selected row. Each number is a cross-sectional average of such pairwise relative MSEs for each stock. 
The formatting is as follows: \textit{number} (\textbf{\textit{number}}) [\underline{\textbf{\textit{number}}}] denotes whether the Diebold-Mariano test of equal predictive accuracy is rejected more than 50\% of the time at the 10\% (5\%) [1\%] level of significance across individual tests for each asset. 
The hypothesis being tested is $\text{H}_{0}: \text{MSE}_{i} = \text{MSE}_{j}$ against a one-sided alternative $\text{H}_{1}: \text{MSE}_{i} > \text{MSE}_{j}$, 
where model $i$ is the label of the selected row, whereas model $j$ is the label of the selected column.}
\end{scriptsize}
\end{center}
\end{scriptsize}
\end{sidewaystable}

Turning next to Table \ref{table:all-1d.tex}, we show the corresponding results for $\mathcal{M}_{ \text{ALL}}$. Here, the supplemental exogenous predictors from Table \ref{table:covariate-list} appear in the information set. The addition of these variables causes a huge impact on the pairwise comparisons. This forcefully demonstrates the complexity and low signal-to-noise ratio in the underlying financial data, as inferred from the MSE ratio of the HAR-X and HAR. As gauged by comparing Table \ref{table:har-1d.tex} and Table \ref{table:all-1d.tex}, there is but a minuscule decrease in MSE of $3.4\%$ moving from $\mathcal{M}_{ \text{HAR}}$ to $\mathcal{M}_{ \text{ALL}}$. Furthermore, the LevHAR, SHAR, and HARQ fare worse than HAR, even though their information set is larger. This supports our argument in Section \ref{section:regularization} that lack of regularization leave the linear HAR models exposed to in-sample overfit. As such, ridge regression is expected to alleviate this problem. This is also confirmed in Table \ref{table:all-1d.tex}. The $\ell_{2}$ regularization enforced by ridge reduces the average MSE of $8.1\%$ overall. The effect of allowing for subset selection\glsadd{Subset selection} can be uncovered by contrasting lasso with ridge. The latter is marginally better than the former, but combining ridge and lasso via the elastic net is best overall.\footnote{The average value of the weight parameter $\alpha$ is about 0.165 in our sample.} The elastic net ranks on top of the linear predictions with a pronounced $8.4\%$ reduction in MSE relative to the HAR.

An appealing observation that can be made from Table \ref{table:all-1d.tex} is that the dimension of the feature space is rather critical to achieve good performance with regression trees. In $\mathcal{M}_{ \text{HAR}}$, their performance were inferior to HAR, but with the expanded database they work better, although bagging and gradient boosting are still worse than a regularized HAR. The best-in-class tree-based approach is the random forest. It delivers a $9.9\%$ reduction in MSE against HAR. This is arguably caused by its ability to decorrelate trees via random selection of the number of explanatory variables, as opposed to bagging and gradient boosting. Comparing random forest to gradient boosting, the latter shows inferior forecast accuracy. Since gradient boosting produces a sequential model based on an ensemble of weak learners\glsadd{Weak learner}, it places more weight on outlying observations in contrast to classical tree-based approaches. Hence, it is unsurprising that random forest does a better job than gradient boosting, given the low signal-noise ratio in the data.\footnote{We also estimated an XGBoost model of \citet*{chen-guestrin:16a} with a refined set of hyperparameters, early stopping and $\ell_{1}$ penalization. However, the results were not much different from gradient boosting.} It should be stressed that even though random forest has the lowest MSE among linear regression and tree-based models, there is only a marginal drop between it and the elastic net, which is moreover typically not statistically significant. This highlights that automatic interactions between variables can be informative about future volatility. Meanwhile, tackling overfitting is a more potent source of forecast improvement in the ``high-dimensional'' $\mathcal{M}_{ \text{ALL}}$ framework compared to $\mathcal{M}_{ \text{HAR}}$.

At last, we compare the neural networks to HAR. It is evident that the average MSE decreases further, such that the neural networks remain in front. However, the enlargement of the feature space benefits several models, and the edge of the neural networks is roughly as before. What is more, when additional predictors are included, the improvement is statistically significant at the 1\% level across more than half of the stocks.

To shed further light on the causes of the decline in MSE for the neural networks, a logical next step is to inspect their complexity. In this aspect the geometric pyramid structure is helpful, as we can readily investigate the effects of changing network architecture. Indeed, our results show that there is no big demand for an overly complex model, because the results are rather stable across network configurations. Initially, adding hidden layers to the network is helpful. Peak performance is attained with a NN$_{2}$ -- NN$_{3}$ architecture, but it starts to deteriorate after that. Moreover, ensemble averaging the forecast over several trained neural networks leads to further declines in MSE, except again for NN$_{4}$.

\begin{figure}[t!]
\begin{center}
\caption{Forecast accuracy over out-of-sample volatility distribution. \label{figure:quantile}}
\begin{tabular}{c}
\includegraphics[height=8cm,width=0.6\textwidth]{{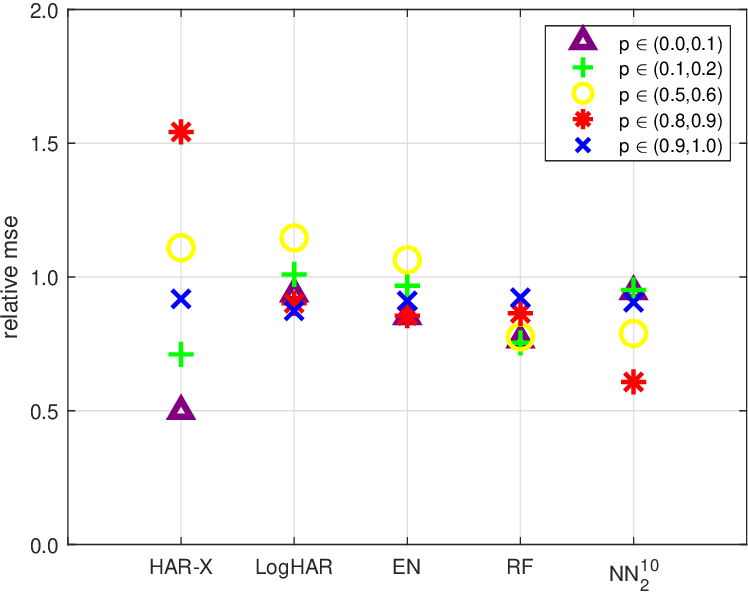}}
\end{tabular}
\begin{scriptsize}
\parbox{\textwidth}{\emph{Note.} We report one-day-ahead out-of-sample forecast MSE relative to the HAR model for HAR-X, LogHAR, elastic net (EN), random forest (RF), and NN$_{2}^{10}$. The sample is split into deciles of realized variance in the test set. $p$ stands for percentile. The dataset is $\mathcal{M}_{ALL}$.}
\end{scriptsize}
\end{center}
\end{figure}

Finally, we highlight that the best HAR model is the LogHAR. At first glance it remains competitive against the ML techniques even in the $\mathcal{M}_{ \text{ALL}}$ setting. Thus, while a regularized HAR outperforms a non-regularized HAR, it is in turn overtaken by LogHAR. As the latter is nonlinear, this implies that handling nonlinearities is a key source to improvement, also evidenced by the ML techniques. This confirms that regularization is less vital than capturing nonlinearities, which may partially be explained by the fact that $\mathcal{M}_{ \text{ALL}}$ is not too vast in our setting.\footnote{We only detect minor improvements in forecast accuracy when regularizing the LogHAR. For example, in $\mathcal{M}_{ \text{ALL}}$ the relative MSE of the LogHAR regularized with elastic net to the unregularized version is 0.995. We believe this is because the log-transformation reduces the influence of outliers and attenuates the need for penalization.} However, in contrast to ML the true story behind the success of the LogHAR is not to be found in a general leap in forecast accuracy, but rather in a more narrow improvement in high states of realized variance, which is a reflection of the log-transformations capability to bring down outliers. This becomes apparent when we take an in-depth look at the forecast accuracy over the entire realized variance distribution in Figure \ref{figure:quantile}.

To sum up, the main improvement in forecasting accuracy in $\mathcal{M}_{ \text{ALL}}$ is due to the inclusion of additional explanatory variables, which generally aids the ML techniques and the LogHAR, whereas the unregularized linear HAR models suffer from overfitting. Within the ML models, the ones that capture non-linearities are superior, and regularization turns out to be more potent when growing the feature space.	

\begin{figure}[t!]
\begin{center}
\caption{Boxplot of cross-sectional out-of-sample relative MSE. \label{figure:boxplot}}
\begin{tabular}{cc}
\small{Panel A: $\mathcal{M}_{ \text{HAR}}$.} & \small{Panel B: $\mathcal{M}_{ \text{ALL}}$.} \\
\includegraphics[height=8cm,width=0.48\textwidth]{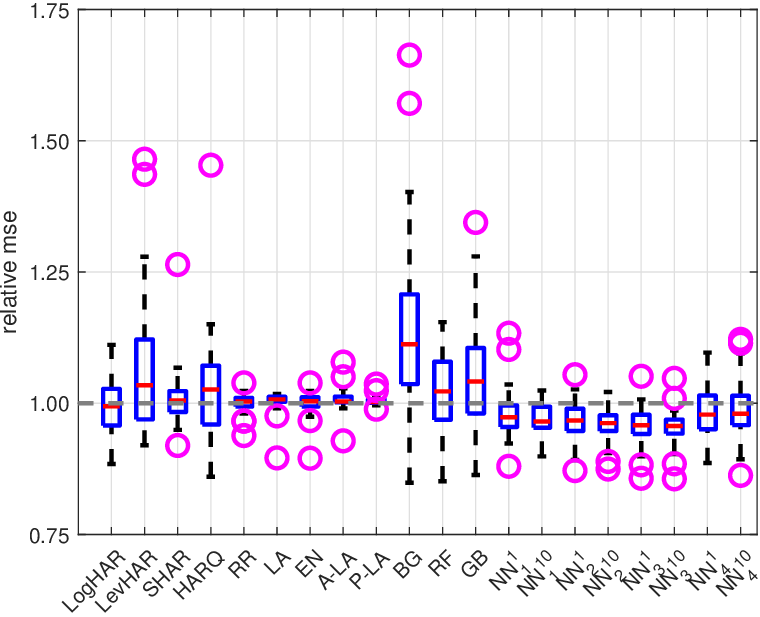} &
\includegraphics[height=8cm,width=0.48\textwidth]{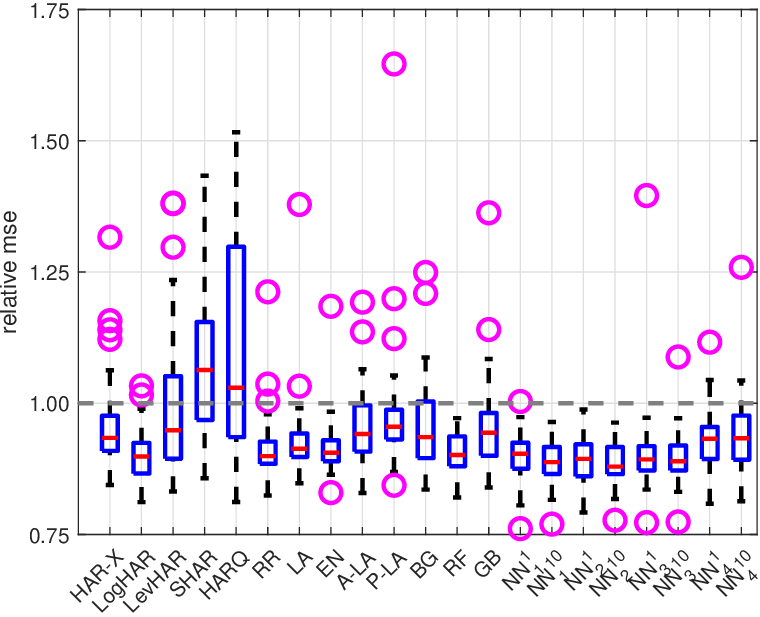} \\
\end{tabular}
\begin{scriptsize}
\parbox{\textwidth}{\emph{Note.} We construct a boxplot for the one-day-ahead out-of-sample forecast MSE relative to the HAR model for HAR-X, LogHAR, LevHAR, SHAR, HARQ, ridge regression (RR), lasso (LA), elastic net (EN), adaptive lasso (A-LA), post lasso (P-LA), bagging (BA), random forest (RF), gradient boosting (GB), and eight neural networks (NN). The sample consists of 29 relative MSEs for each model, corresponding to the number of stocks in our empirical application. The central mark is the median MSE, while the bottom and top edge of the box indicate the interquartile range. The whiskers are the outermost observations not flagged as outliers (the latter are marked with a circle). HAR-X is omitted from Panel A, as it is identical to HAR in $\mathcal{M}_{ \text{HAR}}$.}
\end{scriptsize}
\end{center}
\end{figure}

To examine whether the superior forecast accuracy of the ML methods is driven by a small subset of stocks or it extends to the entire cross-section, we next dissect the aggregate numbers reported in Table \ref{table:har-1d.tex} -- \ref{table:all-1d.tex}. In Figure \ref{figure:boxplot}, we construct a boxplot of the relative MSE for each model versus the HAR across the 29 stocks in our sample. Panel A is for $\mathcal{M}_{ \text{HAR}}$, whereas Panel B is for $\mathcal{M}_{ \text{ALL}}$. The tree-based approaches are slightly inferior to the HAR for $\mathcal{M}_{ \text{HAR}}$, as also noticed above. However, there is a steady and sustained improvement in forecast accuracy of ML as we grow the information set.

\begin{figure}[t!]
\begin{center}
\caption{Inclusion rate in the MCS. \label{figure:epa}}
\begin{tabular}{cc}
\includegraphics[height=8cm,width=0.6\textwidth]{{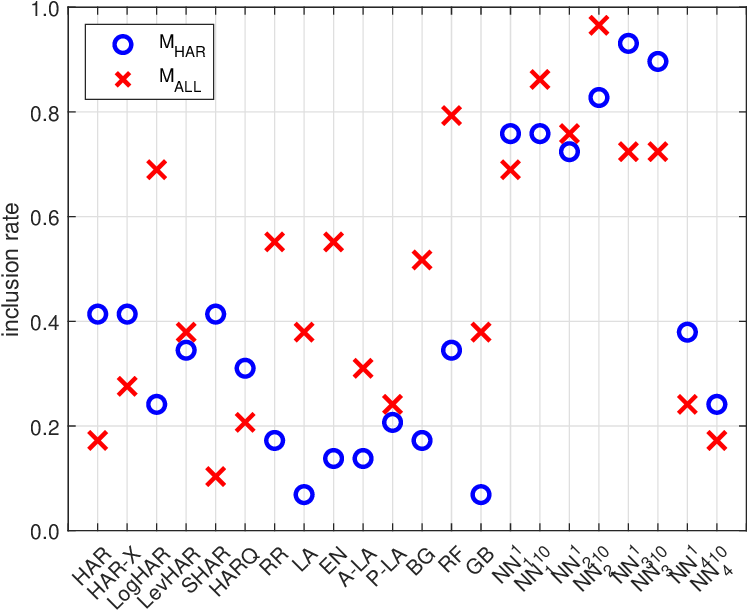}}
\end{tabular}
\begin{scriptsize}
\parbox{\textwidth}{\emph{Note.} We report the outcome of the \citet*{hansen-lunde-nason:11a} MCS procedure, which identifies a subset of best models by running a sequence of tests for equal predictive ability, $H_{0}: \text{MSE}_{ \text{model}_{i}} = \text{MSE}_{\text{model}_{j}}$. The overall confidence level is set to 90\% and the numbers indicate the percentage of times a given model is retained in the MCS.}
\end{scriptsize}
\end{center}
\end{figure}

In Figure \ref{figure:epa}, we compute the MCS at a 90\% confidence level. The figure shows the percentage of times a model was retained in the final set across stocks. In line with our earlier results, the neural networks exhibit an inclusion rate of around $80\%$ for the $\mathcal{M}_{ \text{ALL}}$ dataset, compared to about 20\% -- 40\% for the HAR. The LogHAR is the sole HAR model that is retained at a high rate in the $\mathcal{M}_{ \text{ALL}}$ setting.

To get insights about how good the models forecast across states of the volatility distribution, we split the test set into deciles of the observed values of daily realized variance (i.e., the first subsample holds the $10\%$ smallest values of realized variance in the out-of-sample window and is denoted (0.0,0.1)). For brevity we select a ``preferred'' model from each main category, i.e. we look at HAR-X, LogHAR, elastic net, random forest, and NN$_{2}^{10}$.

In Figure \ref{figure:quantile}, we present the outcome of these calculations for selected low, medium, and high deciles. As indicated, the random forest and neural network yield lower MSE compared to the HAR across the entire support of the volatility distribution. Hence, the key driver for the ML results can be found in the majority of the sample. The MSE is only reduced by $7.38\%$ for the highest 10\% of the out-of-sample realized variancee, suggesting a closer alignment in forecast accuracy during market turmoil, which is of course crucial in practice. The sustained results are not shared with the HAR-X, LogHAR and elastic net. On the one hand, the HAR-X struggles when volatility is high, but has a much better MSE in tranquil markets. On the other hand, LogHAR yields only a minor improvement compared to the HAR, which appears to contradict the conclusion of Table \ref{table:har-1d.tex} and \ref{table:all-1d.tex}. However, a closer look reveals only a sizable improvement of the LogHAR in the highest decile. As realized variance is highly right-skewed with extreme outliers, it is therefore the ability of the LogHAR to downweigh outliers that brings its average squared error markedly down.

\subsection{Marginal effects and variable importance} \label{section:variable-importance}

The autonomy of a neural network or tree-based approach severely limits our ability to explain what is going on ``under the hood'', which often leads ML to be accused of being a black box approach. With their complex and highly nonlinear structure, a complete revelation of the underlying machinery is of course also rather difficult.

Nonetheless, we can illustrate the inner workings of the ML algorithms and shed at least some initial light on the black box by exploiting a recent contribution of \citet*{apley-zhu:20a} that allows to compute an intuitive measure of marginal effect and variable importance in the ML framework. In particular, we calculate an Accumulated Local Effect (ALE) plot, which is an in-sample statistic based on the fitted model. To explain the concept, suppose for ease of exposition that $f(Z_{t})$ is differentiable and that we are interested in analyzing the marginal effect of the $j$th predictor $Z_{jt}$, $j = 1, \dots, J$, from the fitted model $\hat{f}(Z_{t})$. We divide $Z_{t}$ into $Z_{jt}$ and the complement $Z_{ \complement t} = Z_{t} \setminus Z_{jt}$, i.e. the remaining predictors in $Z_{t}$.

The ALE on $\hat{f}(Z_{t})$ of $Z_{jt}$ is the function:
\begin{align}
\begin{split}
f^{ \text{ALE}}(z) &= \text{constant} + \int_{ \min (z_{j})}^{z} E \left[ \frac{ \partial \hat{f}(Z_{jt}, Z_{ \complement t})}{ \partial Z_{jt}} \mid Z_{jt} = x \right] \mathrm{d}x \\
&= \text{constant} + \int_{ \min (z_{j})}^{z} \int \frac{ \partial \hat{f}(x, z_{ \complement t})}{ \partial x} p_{Z_{ \complement t} \mid Z_{jt}}(z_{ \complement t} \mid x) \mathrm{d}z_{ \complement t} \mathrm{d}x,
\end{split}
\end{align}
where $p_{Z_{ \complement t} \mid Z_{jt}}(z_{ \complement t} \mid x)$  is the conditional joint density of $Z_{ \complement t}$ given $Z_{jt} = x$ and the function is defined over the range of observed values of $Z_{jt}$ in the training set, as explained below. The constant is such that $E[f^{ \text{ALE}} (Z_{jt})] = 0$.

The term $\displaystyle \frac{ \partial \hat{f}(x, z_{ \complement t})}{ \partial x}$ is called the local effect of $Z_{jt}$ on $\hat{f}(Z_{t})$. By accumulating the local effect, the ALE can be interpreted as the combined impact of $Z_{jt}$ at the value $z$ compared to the average prediction of the data. The derivative isolates the effect of the target predictor and controls for correlated features. Hence, ALE calculates changes in the prediction and adds them up, rather than computing the prediction directly. As explained in \citet*{apley-zhu:20a}, this circumvents problems with omitted variable biases. Moreover, by exploiting the conditional distribution the ALE is not impaired by unlikely data points, a key weakness of the more common Partial Dependence (PD) function of \citet*{friedman:01a}.

$f^{ \text{ALE}}$ can be estimated via a finite difference approach. The information set used in the estimation is the matrix $Z = (Z_{1}, \ldots, Z_{T_{0}})$, where $T_{0}$ is the number of days in the training sample. The observations of the $j$th predictor correspond to the entries in the $j$th row of $Z$, which we denote $Z_{j} = (z_{j1}, \ldots, z_{jT_{0}})$. $Z_{ \complement}$ is the complement of $Z_{j}$, i.e. the remaining rows in $Z$. We partition the range of $Z_{j}$ such that $\min(Z_{j}) - \epsilon = z_{0} < z_{1} < \dots < z_{K} = \max(Z_{j})$, where $\epsilon$ is a small positive number ensuring that $\min(Z_{j})$ is included in the subsequent calculations.\footnote{In our implementation, we partition $Z_{j}$ into 100 subintervals containing equally many observations.} Next, as a function of $z \in (z_{0}, z_{K}]$ denote by $\text{idx}(z)$ the index of the interval into which $z$ falls, i.e. $z \in ( z_{\text{idx}(z)-1}, z_{\text{idx}(z)}]$. The uncentered ALE is estimated as follows:
\begin{equation} \label{equation:uncentered-ale}
\tilde{f}^{ \text{ALE}} (z) = \sum_{k : z_{k} \leq z} \frac{1}{T_{k}} \sum_{t : z_{jt} \in (z_{k-1}, z_{k}]} \left[ \hat{f}(z_{k},z_{ \complement t}) - \hat{f}(z_{k-1},z_{ \complement t}) \right],
\end{equation}
where $T_{k}$ is the number of data points of $Z_{j}$ that fall in $(z_{k-1}, z_{k}]$ with $T_{0} = \sum_{k=1}^{K} T_{k}$. As readily seen, the second summation approximates the local effect across small intervals by averaging over complement predictors $Z_{ \complement}$ that are associated with the observations of $z_{jt} \in (z_{k-1}, z_{k}]$. The first sum accumulates these effects.

The centered ALE is then:
\begin{equation}
\hat{f}^{ \text{ALE}} (z) = \tilde{f}^{ \text{ALE}} (z) - \frac{1}{T_{0}} \sum_{t=1}^{T_{0}} \tilde{f}^{ \text{ALE}} (z_{jt}).
\end{equation}
An ALE plot is based on the graph of $(z, \hat{f}^{ \text{ALE}} (z))$.

We convert ALE to a variable importance measure following \citet*{greenwell-boehmke-mccarthy:18a} in the PD framework by computing the sample standard deviation of $\hat{f}^{ \text{ALE}}$, i.e.
\begin{equation}
I(Z_{j}) = \sqrt{ \frac{1}{T_{0}-1} \sum_{t=1}^{T_{0}} \left[ \hat{f}^{ \text{ALE}} (z_{jt}) \right]^{2}}.
\end{equation}
Intuitively, if our prediction is only weakly related to $Z_{j}$, once we control for the effects of other predictors, $Z_{j}$ has a low importance. Consistent with this, the importance score is then $I(Z_{j}) \simeq 0$. On the other hand, if $\hat{f}^{ \text{ALE}}$ fluctuates a lot over its domain, it means that $Z_{j}$ has a large influence on the response variable, leading to larger values of $I(Z_{j})$.

Proceeding in this way, we construct a sequence $I(Z_{j})$, for $j = 1, \dots, J$. Our measure of variable importance is based on expressing each $I(Z_{j})$ relative to their sum:
\begin{equation}
VI(Z_{j}) = \frac{I(Z_{j})}{ \sum_{j = 1}^{J} I(Z_{j})}, \quad \text{for } j = 1, \dots, J.
\end{equation}
Note that $VI(Z_{j}) \geq 0$ and $\sum_{j=1}^{J} VI(Z_{j}) = 1$.

\begin{figure}[t!]
\begin{center}
\caption{ALE between explanatory variable and future volatility. \label{figure:ale}}
\begin{tabular}{cc}
\small{Panel A: RVD.} & \small{Panel B: RVW.} \\
\includegraphics[height=8cm,width=0.48\textwidth]{{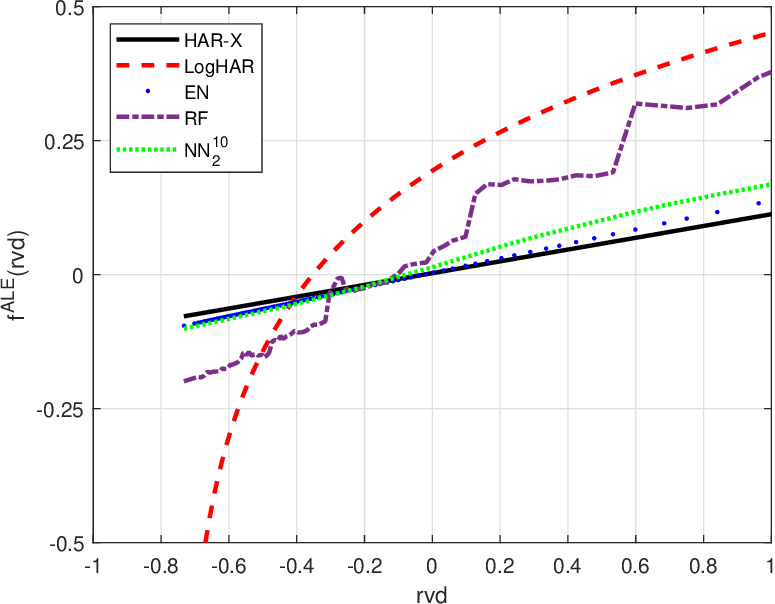}} &
\includegraphics[height=8cm,width=0.48\textwidth]{{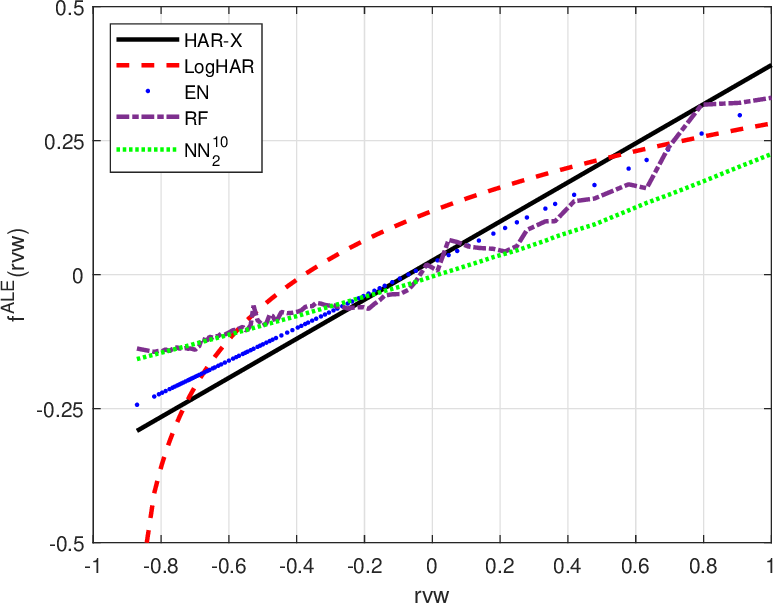}} \\
\small{Panel C: IV.} & \small{Panel D: M1W.} \\
\includegraphics[height=8cm,width=0.48\textwidth]{{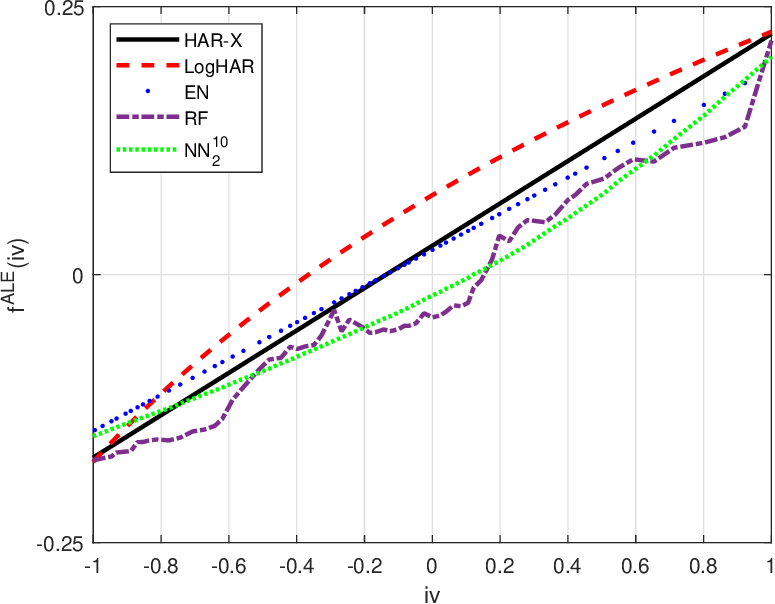}} &
\includegraphics[height=8cm,width=0.48\textwidth]{{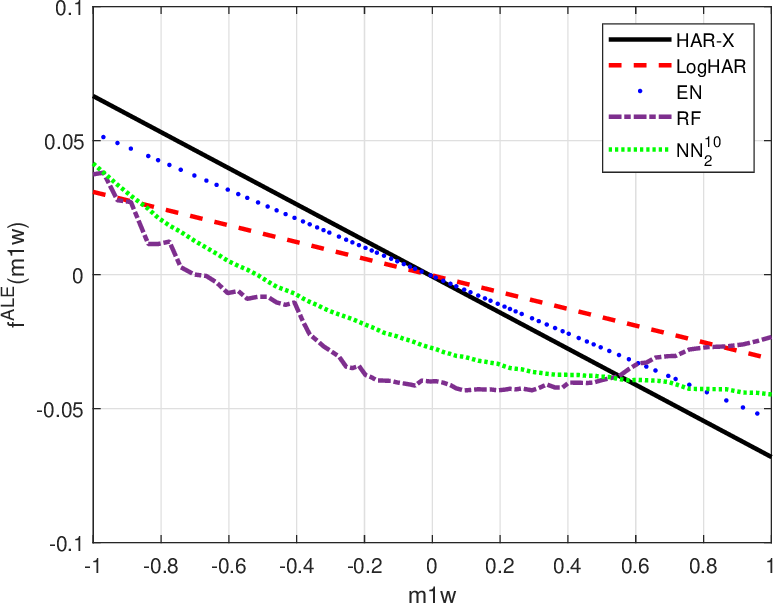}} \\
\end{tabular}
\begin{scriptsize}
\parbox{\textwidth}{\emph{Note.} This figure plots the Accumulated Local Effect (ALE) between $RV_{t}$ and a single explanatory variable, i.e. RVD, RVW, IV, and M1W for Apple's stock price volatility. Note that the covariates are standardized. If the function does not cover the entire domain of the $x$-axis, it is because there are no observations in this area. The $y$-axis is kept constant to illuminate the relative importance of each variable in the prediction.}
\end{scriptsize}
\end{center}
\end{figure}

In Figure \ref{figure:ale}, we plot the ALE on expected realized variance over the interval [-1,1] standard deviation. We restrict attention to a set of selected characteristics that are the most influential drivers of volatility, namely RVD, RVW, IV and M1W. Moreover, we do not explore the marginal effect averaged over stocks, as it smoothes out the unique relationship at the individual asset level. Instead, we arbitrarily select Apple for the illustration. Hence, this part serves as a showcase and may not be representative for the whole sample, although what we observe here is consistent with the broad pattern across stocks. In the construction of Figure \ref{figure:ale}, we again select a single model from each main category.

Firstly, the figure reveals a positive relationship between $RV_{t}$ and RVD, RVW, and IV. In contrast, the marginal association between $RV_{t}$ and M1W is negative. This is corroborated by Table \ref{table:har} in Appendix \ref{appendix:har}, where parameter estimates for the HAR-X model of Apple are reported. Consistent with Figure \ref{figure:ale}, the slope coefficient for M1W is negative and significant. Secondly, the figures illustrate a purely data-driven approach leads to a slightly nonlinear model. Interestingly, the marginal effect for RVD detected by the random forest and NN$_{2}^{10}$, especially the former, are closer to the LogHAR than the HAR. The ML algorithms therefore detect a stronger relationship between $RV_{t}$ and RVD compared to the linear model, whereas the impact of RVW is attenuated. The discrepancy between elastic net and HAR-X is generally caused by the shrinkage of the parameter estimates, such that elastic net ends up suggesting a marginal effect that is more consistent with the random forest.

\begin{figure}[t!]
\begin{center}
\caption{VI measure. \label{figure:variable-importance}}
\begin{tabular}{cc}
\small{Panel A: HAR-X.} & \small{Panel B: Elastic Net.} \\
\includegraphics[height=8cm,width=0.48\textwidth]{{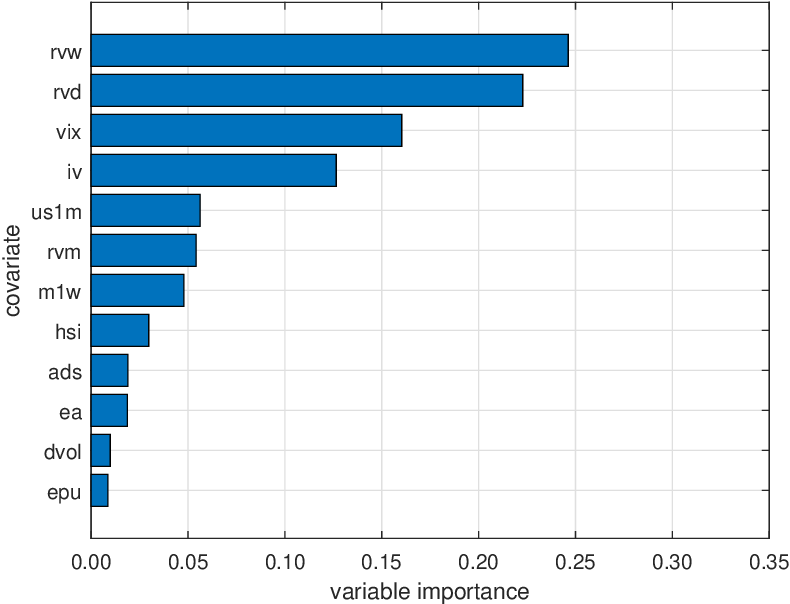}} &
\includegraphics[height=8cm,width=0.48\textwidth]{{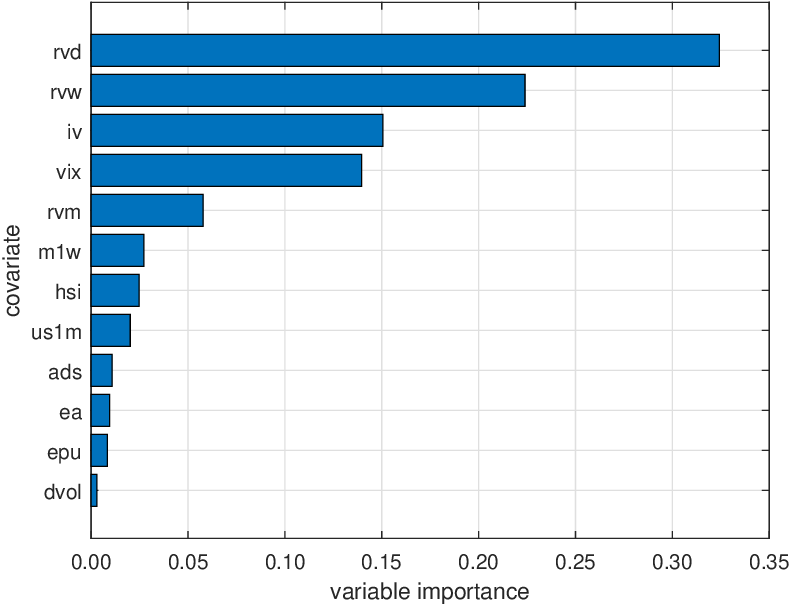}} \\
\small{Panel C: Random Forest.} & \small{Panel D: NN$_{2}^{10}$.} \\
\includegraphics[height=8cm,width=0.48\textwidth]{{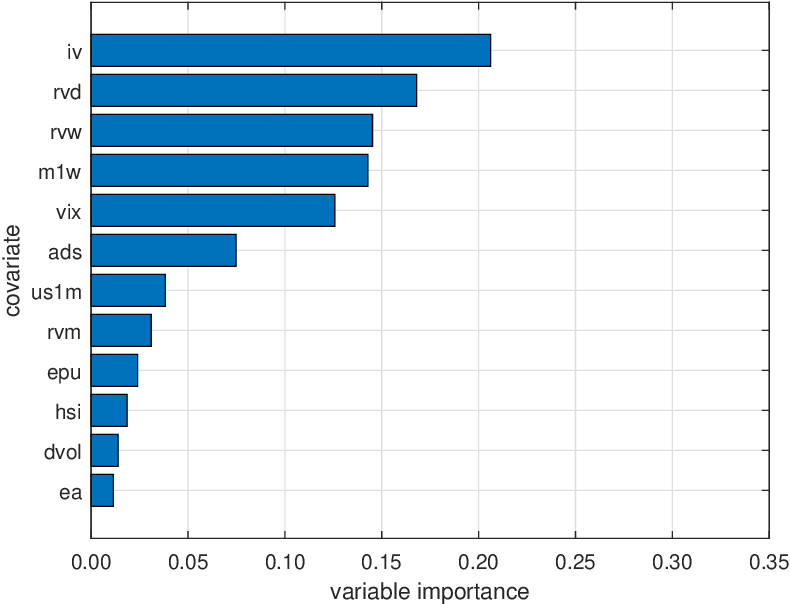}} &
\includegraphics[height=8cm,width=0.48\textwidth]{{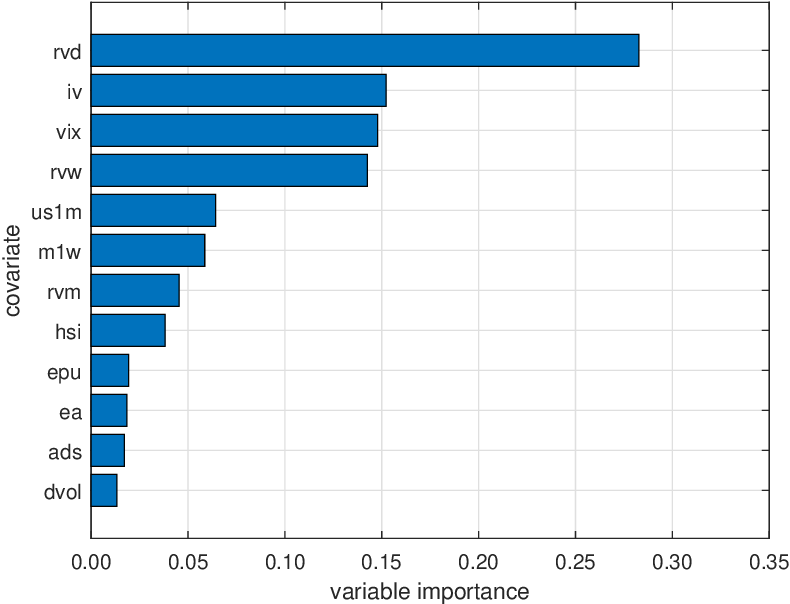}} \\
\end{tabular}
\begin{scriptsize}
\parbox{\textwidth}{\emph{Note.} We report the VI measure for each feature in the $\mathcal{M}_{ \text{ALL}}$ dataset for the HAR-X, elastic net, random forest and NN$_{2}^{10}$ model. The VI measures are sorted in descending order. The figures are based on a cross-sectional average VI measure for each asset.}
\end{scriptsize}
\end{center}
\end{figure}

The cross-sectional average VI measure is reported in Figure \ref{figure:variable-importance}. As illustrated, there is agreement about the dominant set of predictors: RVD, RVW, and IV. Intuitively, these covariates relate to the recent history or future expectation of the stock's volatility. In contrast, less weight is assigned to variables that are not directly related to volatility. However, even if there is consensus about the best set of predictors, there are individual differences in their relative importance. It is interesting to note that IV ranks higher in the random forest and neural network compared to the HAR-X and elastic net. Hence, the ML models weigh market expectations of future volatility higher than past realizations of realized variance in forming their forecast, adding an ML perspective to the literature on the informational content of realized variance versus implied volatility \citep*[e.g.,][]{christensen-prabhala:98a,busch-christensen-nielsen:11a}.

With a nonlinear market structure, valuable information can potentially be extracted from a broader set of covariates. In line with this conjecture, the macroeconomic variable ADS contributes more to the prediction of volatility in the random forest relative to RVM, whereas the ADS variable does not weigh much in the HAR-X. This is again consistent with the parameter estimates of the HAR-X model for Apple in Table \ref{table:har}, where the coefficient in front of ADS is found to be insignificant in the $\mathcal{M}_{ \text{ALL}}$ dataset.

Lasso applies subset selection. Accordingly, it recovers a sparse solution with a median of four beta coefficients set to zero across stocks. Thus, lasso is capable of taming overfitting, but this prevents it from extracting incremental information from several of the explanatory variables, as opposed to elastic net, random forest or neural network.

The EA variable has a low VI score. This highlights a weakness with ALE, since it is mainly a tool for finding the drivers behind the average prediction. Earnings information is released infrequently, but it is arguably very important for making short-run predictions of realized variance in the days surrounding the announcement, see \citet*{siggaard:22a}.

\section{Longer-run forecasting} \label{section:week-month}

In this section, we modify the forecast horizon from one-day-ahead to one-week-ahead and one-month-ahead. We replace the dependent variable in all models from next-day realized variance ($RV_{t+1}$) to next-week average realized variance ($RV_{t+5 \mid t+1}$) and next-month average realized variance ($RV_{t+22 \mid t+1}$), as a function of the information set that is available at time $t$, i.e. $Z_{t}$. Table \ref{table:har-1w.tex} -- \ref{table:all-1w.tex} report the results for the weekly forecasts, whereas the monthly forecast results are available in Table \ref{table:har-1m.tex} -- \ref{table:all-1m.tex}.

\begin{sidewaystable}
\begin{scriptsize}
\setlength{ \tabcolsep}{0.09cm}
\begin{center}
\caption{One-week-ahead relative MSE and Diebold-Mariano test for dataset $\mathcal{M}_{\text{HAR}}$.
\label{table:har-1w.tex}}
\begin{tabular}{lcccccccccccccccccccccc}
\hline \hline
& HAR & HAR-X & LogHAR & LevHAR & SHAR & HARQ & RR & LA & EN & A-LA & P-LA & BG & RF & GB & NN$_{1}^{1}$ & NN$_{1}^{10}$ & NN$_{2}^{1}$ & NN$_{2}^{10}$ & NN$_{3}^{1}$ & NN$_{3}^{10}$ & NN$_{4}^{1}$ & NN$_{4}^{10}$ \\
\hline
HAR  & - & 1.000 & 1.091 & 1.089 & 0.937 & 1.043 & 1.019 & 1.024 & 1.019 & 1.033 & 1.016 & 1.056 & 0.959 & 1.164 & 0.955 & 0.951 & 0.947 & 0.937 & 0.929 & \textit{0.925} & 1.000 & 1.011 \\ 
HAR-X & 1.000  & - & 1.091 & 1.089 & 0.937 & 1.043 & 1.019 & 1.024 & 1.019 & 1.033 & 1.016 & 1.056 & 0.959 & 1.164 & 0.955 & 0.951 & 0.947 & 0.937 & 0.929 & \textit{0.925} & 1.000 & 1.011 \\ 
LogHAR & \textbf{\textit{\underline{0.938}}} & \textbf{\textit{\underline{0.938}}}  & - & 1.007 & \textbf{\textit{0.876}} & 0.960 & \textbf{\textit{0.948}} & \textbf{\textit{0.953}} & \textbf{\textit{0.949}} & \textbf{\textit{0.965}} & \textbf{\textit{\underline{0.952}}} & 0.979 & \textbf{\textit{0.888}} & 1.068 & \textbf{\textit{\underline{0.885}}} & \textbf{\textit{\underline{0.884}}} & \textbf{\textit{\underline{0.879}}} & \textbf{\textit{\underline{0.870}}} & \textbf{\textit{\underline{0.862}}} & \textbf{\textit{\underline{0.859}}} & \textbf{\textit{0.924}} & \textbf{\textit{0.934}} \\ 
LevHAR & 0.942 & 0.942 & 1.016  & - & 0.886 & 0.973 & 0.958 & 0.963 & 0.958 & 0.971 & 0.956 & 0.990 & \textit{0.901} & 1.084 & \textit{0.896} & \textbf{\textit{0.891}} & \textit{0.890} & \textit{0.878} & \textbf{\textit{0.872}} & \textbf{\textit{0.867}} & \textit{0.935} & \textit{0.944} \\ 
SHAR & 1.074 & 1.074 & 1.169 & 1.175  & - & 1.118 & 1.092 & 1.099 & 1.093 & 1.109 & 1.092 & 1.128 & 1.024 & 1.253 & 1.023 & 1.020 & 1.013 & 1.004 & 0.995 & 0.991 & 1.074 & 1.087 \\ 
HARQ & 0.980 & 0.980 & 1.049 & 1.054 & 0.916  & - & 0.989 & 0.994 & 0.990 & 1.006 & 0.994 & 1.021 & 0.928 & 1.112 & 0.925 & 0.923 & 0.918 & 0.909 & 0.899 & 0.897 & 0.965 & 0.975 \\ 
RR & \textbf{\textit{\underline{0.990}}} & \textbf{\textit{\underline{0.990}}} & 1.071 & 1.074 & \textit{0.926} & 1.023  & - & 1.006 & 1.001 & 1.014 & 1.006 & 1.038 & 0.943 & 1.139 & \textbf{\textit{0.939}} & \textbf{\textit{0.938}} & \textit{0.932} & \textit{0.922} & \textbf{\textit{0.914}} & \textbf{\textit{0.911}} & 0.983 & 0.993 \\ 
LA & \textbf{\textit{\underline{0.986}}} & \textbf{\textit{\underline{0.986}}} & 1.066 & 1.069 & \textit{0.921} & 1.018 & 0.995  & - & 0.996 & 1.008 & \textbf{\textit{\underline{1.001}}} & 1.032 & 0.938 & 1.132 & \textbf{\textit{0.934}} & \textbf{\textit{0.933}} & \textit{0.927} & \textbf{\textit{0.917}} & \textbf{\textit{0.909}} & \textbf{\textit{0.906}} & 0.978 & 0.988 \\ 
EN & \textbf{\textit{\underline{0.990}}} & \textbf{\textit{\underline{0.990}}} & 1.071 & 1.074 & \textit{0.925} & 1.022 & 0.999 & 1.005  & - & 1.014 & \textit{1.005} & 1.037 & 0.942 & 1.138 & \textbf{\textit{0.938}} & \textbf{\textit{0.937}} & \textit{0.931} & \textit{0.921} & \textbf{\textit{0.913}} & \textbf{\textit{0.910}} & 0.982 & 0.992 \\ 
A-LA & \textbf{\textit{\underline{0.979}}} & \textbf{\textit{\underline{0.979}}} & 1.062 & 1.062 & \textit{0.916} & 1.014 & 0.989 & 0.993 & 0.989  & - & \textbf{\textit{0.994}} & 1.026 & 0.933 & 1.126 & \textbf{\textit{0.929}} & \textbf{\textit{0.927}} & \textit{0.922} & \textbf{\textit{0.911}} & \textbf{\textit{0.904}} & \textbf{\textit{0.901}} & 0.973 & 0.983 \\ 
P-LA & \textbf{\textit{\underline{0.984}}} & \textbf{\textit{\underline{0.984}}} & 1.072 & 1.069 & 0.923 & 1.024 & 1.002 & 1.008 & 1.003 & 1.016  & - & 1.037 & 0.943 & 1.142 & 0.939 & \textit{0.935} & 0.932 & 0.921 & \textit{0.913} & \textit{0.909} & 0.983 & 0.993 \\ 
BG & 0.977 & 0.977 & 1.052 & 1.057 & 0.909 & 1.005 & 0.988 & 0.993 & 0.988 & 1.002 & 0.991  & - & \textit{0.919} & 1.110 & 0.922 & \textit{0.922} & 0.912 & \textit{0.905} & \textit{0.897} & \textbf{\textit{0.893}} & 0.964 & 0.976 \\ 
RF & 1.060 & 1.060 & 1.142 & 1.149 & 0.986 & 1.092 & 1.072 & 1.078 & 1.073 & 1.089 & 1.076 & 1.097  & - & 1.214 & 1.002 & 1.001 & 0.993 & 0.984 & 0.975 & 0.971 & 1.048 & 1.061 \\ 
GB & 0.906 & 0.906 & 0.967 & 0.971 & \textit{0.848} & 0.923 & 0.911 & 0.915 & 0.912 & 0.924 & 0.918 & 0.935 & \textit{0.855}  & - & \textit{0.853} & \textbf{\textit{0.853}} & \textit{0.846} & \textbf{\textit{0.837}} & \textbf{\textit{0.828}} & \textbf{\textit{0.827}} & 0.887 & 0.898 \\ 
NN$_{1}^{1}$ & 1.058 & 1.058 & 1.140 & 1.144 & 0.988 & 1.090 & 1.071 & 1.076 & 1.071 & 1.087 & 1.074 & 1.104 & 1.005 & 1.216  & - & 0.999 & 0.994 & 0.984 & 0.975 & 0.971 & 1.048 & 1.059 \\ 
NN$_{1}^{10}$ & 1.058 & 1.058 & 1.144 & 1.143 & 0.990 & 1.094 & 1.074 & 1.080 & 1.075 & 1.090 & 1.075 & 1.110 & 1.009 & 1.221 & 1.004  & - & 0.996 & 0.986 & 0.978 & 0.973 & 1.051 & 1.063 \\ 
NN$_{2}^{1}$ & 1.067 & 1.067 & 1.152 & 1.156 & 0.995 & 1.101 & 1.080 & 1.086 & 1.081 & 1.096 & 1.083 & 1.110 & 1.012 & 1.225 & 1.010 & 1.008  & - & 0.991 & 0.983 & \textbf{\textit{0.978}} & 1.056 & 1.068 \\ 
NN$_{2}^{10}$ & 1.076 & 1.076 & 1.162 & 1.163 & 1.005 & 1.111 & 1.090 & 1.095 & 1.091 & 1.105 & 1.092 & 1.124 & 1.023 & 1.236 & 1.020 & 1.017 & 1.011  & - & 0.992 & \textit{0.988} & 1.066 & 1.078 \\ 
NN$_{3}^{1}$ & 1.087 & 1.087 & 1.173 & 1.177 & 1.015 & 1.120 & 1.100 & 1.106 & 1.101 & 1.117 & 1.103 & 1.134 & 1.032 & 1.245 & 1.030 & 1.027 & 1.021 & 1.011  & - & 0.997 & 1.076 & 1.087 \\ 
NN$_{3}^{10}$ & 1.089 & 1.089 & 1.177 & 1.179 & 1.018 & 1.125 & 1.104 & 1.110 & 1.105 & 1.121 & 1.106 & 1.138 & 1.035 & 1.252 & 1.033 & 1.030 & 1.024 & 1.013 & 1.004  & - & 1.079 & 1.091 \\ 
NN$_{4}^{1}$ & 1.017 & 1.017 & 1.093 & 1.095 & 0.952 & 1.044 & 1.028 & 1.034 & 1.029 & 1.045 & 1.032 & 1.061 & 0.965 & 1.158 & \textbf{\textit{0.962}} & \textit{0.960} & \textit{0.954} & \textbf{\textit{\underline{0.944}}} & \textbf{\textit{0.935}} & \textbf{\textit{\underline{0.931}}}  & - & 1.011 \\ 
NN$_{4}^{10}$ & 1.008 & 1.008 & 1.083 & 1.084 & 0.944 & 1.035 & 1.019 & 1.025 & 1.020 & 1.036 & 1.022 & 1.053 & \textit{0.957} & 1.150 & \textbf{\textit{0.954}} & \textbf{\textit{0.952}} & \textbf{\textit{0.946}} & \textbf{\textit{\underline{0.936}}} & \textbf{\textit{\underline{0.927}}} & \textbf{\textit{\underline{0.923}}} & 0.991  & - \\ 
\hline \hline
\end{tabular}
\smallskip
\begin{scriptsize}
\parbox{0.98\textwidth}{\emph{Note.} We report the out-of-sample realized variance forecast MSE of each model in the selected column relative to the benchmark in the selected row. Each number is a cross-sectional average of such pairwise relative MSEs for each stock. 
The formatting is as follows: \textit{number} (\textbf{\textit{number}}) [\underline{\textbf{\textit{number}}}] denotes whether the Diebold-Mariano test of equal predictive accuracy is rejected more than 50\% of the time at the 10\% (5\%) [1\%] level of significance across individual tests for each asset. 
The hypothesis being tested is $\text{H}_{0}: \text{MSE}_{i} = \text{MSE}_{j}$ against a one-sided alternative $\text{H}_{1}: \text{MSE}_{i} > \text{MSE}_{j}$, 
where model $i$ is the label of the selected row, whereas model $j$ is the label of the selected column.}
\end{scriptsize}
\end{center}
\end{scriptsize}
\end{sidewaystable}
\begin{sidewaystable}
\begin{scriptsize}
\setlength{ \tabcolsep}{0.09cm}
\begin{center}
\caption{One-week-ahead relative MSE and Diebold-Mariano test for dataset $\mathcal{M}_{\text{ALL}}$.
\label{table:all-1w.tex}}
\begin{tabular}{lcccccccccccccccccccccc}
\hline \hline
& HAR & HAR-X & LogHAR & LevHAR & SHAR & HARQ & RR & LA & EN & A-LA & P-LA & BG & RF & GB & NN$_{1}^{1}$ & NN$_{1}^{10}$ & NN$_{2}^{1}$ & NN$_{2}^{10}$ & NN$_{3}^{1}$ & NN$_{3}^{10}$ & NN$_{4}^{1}$ & NN$_{4}^{10}$ \\
\hline
HAR  & - & 1.037 & 0.970 & 1.076 & 0.913 & 1.459 & \textit{0.954} & \textit{0.958} & \textbf{\textit{0.925}} & 1.026 & 1.033 & \textbf{\textit{0.840}} & \textbf{\textit{\underline{0.796}}} & \textit{0.930} & \textbf{\textit{\underline{0.877}}} & \textbf{\textit{\underline{0.857}}} & \textbf{\textit{0.865}} & \textbf{\textit{\underline{0.846}}} & \textbf{\textit{0.882}} & \textbf{\textit{\underline{0.855}}} & \textit{0.913} & \textit{0.945} \\ 
HAR-X & 1.004  & - & \textit{0.951} & 1.027 & 0.917 & 1.321 & \textit{0.924} & \textbf{\textit{0.929}} & \textbf{\textit{0.899}} & 0.986 & 1.006 & \textbf{\textit{\underline{0.835}}} & \textbf{\textit{\underline{0.791}}} & \textit{0.919} & \textbf{\textit{\underline{0.870}}} & \textbf{\textit{\underline{0.853}}} & \textbf{\textit{\underline{0.859}}} & \textbf{\textit{\underline{0.842}}} & \textbf{\textit{\underline{0.867}}} & \textbf{\textit{\underline{0.847}}} & \textit{0.898} & \textit{0.923} \\ 
LogHAR & 1.064 & 1.073  & - & 1.104 & 0.966 & 1.439 & 0.986 & 0.992 & 0.958 & 1.059 & 1.076 & \textbf{\textit{\underline{0.884}}} & \textbf{\textit{\underline{0.838}}} & 0.975 & \textbf{\textit{0.924}} & \textbf{\textit{\underline{0.904}}} & \textbf{\textit{0.908}} & \textbf{\textit{0.891}} & \textbf{\textit{0.921}} & \textbf{\textit{\underline{0.898}}} & 0.952 & 0.979 \\ 
LevHAR & 0.988 & 0.977 & \textbf{\textit{0.930}}  & - & 0.902 & 1.272 & \textbf{\textit{0.903}} & \textit{0.909} & \textbf{\textit{0.879}} & 0.963 & 0.985 & \textbf{\textit{\underline{0.819}}} & \textbf{\textit{\underline{0.777}}} & 0.902 & \textbf{\textit{\underline{0.854}}} & \textbf{\textit{\underline{0.838}}} & \textbf{\textit{\underline{0.843}}} & \textbf{\textit{\underline{0.827}}} & \textbf{\textit{\underline{0.850}}} & \textbf{\textit{\underline{0.831}}} & \textit{0.881} & \textit{0.903} \\ 
SHAR & 1.102 & 1.144 & 1.065 & 1.189  & - & 1.622 & 1.051 & 1.055 & 1.018 & 1.131 & 1.139 & \textbf{\textit{0.923}} & \textbf{\textit{\underline{0.874}}} & 1.023 & 0.966 & \textit{0.942} & \textit{0.950} & \textbf{\textit{0.929}} & \textit{0.970} & \textbf{\textit{0.940}} & 1.004 & 1.040 \\ 
HARQ & \textbf{\textit{0.856}} & \textbf{\textit{\underline{0.825}}} & \textbf{\textit{\underline{0.790}}} & \textbf{\textit{0.839}} & \textbf{\textit{0.781}}  & - & \textbf{\textit{\underline{0.765}}} & \textbf{\textit{\underline{0.769}}} & \textbf{\textit{\underline{0.746}}} & \textbf{\textit{\underline{0.811}}} & \textbf{\textit{0.839}} & \textbf{\textit{\underline{0.708}}} & \textbf{\textit{\underline{0.670}}} & \textbf{\textit{\underline{0.771}}} & \textbf{\textit{\underline{0.736}}} & \textbf{\textit{\underline{0.724}}} & \textbf{\textit{\underline{0.727}}} & \textbf{\textit{\underline{0.715}}} & \textbf{\textit{\underline{0.729}}} & \textbf{\textit{\underline{0.715}}} & \textbf{\textit{\underline{0.749}}} & \textbf{\textit{\underline{0.763}}} \\ 
RR & 1.087 & 1.088 & 1.027 & 1.117 & 0.990 & 1.441  & - & 1.008 & \textit{0.973} & 1.073 & 1.094 & \textbf{\textit{0.904}} & \textbf{\textit{\underline{0.856}}} & 0.997 & \textit{0.943} & \textbf{\textit{0.924}} & \textbf{\textit{0.930}} & \textbf{\textit{\underline{0.911}}} & \textbf{\textit{0.938}} & \textbf{\textit{0.917}} & 0.972 & 0.998 \\ 
LA & 1.082 & 1.081 & 1.023 & 1.113 & 0.985 & 1.438 & 0.997  & - & \textit{0.968} & 1.065 & 1.087 & \textbf{\textit{0.900}} & \textbf{\textit{\underline{0.852}}} & 0.989 & \textbf{\textit{0.938}} & \textbf{\textit{0.919}} & \textbf{\textit{0.924}} & \textbf{\textit{\underline{0.906}}} & \textbf{\textit{0.934}} & \textbf{\textit{\underline{0.912}}} & 0.966 & 0.993 \\ 
EN & 1.116 & 1.118 & 1.056 & 1.150 & 1.016 & 1.489 & 1.029 & 1.034  & - & 1.102 & 1.123 & \textbf{\textit{0.928}} & \textbf{\textit{\underline{0.879}}} & 1.022 & 0.968 & 0.948 & 0.953 & \textbf{\textit{0.935}} & \textit{0.963} & \textbf{\textit{0.941}} & 0.998 & 1.025 \\ 
A-LA & 1.029 & 1.020 & 0.971 & 1.048 & 0.937 & 1.344 & 0.943 & 0.946 & \textbf{\textit{0.916}}  & - & 1.028 & \textbf{\textit{\underline{0.854}}} & \textbf{\textit{\underline{0.809}}} & 0.937 & \textbf{\textit{0.891}} & \textbf{\textit{0.873}} & \textbf{\textit{0.878}} & \textbf{\textit{\underline{0.861}}} & \textbf{\textit{0.887}} & \textbf{\textit{\underline{0.866}}} & \textit{0.917} & 0.942 \\ 
P-LA & 0.995 & 0.998 & \textbf{\textit{0.946}} & 1.028 & 0.908 & 1.341 & \textbf{\textit{0.923}} & \textbf{\textit{0.926}} & \textbf{\textit{\underline{0.896}}} & 0.985  & - & \textbf{\textit{\underline{0.830}}} & \textbf{\textit{\underline{0.786}}} & \textit{0.915} & \textbf{\textit{\underline{0.864}}} & \textbf{\textit{\underline{0.847}}} & \textbf{\textit{\underline{0.852}}} & \textbf{\textit{\underline{0.835}}} & \textbf{\textit{\underline{0.862}}} & \textbf{\textit{\underline{0.841}}} & \textbf{\textit{0.893}} & \textbf{\textit{0.918}} \\ 
BG & 1.208 & 1.235 & 1.157 & 1.278 & 1.098 & 1.715 & 1.138 & 1.143 & 1.104 & 1.222 & 1.237  & - & 0.952 & 1.109 & 1.052 & 1.028 & 1.035 & 1.013 & 1.052 & 1.023 & 1.092 & 1.130 \\ 
RF & 1.266 & 1.297 & 1.215 & 1.343 & 1.151 & 1.802 & 1.193 & 1.199 & 1.158 & 1.283 & 1.297 & 1.054  & - & 1.168 & 1.105 & 1.079 & 1.087 & 1.064 & 1.105 & 1.075 & 1.145 & 1.184 \\ 
GB & 1.099 & 1.117 & 1.048 & 1.156 & 0.999 & 1.530 & 1.030 & 1.033 & 0.999 & 1.102 & 1.121 & 0.911 & \textbf{\textit{\underline{0.866}}}  & - & 0.956 & 0.935 & 0.941 & \textit{0.922} & \textit{0.956} & \textit{0.930} & 0.987 & 1.020 \\ 
NN$_{1}^{1}$ & 1.148 & 1.173 & 1.101 & 1.213 & 1.046 & 1.617 & 1.080 & 1.085 & 1.048 & 1.162 & 1.173 & 0.957 & \textbf{\textit{0.908}} & 1.059  & - & 0.978 & 0.986 & 0.965 & 1.001 & 0.974 & 1.037 & 1.072 \\ 
NN$_{1}^{10}$ & 1.175 & 1.206 & 1.130 & 1.249 & 1.070 & 1.679 & 1.109 & 1.114 & 1.076 & 1.194 & 1.204 & 0.980 & \textit{0.930} & 1.087 & 1.024  & - & 1.009 & 0.986 & 1.025 & 0.997 & 1.063 & 1.099 \\ 
NN$_{2}^{1}$ & 1.171 & 1.198 & 1.120 & 1.240 & 1.064 & 1.658 & 1.102 & 1.105 & 1.068 & 1.186 & 1.196 & 0.974 & \textbf{\textit{0.925}} & 1.079 & 1.019 & 0.996  & - & 0.980 & 1.018 & 0.991 & 1.055 & 1.090 \\ 
NN$_{2}^{10}$ & 1.194 & 1.227 & 1.147 & 1.271 & 1.086 & 1.712 & 1.127 & 1.132 & 1.093 & 1.215 & 1.223 & 0.996 & \textbf{\textit{0.944}} & 1.104 & 1.041 & 1.016 & 1.023  & - & 1.039 & 1.012 & 1.079 & 1.115 \\ 
NN$_{3}^{1}$ & 1.158 & 1.175 & 1.104 & 1.213 & 1.054 & 1.606 & 1.081 & 1.086 & 1.049 & 1.163 & 1.176 & 0.963 & \textbf{\textit{0.914}} & 1.066 & 1.007 & 0.983 & 0.990 & 0.969  & - & 0.977 & 1.041 & 1.070 \\ 
NN$_{3}^{10}$ & 1.182 & 1.206 & 1.131 & 1.247 & 1.076 & 1.663 & 1.109 & 1.114 & 1.076 & 1.193 & 1.206 & 0.984 & \textbf{\textit{0.934}} & 1.089 & 1.029 & 1.005 & 1.012 & 0.990 & 1.026  & - & 1.066 & 1.099 \\ 
NN$_{4}^{1}$ & 1.120 & 1.135 & 1.065 & 1.172 & 1.019 & 1.534 & 1.044 & 1.048 & 1.013 & 1.121 & 1.136 & 0.932 & \textbf{\textit{\underline{0.883}}} & 1.026 & 0.973 & \textit{0.952} & \textit{0.957} & \textbf{\textit{0.937}} & \textit{0.970} & \textbf{\textit{0.946}}  & - & 1.031 \\ 
NN$_{4}^{10}$ & 1.094 & 1.100 & 1.035 & 1.133 & 0.996 & 1.467 & 1.012 & 1.017 & 0.983 & 1.087 & 1.104 & \textit{0.910} & \textbf{\textit{\underline{0.862}}} & 1.001 & 0.949 & \textbf{\textit{0.928}} & \textit{0.933} & \textbf{\textit{\underline{0.914}}} & \textbf{\textit{\underline{0.942}}} & \textbf{\textit{\underline{0.921}}} & 0.974  & - \\ 
\hline \hline
\end{tabular}
\smallskip
\begin{scriptsize}
\parbox{0.98\textwidth}{\emph{Note.} We report the out-of-sample realized variance forecast MSE of each model in the selected column relative to the benchmark in the selected row. Each number is a cross-sectional average of such pairwise relative MSEs for each stock. 
The formatting is as follows: \textit{number} (\textbf{\textit{number}}) [\underline{\textbf{\textit{number}}}] denotes whether the Diebold-Mariano test of equal predictive accuracy is rejected more than 50\% of the time at the 10\% (5\%) [1\%] level of significance across individual tests for each asset. 
The hypothesis being tested is $\text{H}_{0}: \text{MSE}_{i} = \text{MSE}_{j}$ against a one-sided alternative $\text{H}_{1}: \text{MSE}_{i} > \text{MSE}_{j}$, 
where model $i$ is the label of the selected row, whereas model $j$ is the label of the selected column.}
\end{scriptsize}
\end{center}
\end{scriptsize}
\end{sidewaystable}
\begin{sidewaystable}
\begin{scriptsize}
\setlength{ \tabcolsep}{0.09cm}
\begin{center}
\caption{One-month-ahead relative MSE and Diebold-Mariano test for dataset $\mathcal{M}_{\text{HAR}}$.
\label{table:har-1m.tex}}
\begin{tabular}{lcccccccccccccccccccccc}
\hline \hline
& HAR & HAR-X & LogHAR & LevHAR & SHAR & HARQ & RR & LA & EN & A-LA & P-LA & BG & RF & GB & NN$_{1}^{1}$ & NN$_{1}^{10}$ & NN$_{2}^{1}$ & NN$_{2}^{10}$ & NN$_{3}^{1}$ & NN$_{3}^{10}$ & NN$_{4}^{1}$ & NN$_{4}^{10}$ \\
\hline
HAR  & - & 1.000 & 1.052 & 1.030 & 1.034 & 0.994 & 1.100 & 1.107 & 1.111 & 1.118 & 1.057 & 1.171 & 1.019 & 1.219 & 0.993 & 0.970 & 0.989 & 0.958 & \textit{0.904} & 0.905 & 1.027 & 1.068 \\ 
HAR-X & 1.000  & - & 1.052 & 1.030 & 1.034 & 0.994 & 1.100 & 1.107 & 1.111 & 1.118 & 1.057 & 1.171 & 1.019 & 1.219 & 0.993 & 0.970 & 0.989 & 0.958 & \textit{0.904} & 0.905 & 1.027 & 1.068 \\ 
LogHAR & 0.988 & 0.988  & - & 1.009 & 1.029 & 0.957 & 1.069 & 1.076 & 1.085 & 1.087 & 1.031 & 1.131 & 0.989 & 1.170 & \textbf{\textit{0.966}} & \textbf{\textit{\underline{0.942}}} & 0.959 & \textit{0.930} & \textbf{\textit{\underline{0.876}}} & \textbf{\textit{\underline{0.877}}} & \textit{0.990} & 1.026 \\ 
LevHAR & 0.984 & 0.984 & 1.028  & - & 1.011 & 0.974 & 1.077 & 1.083 & 1.089 & 1.092 & 1.037 & 1.143 & 0.997 & 1.189 & 0.973 & 0.950 & 0.965 & 0.936 & 0.886 & \textit{0.888} & 1.004 & 1.046 \\ 
SHAR & 0.981 & 0.981 & 1.037 & 1.006  & - & 0.979 & 1.080 & 1.087 & 1.091 & 1.097 & 1.038 & 1.137 & 0.995 & 1.193 & 0.973 & 0.950 & 0.965 & 0.937 & 0.885 & 0.889 & 1.013 & 1.053 \\ 
HARQ & 1.037 & 1.037 & 1.062 & 1.061 & 1.080  & - & 1.121 & 1.127 & 1.137 & 1.139 & 1.079 & 1.172 & 1.028 & 1.230 & 1.015 & 0.991 & 1.000 & 0.973 & \textit{0.916} & \textit{0.918} & 1.034 & 1.074 \\ 
RR & \textbf{\textit{\underline{0.939}}} & \textbf{\textit{\underline{0.939}}} & 0.967 & \textit{0.959} & \textbf{\textit{\underline{0.972}}} & 0.913  & - & 1.007 & 1.021 & 1.016 & \textbf{\textit{\underline{0.966}}} & 1.077 & 0.937 & 1.109 & \textit{0.921} & \textbf{\textit{0.899}} & \textbf{\textit{0.911}} & \textbf{\textit{0.883}} & \textbf{\textit{\underline{0.831}}} & \textbf{\textit{\underline{0.833}}} & 0.937 & 0.975 \\ 
LA & \textbf{\textit{\underline{0.934}}} & \textbf{\textit{\underline{0.934}}} & 0.960 & \textbf{\textit{0.953}} & \textbf{\textit{\underline{0.966}}} & \textit{0.907} & \textit{0.995}  & - & 1.014 & 1.008 & \textbf{\textit{\underline{0.960}}} & 1.065 & 0.928 & 1.100 & \textbf{\textit{0.913}} & \textbf{\textit{\underline{0.892}}} & \textbf{\textit{0.901}} & \textbf{\textit{\underline{0.876}}} & \textbf{\textit{\underline{0.824}}} & \textbf{\textit{\underline{0.827}}} & \textit{0.930} & 0.968 \\ 
EN & \textbf{\textit{\underline{0.911}}} & \textbf{\textit{\underline{0.911}}} & 0.947 & \textbf{\textit{0.934}} & \textbf{\textit{\underline{0.944}}} & 0.896 & 0.986 & 0.992  & - & 1.000 & \textbf{\textit{\underline{0.951}}} & 1.059 & 0.920 & 1.093 & \textbf{\textit{0.898}} & \textbf{\textit{\underline{0.877}}} & \textbf{\textit{0.893}} & \textbf{\textit{\underline{0.865}}} & \textbf{\textit{\underline{0.815}}} & \textbf{\textit{\underline{0.817}}} & \textit{0.920} & 0.957 \\ 
A-LA & \textbf{\textit{\underline{0.932}}} & \textbf{\textit{\underline{0.932}}} & 0.957 & \textbf{\textit{\underline{0.948}}} & \textbf{\textit{\underline{0.963}}} & \textit{0.904} & 0.990 & 0.995 & 1.010  & - & \textbf{\textit{\underline{0.958}}} & 1.061 & 0.924 & 1.094 & \textbf{\textit{0.908}} & \textbf{\textit{\underline{0.888}}} & \textbf{\textit{0.897}} & \textbf{\textit{\underline{0.872}}} & \textbf{\textit{\underline{0.820}}} & \textbf{\textit{\underline{0.823}}} & \textbf{\textit{0.925}} & 0.965 \\ 
P-LA & \textbf{\textit{\underline{0.971}}} & \textbf{\textit{\underline{0.971}}} & 1.003 & 0.994 & \textbf{\textit{1.006}} & 0.947 & 1.041 & 1.047 & 1.061 & 1.059  & - & 1.116 & 0.971 & 1.157 & \textit{0.958} & \textbf{\textit{0.935}} & 0.947 & \textit{0.919} & \textbf{\textit{\underline{0.865}}} & \textbf{\textit{0.867}} & 0.974 & 1.013 \\ 
BG & 1.001 & 1.001 & 1.025 & 1.026 & 1.022 & 0.963 & 1.086 & 1.088 & 1.103 & 1.098 & 1.046  & - & \textit{0.919} & 1.136 & 0.950 & 0.931 & 0.912 & 0.903 & 0.852 & 0.859 & 0.992 & 1.034 \\ 
RF & 1.060 & 1.060 & 1.094 & 1.086 & 1.088 & 1.026 & 1.149 & 1.152 & 1.166 & 1.162 & 1.106 & 1.113  & - & 1.224 & 1.019 & 0.997 & 0.989 & 0.973 & 0.917 & 0.922 & 1.054 & 1.100 \\ 
GB & \textit{0.890} & \textit{0.890} & 0.906 & 0.906 & 0.917 & 0.857 & 0.948 & 0.952 & 0.967 & 0.959 & 0.918 & 0.957 & \textit{0.853}  & - & 0.856 & \textit{0.837} & \textit{0.834} & \textit{0.817} & \textbf{\textit{0.773}} & \textbf{\textit{0.776}} & 0.879 & 0.915 \\ 
NN$_{1}^{1}$ & 1.041 & 1.041 & 1.079 & 1.068 & 1.073 & 1.022 & 1.139 & 1.144 & 1.152 & 1.153 & 1.100 & 1.173 & 1.031 & 1.237  & - & 0.986 & 0.996 & 0.971 & 0.915 & \textit{0.918} & 1.053 & 1.097 \\ 
NN$_{1}^{10}$ & 1.055 & 1.055 & 1.092 & 1.081 & 1.088 & 1.035 & 1.152 & 1.158 & 1.166 & 1.167 & 1.113 & 1.190 & 1.045 & 1.252 & 1.022  & - & 1.010 & 0.984 & 0.928 & \textbf{\textit{0.931}} & 1.066 & 1.110 \\ 
NN$_{2}^{1}$ & 1.079 & 1.079 & 1.111 & 1.103 & 1.107 & 1.046 & 1.173 & 1.176 & 1.189 & 1.185 & 1.133 & 1.161 & 1.035 & 1.252 & 1.033 & 1.011  & - & 0.985 & \textit{0.930} & \textbf{\textit{0.936}} & 1.078 & 1.125 \\ 
NN$_{2}^{10}$ & 1.084 & 1.084 & 1.120 & 1.109 & 1.115 & 1.056 & 1.178 & 1.183 & 1.194 & 1.193 & 1.139 & 1.198 & 1.059 & 1.273 & 1.047 & 1.024 & 1.024  & - & \textbf{\textit{0.944}} & \textbf{\textit{0.948}} & 1.086 & 1.130 \\ 
NN$_{3}^{1}$ & 1.157 & 1.157 & 1.195 & 1.188 & 1.192 & 1.126 & 1.255 & 1.261 & 1.273 & 1.271 & 1.213 & 1.281 & 1.129 & 1.365 & 1.116 & 1.092 & 1.094 & 1.068  & - & 1.007 & 1.154 & 1.200 \\ 
NN$_{3}^{10}$ & 1.148 & 1.148 & 1.184 & 1.179 & 1.188 & 1.117 & 1.247 & 1.253 & 1.264 & 1.263 & 1.205 & 1.280 & 1.125 & 1.356 & 1.108 & 1.085 & 1.090 & 1.062 & 0.997  & - & 1.143 & 1.189 \\ 
NN$_{4}^{1}$ & 1.025 & 1.025 & 1.048 & 1.045 & 1.069 & 0.987 & 1.100 & 1.104 & 1.117 & 1.113 & 1.061 & 1.156 & 1.009 & 1.205 & 0.996 & 0.974 & \textit{0.981} & \textbf{\textit{0.954}} & \textbf{\textit{\underline{0.896}}} & \textbf{\textit{\underline{0.897}}}  & - & 1.043 \\ 
NN$_{4}^{10}$ & \textbf{\textit{\underline{0.984}}} & \textbf{\textit{\underline{0.984}}} & 1.003 & 1.005 & \textbf{\textit{\underline{1.026}}} & 0.947 & 1.057 & 1.063 & 1.073 & 1.072 & 1.020 & 1.117 & \textit{0.975} & 1.159 & \textbf{\textit{\underline{0.960}}} & \textbf{\textit{\underline{0.936}}} & \textbf{\textit{0.947}} & \textbf{\textit{\underline{0.917}}} & \textbf{\textit{\underline{0.862}}} & \textbf{\textit{\underline{0.862}}} & \textbf{\textit{0.963}}  & - \\ 
\hline \hline
\end{tabular}
\smallskip
\begin{scriptsize}
\parbox{0.98\textwidth}{\emph{Note.} We report the out-of-sample realized variance forecast MSE of each model in the selected column relative to the benchmark in the selected row. Each number is a cross-sectional average of such pairwise relative MSEs for each stock. 
The formatting is as follows: \textit{number} (\textbf{\textit{number}}) [\underline{\textbf{\textit{number}}}] denotes whether the Diebold-Mariano test of equal predictive accuracy is rejected more than 50\% of the time at the 10\% (5\%) [1\%] level of significance across individual tests for each asset. 
The hypothesis being tested is $\text{H}_{0}: \text{MSE}_{i} = \text{MSE}_{j}$ against a one-sided alternative $\text{H}_{1}: \text{MSE}_{i} > \text{MSE}_{j}$, 
where model $i$ is the label of the selected row, whereas model $j$ is the label of the selected column.}
\end{scriptsize}
\end{center}
\end{scriptsize}
\end{sidewaystable}
\begin{sidewaystable}
\begin{scriptsize}
\setlength{ \tabcolsep}{0.09cm}
\begin{center}
\caption{One-month-ahead relative MSE and Diebold-Mariano test for dataset $\mathcal{M}_{\text{ALL}}$.
\label{table:all-1m.tex}}
\begin{tabular}{lcccccccccccccccccccccc}
\hline \hline
& HAR & HAR-X & LogHAR & LevHAR & SHAR & HARQ & RR & LA & EN & A-LA & P-LA & BG & RF & GB & NN$_{1}^{1}$ & NN$_{1}^{10}$ & NN$_{2}^{1}$ & NN$_{2}^{10}$ & NN$_{3}^{1}$ & NN$_{3}^{10}$ & NN$_{4}^{1}$ & NN$_{4}^{10}$ \\
\hline
HAR  & - & 1.290 & 0.919 & 1.328 & 1.312 & 1.569 & 1.111 & 1.378 & 1.107 & 1.386 & 1.363 & \textbf{\textit{\underline{0.621}}} & \textbf{\textit{\underline{0.604}}} & \textbf{\textit{0.829}} & \textbf{\textit{\underline{0.800}}} & \textbf{\textit{\underline{0.789}}} & \textbf{\textit{\underline{0.816}}} & \textbf{\textit{\underline{0.786}}} & \textbf{\textit{\underline{0.788}}} & \textbf{\textit{\underline{0.795}}} & \textbf{\textit{0.902}} & 0.964 \\ 
HAR-X & \textbf{\textit{\underline{0.911}}}  & - & \textbf{\textit{\underline{0.790}}} & 1.021 & 1.021 & 1.166 & \textbf{\textit{0.898}} & 1.023 & \textbf{\textit{0.890}} & 1.029 & 1.018 & \textbf{\textit{\underline{0.551}}} & \textbf{\textit{\underline{0.541}}} & \textbf{\textit{\underline{0.738}}} & \textbf{\textit{\underline{0.714}}} & \textbf{\textit{\underline{0.708}}} & \textbf{\textit{\underline{0.736}}} & \textbf{\textit{\underline{0.711}}} & \textbf{\textit{\underline{0.699}}} & \textbf{\textit{\underline{0.707}}} & \textbf{\textit{\underline{0.800}}} & \textbf{\textit{\underline{0.836}}} \\ 
LogHAR & 1.169 & 1.402  & - & 1.438 & 1.430 & 1.672 & 1.212 & 1.470 & 1.203 & 1.476 & 1.459 & \textbf{\textit{\underline{0.703}}} & \textbf{\textit{\underline{0.690}}} & \textbf{\textit{0.938}} & \textbf{\textit{\underline{0.913}}} & \textbf{\textit{\underline{0.905}}} & \textbf{\textit{\underline{0.941}}} & \textbf{\textit{\underline{0.906}}} & \textbf{\textit{\underline{0.894}}} & \textbf{\textit{\underline{0.903}}} & 1.013 & 1.072 \\ 
LevHAR & \textbf{\textit{\underline{0.899}}} & 0.981 & \textbf{\textit{\underline{0.777}}}  & - & 1.000 & 1.142 & \textbf{\textit{0.882}} & 1.003 & \textbf{\textit{0.874}} & 1.008 & 0.998 & \textbf{\textit{\underline{0.543}}} & \textbf{\textit{\underline{0.533}}} & \textbf{\textit{\underline{0.726}}} & \textbf{\textit{\underline{0.703}}} & \textbf{\textit{\underline{0.697}}} & \textbf{\textit{\underline{0.725}}} & \textbf{\textit{\underline{0.701}}} & \textbf{\textit{\underline{0.689}}} & \textbf{\textit{\underline{0.696}}} & \textbf{\textit{\underline{0.788}}} & \textbf{\textit{\underline{0.822}}} \\ 
SHAR & \textbf{\textit{\underline{0.894}}} & 0.982 & \textbf{\textit{\underline{0.777}}} & 1.003  & - & 1.148 & \textit{0.883} & 1.007 & \textbf{\textit{0.875}} & 1.013 & 1.001 & \textbf{\textit{\underline{0.539}}} & \textbf{\textit{\underline{0.529}}} & \textbf{\textit{\underline{0.721}}} & \textbf{\textit{\underline{0.700}}} & \textbf{\textit{\underline{0.693}}} & \textbf{\textit{\underline{0.718}}} & \textbf{\textit{\underline{0.696}}} & \textbf{\textit{\underline{0.686}}} & \textbf{\textit{\underline{0.693}}} & \textbf{\textit{\underline{0.785}}} & \textbf{\textit{0.820}} \\ 
HARQ & \textbf{\textit{\underline{0.832}}} & \textbf{\textit{\underline{0.885}}} & \textbf{\textit{\underline{0.706}}} & \textbf{\textit{0.903}} & \textbf{\textit{\underline{0.905}}}  & - & \textbf{\textit{\underline{0.793}}} & \textbf{\textit{\underline{0.893}}} & \textbf{\textit{\underline{0.788}}} & \textbf{\textit{\underline{0.897}}} & \textbf{\textit{\underline{0.889}}} & \textbf{\textit{\underline{0.501}}} & \textbf{\textit{\underline{0.491}}} & \textbf{\textit{\underline{0.666}}} & \textbf{\textit{\underline{0.653}}} & \textbf{\textit{\underline{0.650}}} & \textbf{\textit{\underline{0.682}}} & \textbf{\textit{\underline{0.658}}} & \textbf{\textit{\underline{0.635}}} & \textbf{\textit{\underline{0.644}}} & \textbf{\textit{\underline{0.720}}} & \textbf{\textit{\underline{0.747}}} \\ 
RR & 1.012 & 1.140 & \textbf{\textit{\underline{0.874}}} & 1.166 & 1.165 & 1.327  & - & 1.168 & 0.997 & 1.176 & 1.160 & \textbf{\textit{\underline{0.615}}} & \textbf{\textit{\underline{0.603}}} & \textbf{\textit{\underline{0.822}}} & \textbf{\textit{\underline{0.793}}} & \textbf{\textit{\underline{0.788}}} & \textbf{\textit{\underline{0.825}}} & \textbf{\textit{\underline{0.797}}} & \textbf{\textit{\underline{0.775}}} & \textbf{\textit{\underline{0.787}}} & \textit{0.883} & \textbf{\textit{0.927}} \\ 
LA & \textbf{\textit{\underline{0.937}}} & 0.996 & \textbf{\textit{\underline{0.793}}} & 1.017 & 1.020 & 1.146 & \textbf{\textit{\underline{0.894}}}  & - & \textbf{\textit{\underline{0.885}}} & 1.005 & 0.999 & \textbf{\textit{\underline{0.563}}} & \textbf{\textit{\underline{0.552}}} & \textbf{\textit{\underline{0.750}}} & \textbf{\textit{\underline{0.728}}} & \textbf{\textit{\underline{0.725}}} & \textbf{\textit{\underline{0.760}}} & \textbf{\textit{\underline{0.734}}} & \textbf{\textit{\underline{0.712}}} & \textbf{\textit{\underline{0.722}}} & \textbf{\textit{\underline{0.812}}} & \textbf{\textit{\underline{0.845}}} \\ 
EN & 1.019 & 1.143 & \textbf{\textit{\underline{0.877}}} & 1.169 & 1.168 & 1.336 & 1.009 & 1.171  & - & 1.178 & 1.165 & \textbf{\textit{\underline{0.617}}} & \textbf{\textit{\underline{0.606}}} & \textbf{\textit{\underline{0.825}}} & \textbf{\textit{\underline{0.796}}} & \textbf{\textit{\underline{0.790}}} & \textbf{\textit{\underline{0.823}}} & \textbf{\textit{\underline{0.795}}} & \textbf{\textit{\underline{0.780}}} & \textbf{\textit{\underline{0.789}}} & 0.888 & 0.929 \\ 
A-LA & \textbf{\textit{\underline{0.940}}} & \textit{0.997} & \textbf{\textit{\underline{0.793}}} & \textbf{\textit{1.018}} & \textit{1.021} & 1.146 & \textbf{\textit{\underline{0.897}}} & 0.999 & \textbf{\textit{\underline{0.885}}}  & - & 0.997 & \textbf{\textit{\underline{0.565}}} & \textbf{\textit{\underline{0.553}}} & \textbf{\textit{\underline{0.753}}} & \textbf{\textit{\underline{0.731}}} & \textbf{\textit{\underline{0.728}}} & \textbf{\textit{\underline{0.763}}} & \textbf{\textit{\underline{0.737}}} & \textbf{\textit{\underline{0.716}}} & \textbf{\textit{\underline{0.724}}} & \textbf{\textit{\underline{0.813}}} & \textbf{\textit{\underline{0.845}}} \\ 
P-LA & \textbf{\textit{\underline{0.929}}} & 0.997 & \textbf{\textit{\underline{0.793}}} & 1.018 & 1.020 & 1.148 & \textbf{\textit{\underline{0.894}}} & 1.005 & \textbf{\textit{\underline{0.886}}} & 1.010  & - & \textbf{\textit{\underline{0.561}}} & \textbf{\textit{\underline{0.550}}} & \textbf{\textit{\underline{0.749}}} & \textbf{\textit{\underline{0.727}}} & \textbf{\textit{\underline{0.722}}} & \textbf{\textit{\underline{0.756}}} & \textbf{\textit{\underline{0.730}}} & \textbf{\textit{\underline{0.710}}} & \textbf{\textit{\underline{0.719}}} & \textbf{\textit{0.809}} & \textbf{\textit{\underline{0.842}}} \\ 
BG & 1.693 & 2.149 & 1.512 & 2.211 & 2.179 & 2.618 & 1.855 & 2.296 & 1.842 & 2.311 & 2.276  & - & 0.988 & 1.337 & 1.326 & 1.305 & 1.340 & 1.302 & 1.305 & 1.315 & 1.495 & 1.601 \\ 
RF & 1.705 & 2.191 & 1.533 & 2.256 & 2.223 & 2.672 & 1.885 & 2.346 & 1.874 & 2.359 & 2.325 & 1.021  & - & 1.354 & 1.339 & 1.321 & 1.356 & 1.316 & 1.321 & 1.330 & 1.510 & 1.618 \\ 
GB & 1.299 & 1.679 & 1.158 & 1.729 & 1.701 & 2.041 & 1.434 & 1.796 & 1.425 & 1.809 & 1.783 & \textbf{\textit{\underline{0.768}}} & \textbf{\textit{\underline{0.753}}}  & - & 1.009 & 0.999 & 1.020 & 0.998 & 1.003 & 1.007 & 1.139 & 1.223 \\ 
NN$_{1}^{1}$ & 1.289 & 1.647 & 1.160 & 1.693 & 1.674 & 2.010 & 1.411 & 1.758 & 1.402 & 1.770 & 1.744 & \textbf{\textit{\underline{0.782}}} & \textbf{\textit{\underline{0.765}}} & 1.038  & - & 0.993 & 1.020 & 0.990 & 0.995 & 1.003 & 1.141 & 1.222 \\ 
NN$_{1}^{10}$ & 1.298 & 1.655 & 1.171 & 1.702 & 1.680 & 2.018 & 1.422 & 1.767 & 1.413 & 1.779 & 1.751 & \textbf{\textit{\underline{0.785}}} & \textbf{\textit{\underline{0.770}}} & 1.048 & 1.013  & - & 1.028 & 0.998 & 1.002 & 1.011 & 1.151 & 1.232 \\ 
NN$_{2}^{1}$ & 1.294 & 1.686 & 1.167 & 1.734 & 1.707 & 2.075 & 1.437 & 1.818 & 1.425 & 1.829 & 1.801 & \textbf{\textit{\underline{0.779}}} & \textbf{\textit{\underline{0.763}}} & 1.032 & 1.004 & 0.992  & - & 0.978 & 0.990 & 0.995 & 1.137 & 1.216 \\ 
NN$_{2}^{10}$ & 1.318 & 1.715 & 1.189 & 1.765 & 1.740 & 2.103 & 1.463 & 1.847 & 1.452 & 1.857 & 1.829 & \textbf{\textit{\underline{0.797}}} & \textbf{\textit{\underline{0.780}}} & 1.061 & 1.028 & 1.015 & 1.031  & - & 1.010 & 1.017 & 1.162 & 1.242 \\ 
NN$_{3}^{1}$ & 1.323 & 1.686 & 1.183 & 1.734 & 1.715 & 2.059 & 1.441 & 1.808 & 1.434 & 1.818 & 1.792 & \textbf{\textit{\underline{0.800}}} & \textbf{\textit{\underline{0.783}}} & 1.067 & 1.033 & 1.021 & 1.048 & 1.012  & - & 1.018 & 1.163 & 1.240 \\ 
NN$_{3}^{10}$ & 1.300 & 1.656 & 1.161 & 1.703 & 1.681 & 2.020 & 1.418 & 1.774 & 1.408 & 1.782 & 1.757 & \textbf{\textit{\underline{0.785}}} & \textbf{\textit{\underline{0.769}}} & 1.045 & 1.015 & 1.003 & 1.024 & 0.993 & 0.990  & - & 1.141 & 1.212 \\ 
NN$_{4}^{1}$ & 1.147 & 1.456 & 1.015 & 1.497 & 1.480 & 1.757 & 1.241 & 1.553 & 1.233 & 1.560 & 1.537 & \textbf{\textit{\underline{0.695}}} & \textbf{\textit{\underline{0.680}}} & \textit{0.924} & \textbf{\textit{0.899}} & \textbf{\textit{\underline{0.889}}} & \textbf{\textit{\underline{0.913}}} & \textbf{\textit{\underline{0.884}}} & \textbf{\textit{\underline{0.880}}} & \textbf{\textit{\underline{0.888}}}  & - & 1.065 \\ 
NN$_{4}^{10}$ & 1.099 & 1.346 & 0.963 & 1.382 & 1.368 & 1.607 & 1.159 & 1.423 & 1.149 & 1.428 & 1.408 & \textbf{\textit{\underline{0.665}}} & \textbf{\textit{\underline{0.652}}} & \textbf{\textit{0.886}} & \textbf{\textit{\underline{0.861}}} & \textbf{\textit{\underline{0.851}}} & \textbf{\textit{\underline{0.872}}} & \textbf{\textit{\underline{0.845}}} & \textbf{\textit{\underline{0.840}}} & \textbf{\textit{\underline{0.845}}} & 0.955  & - \\ 
\hline \hline
\end{tabular}
\smallskip
\begin{scriptsize}
\parbox{0.98\textwidth}{\emph{Note.} We report the out-of-sample realized variance forecast MSE of each model in the selected column relative to the benchmark in the selected row. Each number is a cross-sectional average of such pairwise relative MSEs for each stock. 
The formatting is as follows: \textit{number} (\textbf{\textit{number}}) [\underline{\textbf{\textit{number}}}] denotes whether the Diebold-Mariano test of equal predictive accuracy is rejected more than 50\% of the time at the 10\% (5\%) [1\%] level of significance across individual tests for each asset. 
The hypothesis being tested is $\text{H}_{0}: \text{MSE}_{i} = \text{MSE}_{j}$ against a one-sided alternative $\text{H}_{1}: \text{MSE}_{i} > \text{MSE}_{j}$, 
where model $i$ is the label of the selected row, whereas model $j$ is the label of the selected column.}
\end{scriptsize}
\end{center}
\end{scriptsize}
\end{sidewaystable}

The results are striking. On the one hand, changing the forecast horizon does not alter the conclusions for $\mathcal{M}_{ \text{HAR}}$ in any discernible way. The neural networks are still dominating and perhaps marginally better than before. On the other hand, the picture changes for $\mathcal{M}_{ \text{ALL}}$, where regression trees and neural networks now provide outstanding improvements in forecast accuracy. Moreover, they appear to get better as we proceed from the weekly to monthly horizon. Whereas neural networks were ahead at the one-day-ahead forecast horizon, we observe that regression trees start to be competitive at the one-week-ahead, and even surpass neural networks at the one-month-ahead, horizon. This shows that while nonlinearities remain important at longer horizons, functional form and interaction effects inherent in the regression trees are more crucial when expanding the predictor base and forecast horizon. The size of the reduction in relative MSE is large with a 40\% or more improvement in forecast accuracy on several occasions, although one should observe that this is partly driven by a deterioration in performance for some HAR models. Even gradient boosting is outperforming most of the HAR models. The differences are often significant at the 1\% level for more than 50\% of the included stocks.

The LogHAR remains ahead of the HAR pack, but the Diebold-Mariano test of equal forecast accuracy with this model as the benchmark is now routinely rejected in favor of ML models. As indicated in Figure \ref{figure:quantile}, the main driver behind the LogHAR model is it ability to handle outliers. As longer-run average realized variance is smoother, it is not surprising that LogHAR cannot compete with more sophisticated models, such as random forest and neural networks. This also explains the mediocre results for regularization, since this is less critical here.

\begin{figure}[t!]
\begin{center}
\caption{In-sample autocorrelation function. \label{figure:acf}}
\begin{tabular}{cc}
\small{Panel A: One-day-ahead forecast.} & \small{Panel B: One-month-ahead forecast.} \\
\includegraphics[height=8cm,width=0.48\textwidth]{{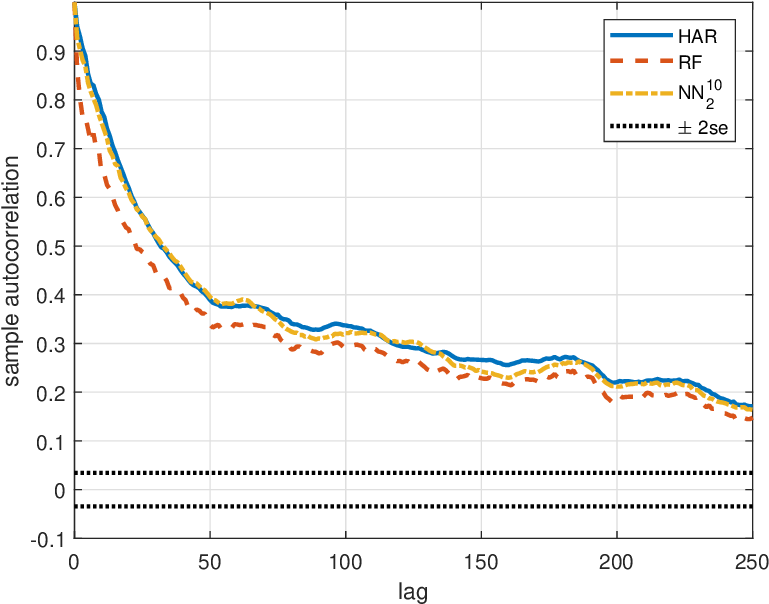}} &
\includegraphics[height=8cm,width=0.48\textwidth]{{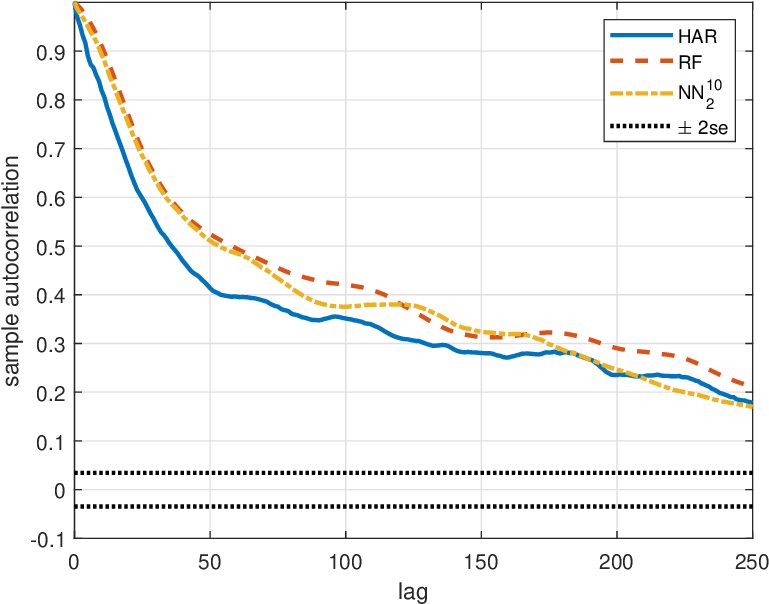}} \\
\end{tabular}
\begin{scriptsize}
\parbox{\textwidth}{\emph{Note.} We report the acf of the in-sample fitted realized variance series at the one-day-ahead (left panel) and one-month-ahead (right panel) horizon for the HAR, random forest, and neural network in $\mathcal{M}_{ \text{ALL}}$. The dotted lines are white noise standard error bands.}
\end{scriptsize}
\end{center}
\end{figure}

Why does random forest suddenly emerge as the lead forecaster of future volatility at the one-month horizon? An understanding of this can help to shed light on where the improved forecast gains are extracted from. In Figure \ref{figure:acf}, we plot the in-sample acf of the fitted realized variance series extracted from the initial training set for the HAR, random forecast, and NN$_{2}^{10}$. The acf embodies how each model perceives the persistence of the realized variance series. In Panel A, we plot the acf corresponding to one-day-ahead forecasts, whereas Panel B is for the one-month ahead forecasts. We observe that at the short horizon, the models are roughly equally persistent, albeit a bit lower for random forest. However, at the monthly horizon random forecast and neural networks display a visibly higher acf, suggesting they are better at approximating an underlying long-memory structure in realized variance, which is harder to detect at the one-day-ahead window, but becomes more pronounced at longer horizons.

\section{A Value-at-Risk analysis} \label{section:value-at-risk}

We now turn our attention toward a more economically motivated application by looking at volatility forecasting through the lens of a Value-at-Risk (VaR) framework. We follow \citet*{audrino-huang-okhrin:19a} and construct one-day-ahead forecasts $\widehat{ \text{VaR}}_{t+1}^{ \alpha}$ of $\text{VaR}_{t+1}^{ \alpha}$, where $\mathbb{P}(r_{t+1} \leq \text{VaR}_{t+1}^{ \alpha}) = \alpha$, which are obtained by means of filtered historical simulation \citep*{barone-adesi-bourgoin-giannopoulos:98a, barone-adesi-giannopoulos-vosper:99a}. The VaR forecasts are evaluated using a loss function that is commonly applied in quantile regression \citep*[e.g.,][]{koenker-bassett:78a}, namely
\begin{equation}
\mathcal{L} = (\alpha - d_{t+1}) \left(r_{t+1} - \widehat{ \text{VaR}}_{t+1}^{ \alpha} \right),
\end{equation}
where $d_{t+1} = 1_{ \{ r_{t+1} <  \widehat{ \text{VaR}}_{t+1}^{ \alpha} \}}$ is the ``hit'' function. This loss function is asymmetric and associates a weight $1- \alpha$ to observations for which the daily log-return is beneath the forecasted quantile, which is arguably more critical in practice, especially for negative returns. Otherwise, a weight $\alpha$ is assigned. Throughout, we describe the results for $\alpha = 0.05$. We also investigated $\alpha = 0.01$ without any major shift in the outcome.

We also examine the exceedance probabilities of the hit sequence by conducting the likelihood ratio tests of \citet*{kupiec:95a} and \citet*{christoffersen:98a} to inspect the unconditional coverage and independence property of the VaR forecasts. The first inspects whether the empirical rejection rates are statistically speaking equal to $\alpha$, on average, whereas the second test checks this assumption conditionally on the previous outcome.

\begin{figure}[t!]
\begin{center}
\caption{Value-at-Risk for Apple. \label{figure:VaR}}
\begin{tabular}{cc}
\small{Panel A: HAR and HAR-X. } & \small{Panel B: RF and NN$_{2}^{10}$.} \\
\includegraphics[height=8cm,width=0.48\textwidth]{{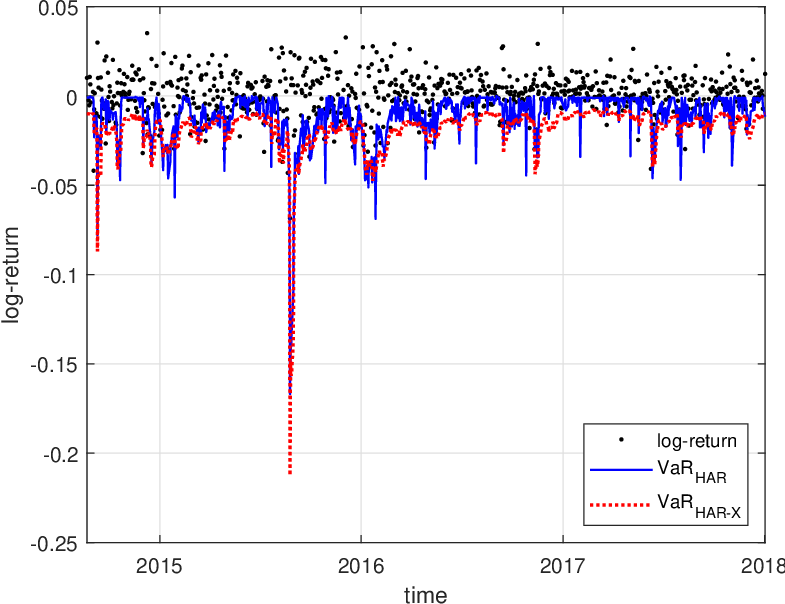}} &
\includegraphics[height=8cm,width=0.48\textwidth]{{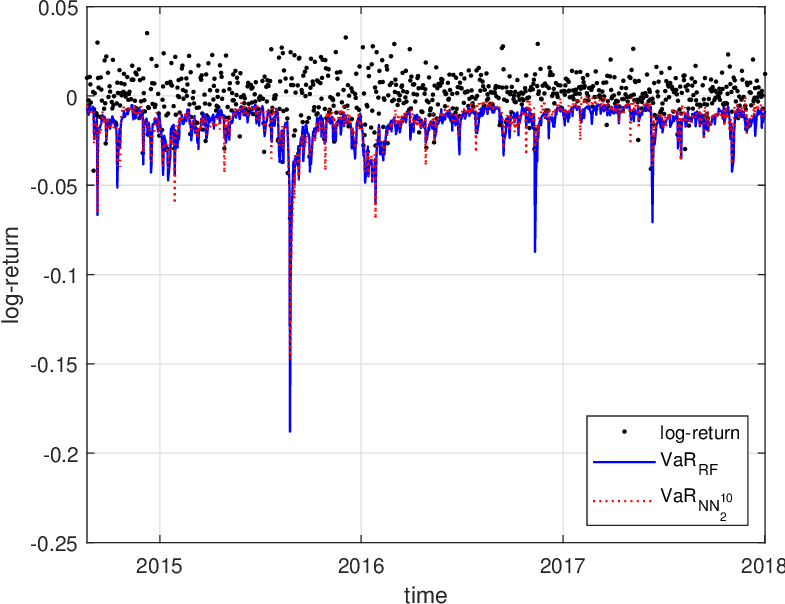}} \\
\end{tabular}
\begin{scriptsize}
\parbox{\textwidth}{\emph{Note.} We plot the open-to-close log-returns for Apple in the out-of-sample window. We superimpose the forecasted VaR at the 95\% confidence level for HAR and HAR-X in Panel A, and for random forecast and the NN$_{2}^{10}$ neural network in Panel B, for $\mathcal{M}_{ \text{ALL}}$.}
\end{scriptsize}
\end{center}
\end{figure}

A graphical representation of the VaR forecasts for Apple are presented in Figure \ref{figure:VaR} for a few models. As evident, they capture the temporal clustering of volatility. The forecast comparison and exceedance probability tests are reported in Table \ref{table:har-1d-var.tex} -- \ref{table:all-1d-var.tex}. Overall, the main conclusions from the previous sections are still valid, although the VaR loss differentials are smaller and less statistically significant than before. Interestingly, the basic HAR model is remarkably solid and tough to beat in the VaR framework. The ML techiques are at best on par with the HAR in the $\mathcal{M}_{ \text{HAR}}$ setting, but we do again observe some more substantial improvements in the $\mathcal{M}_{ \text{ALL}}$ dataset, where many of the unregularized HAR models continue to struggle. Moreover, from the unconditional coverage test we see that ML is slightly worse than the HAR and LogHAR, whereas conditional on the previous outcome they are equally good. The overall impression of the results is that ML delivers precise measures of the Value-at-Risk.

\begin{sidewaystable}
\begin{scriptsize}
\setlength{ \tabcolsep}{0.09cm}
\begin{center}
\caption{One-day-ahead relative VaR and Diebold-Mariano test for dataset $\mathcal{M}_{\text{HAR}}$.
\label{table:har-1d-var.tex}}
\begin{tabular}{lcccccccccccccccccccccc}
\hline \hline
& HAR & HAR-X & LogHAR & LevHAR & SHAR & HARQ & RR & LA & EN & A-LA & P-LA & BG & RF & GB & NN$_{1}^{1}$ & NN$_{1}^{10}$ & NN$_{2}^{1}$ & NN$_{2}^{10}$ & NN$_{3}^{1}$ & NN$_{3}^{10}$ & NN$_{4}^{1}$ & NN$_{4}^{10}$ \\
\hline
HAR  & - & 1.000 & 1.025 & 1.217 & 1.013 & 1.084 & 1.012 & 1.015 & 1.012 & 1.022 & 1.000 & 1.048 & 1.026 & 1.005 & 1.019 & 1.017 & 1.009 & 1.000 & \textit{0.993} & 0.985 & 0.994 & 0.998 \\ 
HAR-X & 1.000  & - & 1.025 & 1.217 & 1.013 & 1.084 & 1.012 & 1.015 & 1.012 & 1.022 & 1.000 & 1.048 & 1.026 & 1.005 & 1.019 & 1.017 & 1.009 & 1.000 & \textit{0.993} & 0.985 & 0.994 & 0.998 \\ 
LogHAR & \textbf{\textit{0.977}} & \textbf{\textit{0.977}}  & - & 1.185 & 0.990 & 1.057 & \textbf{\textit{0.989}} & \textbf{\textit{0.992}} & \textbf{\textit{0.989}} & \textit{0.999} & \textbf{\textit{0.977}} & 1.023 & 1.002 & 0.982 & 0.995 & 0.993 & 0.985 & \textbf{\textit{0.977}} & \textbf{\textit{0.969}} & \textbf{\textit{\underline{0.961}}} & \textbf{\textit{0.970}} & \textit{0.974} \\ 
LevHAR & \textbf{\textit{\underline{0.841}}} & \textbf{\textit{\underline{0.841}}} & \textbf{\textit{\underline{0.859}}}  & - & \textbf{\textit{\underline{0.852}}} & \textbf{\textit{0.904}} & \textbf{\textit{\underline{0.852}}} & \textbf{\textit{\underline{0.855}}} & \textbf{\textit{\underline{0.851}}} & \textbf{\textit{\underline{0.860}}} & \textbf{\textit{\underline{0.841}}} & \textbf{\textit{\underline{0.880}}} & \textbf{\textit{\underline{0.862}}} & \textbf{\textit{\underline{0.846}}} & \textbf{\textit{\underline{0.857}}} & \textbf{\textit{\underline{0.855}}} & \textbf{\textit{\underline{0.848}}} & \textbf{\textit{\underline{0.840}}} & \textbf{\textit{\underline{0.834}}} & \textbf{\textit{\underline{0.828}}} & \textbf{\textit{\underline{0.834}}} & \textbf{\textit{\underline{0.837}}} \\ 
SHAR & 0.988 & 0.988 & 1.012 & 1.202  & - & 1.071 & 1.000 & 1.003 & 1.000 & 1.009 & 0.987 & 1.035 & 1.014 & 0.993 & 1.007 & 1.004 & 0.996 & 0.988 & 0.981 & \textbf{\textit{0.973}} & 0.982 & 0.986 \\ 
HARQ & \textbf{\textit{\underline{0.930}}} & \textbf{\textit{\underline{0.930}}} & \textbf{\textit{0.950}} & 1.121 & \textbf{\textit{\underline{0.941}}}  & - & \textbf{\textit{0.941}} & \textbf{\textit{0.944}} & \textbf{\textit{0.941}} & \textbf{\textit{\underline{0.951}}} & \textbf{\textit{\underline{0.929}}} & 0.972 & \textbf{\textit{0.952}} & \textbf{\textit{\underline{0.935}}} & \textbf{\textit{\underline{0.945}}} & \textbf{\textit{\underline{0.944}}} & \textbf{\textit{\underline{0.936}}} & \textbf{\textit{\underline{0.928}}} & \textbf{\textit{\underline{0.921}}} & \textbf{\textit{\underline{0.914}}} & \textbf{\textit{\underline{0.922}}} & \textbf{\textit{\underline{0.926}}} \\ 
RR & 0.990 & 0.990 & 1.015 & 1.204 & 1.003 & 1.073  & - & 1.003 & 1.000 & 1.009 & 0.990 & 1.037 & 1.016 & 0.993 & 1.009 & 1.006 & 0.998 & 0.990 & 0.983 & 0.975 & 0.984 & 0.987 \\ 
LA & 0.987 & 0.987 & 1.012 & 1.202 & 1.000 & 1.071 & 0.997  & - & 0.997 & 1.006 & 0.987 & 1.034 & 1.013 & 0.991 & 1.006 & 1.004 & 0.995 & 0.987 & 0.980 & 0.972 & 0.981 & 0.985 \\ 
EN & 0.990 & 0.990 & 1.015 & 1.205 & 1.003 & 1.074 & 1.000 & 1.003  & - & 1.009 & 0.990 & 1.037 & 1.016 & 0.994 & 1.009 & 1.007 & 0.998 & 0.990 & 0.983 & 0.975 & 0.984 & 0.988 \\ 
A-LA & 0.982 & 0.982 & 1.006 & 1.196 & 0.995 & 1.065 & 0.991 & 0.994 & 0.991  & - & 0.982 & 1.028 & 1.008 & 0.985 & 1.001 & 0.998 & 0.990 & 0.982 & \textit{0.975} & \textit{0.967} & 0.976 & 0.980 \\ 
P-LA & 1.000 & 1.000 & 1.025 & 1.217 & 1.013 & 1.085 & 1.013 & 1.016 & 1.013 & 1.022  & - & 1.048 & 1.027 & 1.006 & 1.020 & 1.017 & 1.009 & 1.001 & 0.993 & 0.985 & 0.994 & 0.998 \\ 
BG & \textbf{\textit{0.957}} & \textbf{\textit{0.957}} & 0.979 & 1.163 & \textit{0.969} & 1.035 & \textbf{\textit{0.968}} & \textbf{\textit{0.971}} & \textbf{\textit{0.968}} & \textbf{\textit{0.977}} & \textbf{\textit{0.956}}  & - & \textbf{\textit{0.980}} & \textbf{\textit{0.961}} & 0.974 & \textit{0.972} & \textbf{\textit{0.964}} & \textbf{\textit{\underline{0.956}}} & \textbf{\textit{\underline{0.949}}} & \textbf{\textit{\underline{0.941}}} & \textbf{\textit{0.950}} & \textit{0.954} \\ 
RF & \textit{0.976} & \textit{0.976} & 0.999 & 1.186 & 0.988 & 1.056 & 0.988 & 0.991 & 0.988 & 0.997 & \textit{0.975} & 1.020  & - & \textit{0.981} & 0.993 & 0.991 & 0.983 & \textbf{\textit{0.975}} & \textbf{\textit{0.968}} & \textbf{\textit{0.960}} & 0.969 & 0.973 \\ 
GB & 0.997 & 0.997 & 1.021 & 1.213 & 1.009 & 1.081 & 1.007 & 1.010 & 1.007 & 1.016 & 0.996 & 1.043 & 1.023  & - & 1.015 & 1.013 & 1.005 & 0.997 & 0.989 & 0.981 & 0.990 & 0.994 \\ 
NN$_{1}^{1}$ & \textit{0.983} & \textit{0.983} & 1.007 & 1.196 & 0.996 & 1.064 & 0.995 & 0.998 & 0.995 & 1.005 & \textit{0.983} & 1.028 & 1.008 & 0.988  & - & 0.998 & \textit{0.990} & \textbf{\textit{0.982}} & \textbf{\textit{0.975}} & \textbf{\textit{\underline{0.967}}} & \textit{0.976} & 0.980 \\ 
NN$_{1}^{10}$ & \textbf{\textit{0.985}} & \textbf{\textit{0.985}} & 1.008 & 1.197 & 0.997 & 1.066 & 0.997 & 1.000 & 0.997 & \textit{1.006} & \textbf{\textit{0.984}} & 1.030 & 1.010 & 0.990 & 1.002  & - & \textit{0.992} & \textbf{\textit{0.984}} & \textbf{\textit{0.977}} & \textbf{\textit{\underline{0.969}}} & \textit{0.978} & 0.982 \\ 
NN$_{2}^{1}$ & 0.993 & 0.993 & 1.017 & 1.207 & 1.006 & 1.075 & 1.005 & 1.008 & 1.005 & 1.014 & 0.993 & 1.039 & 1.018 & 0.998 & 1.011 & 1.009  & - & 0.992 & 0.985 & \textit{0.977} & 0.986 & 0.990 \\ 
NN$_{2}^{10}$ & 1.001 & 1.001 & 1.025 & 1.215 & 1.014 & 1.083 & 1.013 & 1.016 & 1.012 & 1.022 & 1.000 & 1.047 & 1.026 & 1.005 & 1.019 & 1.017 & 1.008  & - & \textbf{\textit{0.993}} & \textbf{\textit{0.984}} & 0.994 & 0.998 \\ 
NN$_{3}^{1}$ & 1.009 & 1.009 & 1.033 & 1.226 & 1.022 & 1.092 & 1.021 & 1.024 & 1.021 & 1.031 & 1.009 & 1.055 & 1.034 & 1.014 & 1.027 & 1.025 & 1.017 & 1.008  & - & 0.992 & 1.002 & 1.006 \\ 
NN$_{3}^{10}$ & 1.017 & 1.017 & 1.041 & 1.236 & 1.030 & 1.101 & 1.029 & 1.032 & 1.029 & 1.039 & 1.017 & 1.064 & 1.042 & 1.022 & 1.035 & 1.033 & 1.025 & 1.016 & 1.008  & - & 1.010 & 1.014 \\ 
NN$_{4}^{1}$ & 1.007 & 1.007 & 1.031 & 1.223 & 1.020 & 1.091 & 1.020 & 1.023 & 1.019 & 1.029 & 1.007 & 1.054 & 1.033 & 1.012 & 1.026 & 1.024 & 1.015 & 1.007 & 0.999 & 0.991  & - & 1.004 \\ 
NN$_{4}^{10}$ & 1.004 & 1.004 & 1.028 & 1.217 & 1.017 & 1.087 & 1.016 & 1.019 & 1.016 & 1.026 & 1.004 & 1.050 & 1.030 & 1.009 & 1.022 & 1.020 & 1.012 & 1.003 & 0.996 & 0.988 & 0.996  & - \\ 
\hline
Prb. & 0.053 & 0.053 & 0.047 & 0.110 & 0.057 & 0.086 & 0.041 & 0.044 & 0.041 & 0.046 & 0.050 & 0.070 & 0.069 & 0.040 & 0.073 & 0.071 & 0.067 & 0.065 & 0.062 & 0.056 & 0.039 & 0.032 \\ 
Unc. & 5 & 5 & 3 & 26 & 10 & 22 & 12 & 9 & 12 & 11 & 6 & 20 & 20 & 10 & 20 & 19 & 18 & 15 & 11 & 7 & 11 & 18 \\ 
Cond. & 4 & 4 & 1 & 20 & 2 & 2 & 4 & 4 & 4 & 4 & 4 & 0 & 2 & 2 & 3 & 2 & 3 & 2 & 2 & 2 & 2 & 1 \\ 
\hline \hline
\end{tabular}
\smallskip
\begin{scriptsize}
\parbox{0.98\textwidth}{\emph{Note.} We report the VaR forecast from each model in the selected column relative to the benchmark in the selected row. Each number is a cross-sectional average of such pairwise relative VaRs for each stock. 
The VaR confidence level is $1-\alpha = 0.95$. 
The formatting is as follows: \textit{number} (\textbf{\textit{number}}) [\underline{\textbf{\textit{number}}}] denotes whether the Diebold-Mariano test of equal predictive accuracy is rejected more than 50\% of the time at the 10\% (5\%) [1\%] level of significance across individual tests for each asset. 
The hypothesis being tested is $\text{H}_{0}: \mathbb{E}( \mathcal{L}_{ \text{VaR}_{i}}) = \mathbb{E}( \mathcal{L}_{ \text{VaR}_{j}})$ against a one-sided alternative $\text{H}_{1}: \mathbb{E}( \mathcal{L}_{ \text{VaR}_{i}}) > \mathbb{E}( \mathcal{L}_{ \text{VaR}_{j}})$, 
where model $i$ is the label of the selected row, whereas model $j$ is the label of the selected column.
Prb. is the exceedance rate, Unc. (Cond.) is the number of times the unconditional (conditional) coverage test was rejected at the 5\% level of significance.
}
\end{scriptsize}
\end{center}
\end{scriptsize}
\end{sidewaystable}
\begin{sidewaystable}
\begin{scriptsize}
\setlength{ \tabcolsep}{0.09cm}
\begin{center}
\caption{One-day-ahead relative VaR and Diebold-Mariano test for dataset $\mathcal{M}_{\text{ALL}}$.
\label{table:all-1d-var.tex}}
\begin{tabular}{lcccccccccccccccccccccc}
\hline \hline
& HAR & HAR-X & LogHAR & LevHAR & SHAR & HARQ & RR & LA & EN & A-LA & P-LA & BG & RF & GB & NN$_{1}^{1}$ & NN$_{1}^{10}$ & NN$_{2}^{1}$ & NN$_{2}^{10}$ & NN$_{3}^{1}$ & NN$_{3}^{10}$ & NN$_{4}^{1}$ & NN$_{4}^{10}$ \\
\hline
HAR  & - & 1.320 & 0.995 & 1.397 & 1.523 & 1.294 & 1.067 & 1.077 & 1.043 & 1.119 & 1.348 & 0.999 & 0.988 & 0.973 & 1.010 & 0.999 & 0.980 & \textit{0.965} & 0.981 & \textit{0.965} & 0.976 & 0.980 \\ 
HAR-X & \textbf{\textit{\underline{0.773}}}  & - & \textbf{\textit{\underline{0.767}}} & 1.057 & 1.164 & 0.981 & \textbf{\textit{\underline{0.818}}} & \textbf{\textit{\underline{0.827}}} & \textbf{\textit{\underline{0.802}}} & \textbf{\textit{\underline{0.858}}} & 1.020 & \textbf{\textit{\underline{0.771}}} & \textbf{\textit{\underline{0.763}}} & \textbf{\textit{\underline{0.752}}} & \textbf{\textit{\underline{0.778}}} & \textbf{\textit{\underline{0.769}}} & \textbf{\textit{\underline{0.756}}} & \textbf{\textit{\underline{0.745}}} & \textbf{\textit{\underline{0.755}}} & \textbf{\textit{\underline{0.744}}} & \textbf{\textit{\underline{0.751}}} & \textbf{\textit{\underline{0.755}}} \\ 
LogHAR & 1.008 & 1.327  & - & 1.402 & 1.533 & 1.298 & 1.071 & 1.082 & 1.048 & 1.125 & 1.355 & 1.005 & 0.994 & 0.979 & 1.015 & 1.004 & 0.985 & \textit{0.971} & 0.985 & \textbf{\textit{0.970}} & 0.980 & 0.985 \\ 
LevHAR & \textbf{\textit{\underline{0.733}}} & \textbf{\textit{0.947}} & \textbf{\textit{\underline{0.727}}}  & - & 1.105 & \textbf{\textit{0.929}} & \textbf{\textit{\underline{0.775}}} & \textbf{\textit{\underline{0.784}}} & \textbf{\textit{\underline{0.760}}} & \textbf{\textit{\underline{0.814}}} & 0.966 & \textbf{\textit{\underline{0.731}}} & \textbf{\textit{\underline{0.723}}} & \textbf{\textit{\underline{0.713}}} & \textbf{\textit{\underline{0.738}}} & \textbf{\textit{\underline{0.730}}} & \textbf{\textit{\underline{0.717}}} & \textbf{\textit{\underline{0.706}}} & \textbf{\textit{\underline{0.715}}} & \textbf{\textit{\underline{0.706}}} & \textbf{\textit{\underline{0.712}}} & \textbf{\textit{\underline{0.715}}} \\ 
SHAR & \textbf{\textit{\underline{0.664}}} & \textbf{\textit{\underline{0.868}}} & \textbf{\textit{\underline{0.661}}} & \textbf{\textit{\underline{0.919}}}  & - & \textbf{\textit{\underline{0.852}}} & \textbf{\textit{\underline{0.707}}} & \textbf{\textit{\underline{0.712}}} & \textbf{\textit{\underline{0.691}}} & \textbf{\textit{\underline{0.740}}} & \textbf{\textit{\underline{0.885}}} & \textbf{\textit{\underline{0.663}}} & \textbf{\textit{\underline{0.656}}} & \textbf{\textit{\underline{0.646}}} & \textbf{\textit{\underline{0.669}}} & \textbf{\textit{\underline{0.661}}} & \textbf{\textit{\underline{0.650}}} & \textbf{\textit{\underline{0.640}}} & \textbf{\textit{\underline{0.650}}} & \textbf{\textit{\underline{0.640}}} & \textbf{\textit{\underline{0.647}}} & \textbf{\textit{\underline{0.651}}} \\ 
HARQ & \textbf{\textit{\underline{0.791}}} & 1.025 & \textbf{\textit{\underline{0.784}}} & 1.083 & 1.193  & - & \textbf{\textit{\underline{0.835}}} & \textbf{\textit{\underline{0.845}}} & \textbf{\textit{\underline{0.819}}} & \textbf{\textit{\underline{0.879}}} & 1.044 & \textbf{\textit{\underline{0.788}}} & \textbf{\textit{\underline{0.780}}} & \textbf{\textit{\underline{0.768}}} & \textbf{\textit{\underline{0.794}}} & \textbf{\textit{\underline{0.786}}} & \textbf{\textit{\underline{0.772}}} & \textbf{\textit{\underline{0.761}}} & \textbf{\textit{\underline{0.770}}} & \textbf{\textit{\underline{0.760}}} & \textbf{\textit{\underline{0.766}}} & \textbf{\textit{\underline{0.770}}} \\ 
RR & \textit{0.944} & 1.237 & \textbf{\textit{0.937}} & 1.308 & 1.434 & 1.209  & - & 1.012 & 0.980 & 1.054 & 1.262 & \textbf{\textit{0.941}} & \textbf{\textit{0.931}} & \textbf{\textit{0.918}} & \textit{0.951} & \textbf{\textit{0.941}} & \textbf{\textit{\underline{0.923}}} & \textbf{\textit{\underline{0.910}}} & \textbf{\textit{\underline{0.923}}} & \textbf{\textit{\underline{0.909}}} & \textbf{\textit{\underline{0.917}}} & \textbf{\textit{0.922}} \\ 
LA & \textbf{\textit{0.933}} & 1.225 & \textbf{\textit{0.927}} & 1.296 & 1.415 & 1.199 & 0.992  & - & \textbf{\textit{0.970}} & 1.040 & 1.250 & \textbf{\textit{0.931}} & \textbf{\textit{\underline{0.921}}} & \textbf{\textit{\underline{0.907}}} & \textbf{\textit{\underline{0.940}}} & \textbf{\textit{\underline{0.929}}} & \textbf{\textit{\underline{0.912}}} & \textbf{\textit{\underline{0.899}}} & \textbf{\textit{\underline{0.912}}} & \textbf{\textit{\underline{0.899}}} & \textbf{\textit{\underline{0.908}}} & \textbf{\textit{\underline{0.912}}} \\ 
EN & \textit{0.963} & 1.264 & \textit{0.956} & 1.337 & 1.462 & 1.236 & 1.022 & 1.033  & - & 1.074 & 1.290 & 0.960 & \textbf{\textit{0.950}} & \textbf{\textit{0.936}} & 0.970 & \textbf{\textit{0.959}} & \textbf{\textit{\underline{0.941}}} & \textbf{\textit{\underline{0.927}}} & \textbf{\textit{\underline{0.941}}} & \textbf{\textit{\underline{0.927}}} & \textbf{\textit{0.936}} & \textbf{\textit{0.941}} \\ 
A-LA & \textbf{\textit{\underline{0.900}}} & 1.180 & \textbf{\textit{\underline{0.894}}} & 1.248 & 1.364 & 1.157 & \textbf{\textit{0.958}} & \textbf{\textit{\underline{0.965}}} & \textbf{\textit{\underline{0.936}}}  & - & 1.205 & \textbf{\textit{\underline{0.897}}} & \textbf{\textit{\underline{0.888}}} & \textbf{\textit{\underline{0.875}}} & \textbf{\textit{\underline{0.907}}} & \textbf{\textit{\underline{0.896}}} & \textbf{\textit{\underline{0.880}}} & \textbf{\textit{\underline{0.867}}} & \textbf{\textit{\underline{0.880}}} & \textbf{\textit{\underline{0.867}}} & \textbf{\textit{\underline{0.876}}} & \textbf{\textit{\underline{0.880}}} \\ 
P-LA & \textbf{\textit{\underline{0.760}}} & 0.982 & \textbf{\textit{\underline{0.754}}} & 1.038 & 1.143 & 0.962 & \textbf{\textit{\underline{0.804}}} & \textbf{\textit{\underline{0.813}}} & \textbf{\textit{\underline{0.787}}} & \textbf{\textit{\underline{0.844}}}  & - & \textbf{\textit{\underline{0.758}}} & \textbf{\textit{\underline{0.750}}} & \textbf{\textit{\underline{0.739}}} & \textbf{\textit{\underline{0.765}}} & \textbf{\textit{\underline{0.756}}} & \textbf{\textit{\underline{0.744}}} & \textbf{\textit{\underline{0.732}}} & \textbf{\textit{\underline{0.741}}} & \textbf{\textit{\underline{0.731}}} & \textbf{\textit{\underline{0.738}}} & \textbf{\textit{\underline{0.742}}} \\ 
BG & 1.004 & 1.323 & 0.997 & 1.399 & 1.526 & 1.295 & 1.068 & 1.078 & 1.044 & 1.120 & 1.351  & - & \textit{0.989} & \textit{0.975} & 1.011 & 1.000 & 0.980 & \textbf{\textit{0.966}} & \textit{0.981} & \textbf{\textit{0.966}} & \textit{0.976} & 0.981 \\ 
RF & 1.014 & 1.338 & 1.007 & 1.415 & 1.543 & 1.309 & 1.080 & 1.090 & 1.056 & 1.133 & 1.366 & 1.011  & - & 0.985 & 1.022 & 1.011 & 0.991 & \textit{0.977} & 0.992 & \textit{0.977} & 0.987 & 0.992 \\ 
GB & 1.029 & 1.358 & 1.023 & 1.436 & 1.566 & 1.329 & 1.097 & 1.107 & 1.072 & 1.150 & 1.386 & 1.027 & 1.016  & - & 1.038 & 1.026 & 1.006 & 0.992 & 1.007 & 0.992 & 1.002 & 1.007 \\ 
NN$_{1}^{1}$ & 0.994 & 1.307 & 0.987 & 1.383 & 1.509 & 1.278 & 1.057 & 1.067 & 1.033 & 1.109 & 1.334 & 0.990 & 0.980 & 0.966  & - & 0.989 & 0.971 & \textbf{\textit{0.957}} & \textbf{\textit{0.972}} & \textbf{\textit{0.957}} & 0.966 & \textit{0.971} \\ 
NN$_{1}^{10}$ & 1.005 & 1.323 & 0.998 & 1.400 & 1.525 & 1.295 & 1.070 & 1.079 & 1.045 & 1.121 & 1.350 & 1.002 & 0.991 & 0.976 & 1.012  & - & 0.982 & \textbf{\textit{0.967}} & 0.982 & \textbf{\textit{0.967}} & 0.978 & 0.983 \\ 
NN$_{2}^{1}$ & 1.025 & 1.351 & 1.018 & 1.430 & 1.559 & 1.321 & 1.091 & 1.101 & 1.066 & 1.144 & 1.380 & 1.021 & 1.010 & 0.996 & 1.032 & 1.020  & - & \textit{0.986} & 1.001 & 0.986 & 0.996 & 1.001 \\ 
NN$_{2}^{10}$ & 1.039 & 1.369 & 1.032 & 1.449 & 1.579 & 1.340 & 1.106 & 1.116 & 1.081 & 1.160 & 1.398 & 1.036 & 1.024 & 1.009 & 1.047 & 1.035 & 1.015  & - & 1.016 & 1.000 & 1.011 & 1.016 \\ 
NN$_{3}^{1}$ & 1.024 & 1.347 & 1.017 & 1.425 & 1.558 & 1.317 & 1.089 & 1.099 & 1.065 & 1.143 & 1.375 & 1.021 & 1.010 & 0.995 & 1.032 & 1.020 & 1.000 & 0.986  & - & 0.986 & 0.996 & 1.000 \\ 
NN$_{3}^{10}$ & 1.039 & 1.368 & 1.032 & 1.447 & 1.580 & 1.338 & 1.106 & 1.116 & 1.081 & 1.160 & 1.396 & 1.036 & 1.025 & 1.009 & 1.047 & 1.035 & 1.015 & 1.000 & 1.015  & - & 1.011 & 1.015 \\ 
NN$_{4}^{1}$ & 1.030 & 1.354 & 1.022 & 1.433 & 1.566 & 1.323 & 1.094 & 1.106 & 1.071 & 1.150 & 1.383 & 1.027 & 1.016 & 1.000 & 1.037 & 1.026 & 1.006 & 0.992 & 1.006 & 0.991  & - & 1.005 \\ 
NN$_{4}^{10}$ & 1.026 & 1.349 & 1.018 & 1.427 & 1.562 & 1.317 & 1.090 & 1.102 & 1.066 & 1.146 & 1.377 & 1.022 & 1.012 & 0.996 & 1.033 & 1.022 & 1.002 & 0.988 & 1.002 & 0.987 & 0.996  & - \\ 
\hline
Prb. & 0.053 & 0.160 & 0.047 & 0.172 & 0.247 & 0.106 & 0.082 & 0.089 & 0.079 & 0.102 & 0.168 & 0.069 & 0.066 & 0.046 & 0.073 & 0.073 & 0.060 & 0.057 & 0.057 & 0.050 & 0.037 & 0.030 \\ 
Unc. & 5 & 29 & 4 & 29 & 29 & 27 & 27 & 25 & 25 & 27 & 29 & 20 & 14 & 3 & 17 & 17 & 10 & 8 & 9 & 10 & 15 & 18 \\ 
Cond. & 4 & 5 & 0 & 8 & 3 & 2 & 5 & 5 & 2 & 9 & 11 & 1 & 2 & 2 & 4 & 2 & 4 & 3 & 2 & 2 & 0 & 0 \\ 
\hline \hline
\end{tabular}
\smallskip
\begin{scriptsize}
\parbox{0.98\textwidth}{\emph{Note.} We report the VaR forecast from each model in the selected column relative to the benchmark in the selected row. Each number is a cross-sectional average of such pairwise relative VaRs for each stock. 
The VaR confidence level is $1-\alpha = 0.95$. 
The formatting is as follows: \textit{number} (\textbf{\textit{number}}) [\underline{\textbf{\textit{number}}}] denotes whether the Diebold-Mariano test of equal predictive accuracy is rejected more than 50\% of the time at the 10\% (5\%) [1\%] level of significance across individual tests for each asset. 
The hypothesis being tested is $\text{H}_{0}: \mathbb{E}( \mathcal{L}_{ \text{VaR}_{i}}) = \mathbb{E}( \mathcal{L}_{ \text{VaR}_{j}})$ against a one-sided alternative $\text{H}_{1}: \mathbb{E}( \mathcal{L}_{ \text{VaR}_{i}}) > \mathbb{E}( \mathcal{L}_{ \text{VaR}_{j}})$, 
where model $i$ is the label of the selected row, whereas model $j$ is the label of the selected column.
Prb. is the exceedance rate, Unc. (Cond.) is the number of times the unconditional (conditional) coverage test was rejected at the 5\% level of significance.
}
\end{scriptsize}
\end{center}
\end{scriptsize}
\end{sidewaystable}

\section{Conclusion} \label{section:conclusion}

This paper synthesizes the field of machine learning (ML) with volatility forecasting. We conduct an extensive out-of-sample forecast comparison of current ML algoritms with a broad suite of HAR models for forecasting realized variance. We show that ML is better at handling the highly nonlinear structure governing financial markets, and that they are able to extract valuable incremental information about future volatility in a data-rich environment with a large feature space; a setting in which standard HAR models struggle. The random forest and neural networks are preferred among ML algorithms and significantly improve prediction relative to HAR. The improvements increase at longer horizons and are often statistically significant. We dissect the underlying structure of the data via ALE plots in order to pin down the main covariates leading to the improved prediction. In an economic Value-at-Risk application, we continue to observe reduced loss with ML models, although it is less statistically significant.

To the best of our knowledge, this study is the first to extensively implement a broad selection of the most widely applied ML methods for the problem of volatility prediction and compare them to a broad suite of HAR models. We envision the paper can serve as a starting point for applying ML to forecast stochastic volatility and perhaps even as a benchmark when building more complicated models. Moreover, we look forward to behold how much additional improvement that can be had by doing more extensive hyperparameter tuning than what we do in this paper.

\pagebreak


\bibliographystyle{rfs}
\bibliography{userref}

\pagebreak

\appendix

\section{Appendix} \label{section:appendix}

\subsection{Span of test set} \label{appendix:test-set}

This section reports the out-of-sample relative MSE of each volatility model when the length of the training set\glsadd{Test Set} is shortened from 70\% of the sample, or 2,964 days, to 1,000 days (Table \ref{table:har-1d-1000.tex} and \ref{table:all-1d-1000.tex}) and 2,000 days (Table \ref{table:har-1d-2000.tex} and \ref{table:all-1d-2000.tex}), while keeping the validation set fixed at 200 days, yielding a one-for-one extension of the out-of-sample test set.

\clearpage

\noindent \underline{\textbf{Training set = 1,000 days}}:

\begin{sidewaystable}
\begin{scriptsize}
\setlength{ \tabcolsep}{0.09cm}
\begin{center}
\caption{One-day-ahead relative MSE and Diebold-Mariano test for dataset $\mathcal{M}_{\text{HAR}}$.
\label{table:har-1d-1000.tex}}
\begin{tabular}{lcccccccccccccccccccccc}
\hline \hline
& HAR & HAR-X & LogHAR & LevHAR & SHAR & HARQ & RR & LA & EN & A-LA & P-LA & BG & RF & GB & NN$_{1}^{1}$ & NN$_{1}^{10}$ & NN$_{2}^{1}$ & NN$_{2}^{10}$ & NN$_{3}^{1}$ & NN$_{3}^{10}$ & NN$_{4}^{1}$ & NN$_{4}^{10}$ \\
\hline
HAR  & - & 1.000 & 0.905 & \textit{0.940} & 1.011 & 0.870 & 0.865 & 0.866 & 0.864 & 0.869 & 0.865 & 1.050 & 0.958 & 0.937 & 0.866 & 0.835 & 0.867 & 0.844 & 0.865 & 0.861 & 0.900 & 0.899 \\ 
HAR-X & 1.000  & - & 0.905 & \textit{0.940} & 1.011 & 0.870 & 0.865 & 0.866 & 0.864 & 0.869 & 0.865 & 1.050 & 0.958 & 0.937 & 0.866 & 0.835 & 0.867 & 0.844 & 0.865 & 0.861 & 0.900 & 0.899 \\ 
LogHAR & 1.169 & 1.169  & - & 1.096 & 1.187 & 0.993 & 0.984 & 0.986 & 0.983 & 0.989 & 0.985 & 1.201 & 1.094 & 1.060 & 0.979 & 0.947 & 0.978 & 0.954 & 0.975 & 0.972 & 1.014 & 1.012 \\ 
LevHAR & 1.066 & 1.066 & 0.965  & - & 1.076 & 0.926 & 0.920 & 0.921 & 0.919 & 0.924 & 0.920 & 1.118 & 1.020 & 0.996 & 0.921 & 0.888 & 0.924 & 0.899 & 0.920 & 0.917 & 0.958 & 0.956 \\ 
SHAR & 0.994 & 0.994 & 0.903 & 0.934  & - & 0.867 & 0.861 & 0.862 & 0.860 & 0.865 & 0.862 & 1.044 & 0.954 & 0.933 & 0.865 & 0.832 & 0.865 & 0.842 & 0.862 & 0.859 & 0.897 & 0.896 \\ 
HARQ & 1.180 & 1.180 & 1.043 & 1.105 & 1.196  & - & 1.003 & 1.004 & 1.002 & 1.008 & 1.003 & 1.216 & 1.111 & 1.079 & 0.996 & 0.965 & 1.003 & 0.975 & 0.998 & 0.994 & 1.036 & 1.035 \\ 
RR & 1.189 & 1.189 & 1.049 & 1.115 & 1.205 & 1.020  & - & 1.002 & 0.999 & 1.005 & 1.001 & 1.233 & 1.125 & 1.090 & 1.007 & 0.972 & 1.009 & 0.982 & 1.005 & 1.001 & 1.045 & 1.043 \\ 
LA & 1.186 & 1.186 & 1.048 & 1.112 & 1.203 & 1.017 & 0.998  & - & 0.998 & 1.003 & 0.999 & 1.231 & 1.123 & 1.088 & 1.005 & 0.970 & 1.007 & 0.980 & 1.003 & 0.999 & 1.043 & 1.041 \\ 
EN & 1.190 & 1.190 & 1.050 & 1.115 & 1.206 & 1.020 & 1.001 & 1.003  & - & 1.006 & 1.002 & 1.234 & 1.126 & 1.091 & 1.008 & 0.973 & 1.010 & 0.983 & 1.006 & 1.002 & 1.046 & 1.044 \\ 
A-LA & 1.182 & 1.182 & 1.044 & 1.109 & 1.198 & 1.014 & 0.995 & 0.997 & 0.994  & - & 0.996 & 1.228 & 1.119 & 1.085 & 1.002 & 0.967 & 1.004 & 0.977 & 1.000 & 0.996 & 1.040 & 1.038 \\ 
P-LA & 1.188 & 1.188 & 1.049 & 1.114 & 1.204 & 1.018 & 1.000 & 1.001 & 0.999 & 1.004  & - & 1.233 & 1.124 & 1.089 & 1.006 & 0.971 & 1.008 & 0.981 & 1.004 & 1.000 & 1.044 & 1.042 \\ 
BG & 0.977 & 0.977 & \textit{0.861} & \textit{0.915} & 0.988 & \textbf{\textit{0.834}} & \textit{0.833} & \textit{0.834} & \textit{0.832} & \textit{0.837} & \textit{0.833}  & - & \textbf{\textit{0.915}} & 0.894 & \textbf{\textit{0.829}} & \textbf{\textit{0.801}} & \textit{0.830} & \textit{0.808} & \textit{0.827} & 0.824 & 0.860 & 0.859 \\ 
RF & 1.065 & 1.065 & 0.940 & 0.999 & 1.078 & 0.911 & 0.908 & 0.909 & 0.907 & 0.913 & 0.909 & 1.095  & - & 0.976 & 0.905 & 0.873 & 0.905 & 0.882 & 0.902 & 0.899 & 0.938 & 0.937 \\ 
GB & 1.103 & 1.103 & \textit{0.966} & 1.033 & 1.119 & 0.937 & 0.932 & 0.933 & 0.931 & 0.937 & 0.932 & 1.134 & 1.035  & - & 0.927 & \textit{0.897} & 0.929 & 0.905 & 0.926 & 0.922 & 0.962 & 0.960 \\ 
NN$_{1}^{1}$ & 1.203 & 1.203 & 1.055 & 1.128 & 1.224 & 1.023 & 1.017 & 1.018 & 1.016 & 1.022 & 1.017 & 1.240 & 1.132 & 1.096  & - & 0.977 & 1.013 & 0.986 & 1.010 & 1.006 & 1.049 & 1.047 \\ 
NN$_{1}^{10}$ & 1.228 & 1.228 & 1.079 & 1.152 & 1.248 & 1.050 & 1.040 & 1.041 & 1.039 & 1.045 & 1.040 & 1.270 & 1.158 & 1.122 & 1.034  & - & 1.036 & 1.010 & 1.034 & 1.030 & 1.075 & 1.073 \\ 
NN$_{2}^{1}$ & 1.197 & 1.197 & 1.043 & 1.124 & 1.216 & 1.023 & 1.012 & 1.014 & 1.011 & 1.017 & 1.013 & 1.234 & 1.125 & 1.089 & 1.005 & 0.971  & - & 0.980 & 1.002 & 0.999 & 1.042 & 1.040 \\ 
NN$_{2}^{10}$ & 1.220 & 1.220 & 1.066 & 1.144 & 1.239 & 1.041 & 1.031 & 1.032 & 1.030 & 1.036 & 1.031 & 1.258 & 1.147 & 1.111 & 1.025 & 0.991 & 1.026  & - & 1.023 & 1.019 & 1.063 & 1.062 \\ 
NN$_{3}^{1}$ & 1.194 & 1.194 & 1.041 & 1.120 & 1.212 & 1.018 & 1.008 & 1.009 & 1.007 & 1.013 & 1.008 & 1.230 & 1.122 & 1.086 & 1.002 & \textit{0.970} & 1.003 & 0.978  & - & 0.996 & 1.039 & 1.038 \\ 
NN$_{3}^{10}$ & 1.198 & 1.198 & 1.045 & 1.124 & 1.217 & 1.022 & 1.012 & 1.013 & 1.011 & 1.017 & 1.012 & 1.235 & 1.126 & 1.090 & 1.006 & 0.973 & 1.007 & \textit{0.981} & 1.004  & - & 1.043 & 1.041 \\ 
NN$_{4}^{1}$ & 1.152 & 1.152 & 1.003 & 1.080 & 1.170 & 0.979 & 0.972 & 0.973 & 0.971 & 0.977 & 0.972 & 1.185 & 1.081 & 1.046 & 0.965 & \textbf{\textit{0.934}} & \textit{0.966} & \textbf{\textit{0.942}} & \textbf{\textit{0.963}} & \textbf{\textit{0.959}}  & - & 0.999 \\ 
NN$_{4}^{10}$ & 1.154 & 1.154 & 1.004 & 1.082 & 1.172 & 0.981 & 0.972 & 0.974 & 0.972 & 0.977 & 0.973 & 1.188 & 1.083 & 1.047 & 0.966 & \textit{0.935} & \textit{0.967} & \textbf{\textit{0.943}} & \textit{0.965} & \textbf{\textit{\underline{0.961}}} & 1.002  & - \\ 
\hline \hline
\end{tabular}
\smallskip
\begin{scriptsize}
\parbox{0.98\textwidth}{\emph{Note.} We report the out-of-sample realized variance forecast MSE of each model in the selected column relative to the benchmark in the selected row. Each number is a cross-sectional average of such pairwise relative MSEs for each stock. 
The formatting is as follows: \textit{number} (\textbf{\textit{number}}) [\underline{\textbf{\textit{number}}}] denotes whether the Diebold-Mariano test of equal predictive accuracy is rejected more than 50\% of the time at the 10\% (5\%) [1\%] level of significance across individual tests for each asset. 
The hypothesis being tested is $\text{H}_{0}: \text{MSE}_{i} = \text{MSE}_{j}$ against a one-sided alternative $\text{H}_{1}: \text{MSE}_{i} > \text{MSE}_{j}$, 
where model $i$ is the label of the selected row, whereas model $j$ is the label of the selected column.}
\end{scriptsize}
\end{center}
\end{scriptsize}
\end{sidewaystable}
\begin{sidewaystable}
\begin{scriptsize}
\setlength{ \tabcolsep}{0.09cm}
\begin{center}
\caption{One-day-ahead relative MSE and Diebold-Mariano test for dataset $\mathcal{M}_{\text{ALL}}$.
\label{table:all-1d-1000.tex}}
\begin{tabular}{lcccccccccccccccccccccc}
\hline \hline
& HAR & HAR-X & LogHAR & LevHAR & SHAR & HARQ & RR & LA & EN & A-LA & P-LA & BG & RF & GB & NN$_{1}^{1}$ & NN$_{1}^{10}$ & NN$_{2}^{1}$ & NN$_{2}^{10}$ & NN$_{3}^{1}$ & NN$_{3}^{10}$ & NN$_{4}^{1}$ & NN$_{4}^{10}$ \\
\hline
HAR  & - & 0.937 & 0.970 & 0.917 & 0.969 & 0.905 & 0.829 & 0.835 & 0.827 & 0.837 & 1.731 & 0.919 & 0.849 & 0.904 & \textit{0.835} & \textit{0.802} & 0.843 & 0.849 & 0.862 & 0.827 & 0.930 & 0.871 \\ 
HAR-X & 1.075  & - & 1.033 & 0.978 & 1.033 & 0.965 & 0.888 & 0.894 & 0.886 & 0.896 & 1.845 & 0.983 & 0.908 & 0.965 & 0.897 & 0.860 & 0.902 & 0.912 & 0.922 & 0.886 & 0.996 & 0.932 \\ 
LogHAR & 1.169 & 1.086  & - & 1.063 & 1.131 & 1.032 & 0.943 & 0.950 & 0.941 & 0.953 & 1.898 & 1.057 & 0.973 & 1.025 & 0.967 & 0.919 & 0.958 & 0.981 & 0.980 & 0.944 & 1.069 & 0.992 \\ 
LevHAR & 1.101 & 1.024 & 1.061  & - & 1.057 & 0.988 & 0.908 & 0.915 & 0.906 & 0.917 & 1.888 & 1.005 & 0.929 & 0.987 & 0.918 & 0.880 & 0.923 & 0.932 & 0.944 & 0.907 & 1.020 & 0.954 \\ 
SHAR & 1.046 & 0.973 & 1.011 & 0.951  & - & 0.941 & 0.866 & 0.872 & 0.863 & 0.874 & 1.805 & 0.957 & 0.885 & 0.941 & 0.876 & 0.839 & 0.882 & 0.889 & 0.899 & 0.864 & 0.973 & 0.909 \\ 
HARQ & 1.138 & 1.056 & 1.074 & 1.032 & 1.095  & - & 0.927 & 0.933 & 0.925 & 0.935 & 1.898 & 1.030 & 0.949 & 1.006 & 0.933 & 0.895 & 0.937 & 0.950 & 0.960 & 0.922 & 1.035 & 0.970 \\ 
RR & 1.238 & 1.157 & 1.169 & 1.131 & 1.200 & 1.105  & - & 1.007 & 0.997 & 1.011 & 2.063 & 1.125 & 1.037 & 1.097 & 1.020 & 0.976 & 1.021 & 1.028 & 1.048 & 1.003 & 1.134 & 1.055 \\ 
LA & 1.231 & 1.151 & 1.161 & 1.125 & 1.194 & 1.099 & 0.994  & - & \textbf{\textit{\underline{0.991}}} & 1.004 & 2.049 & 1.118 & 1.031 & 1.091 & 1.013 & 0.970 & 1.014 & 1.020 & 1.041 & 0.997 & 1.126 & 1.048 \\ 
EN & 1.242 & 1.161 & 1.172 & 1.134 & 1.204 & 1.108 & 1.003 & 1.009  & - & 1.013 & 2.068 & 1.127 & 1.040 & 1.100 & 1.023 & 0.979 & 1.024 & 1.031 & 1.051 & 1.006 & 1.137 & 1.058 \\ 
A-LA & 1.225 & 1.145 & 1.158 & 1.119 & 1.188 & 1.093 & 0.990 & 0.997 & \textit{0.988}  & - & 2.043 & 1.113 & 1.026 & 1.086 & 1.009 & 0.966 & 1.011 & 1.016 & 1.037 & 0.993 & 1.121 & 1.044 \\ 
P-LA & \textbf{\textit{\underline{0.657}}} & \textbf{\textit{\underline{0.605}}} & \textbf{\textit{\underline{0.582}}} & \textbf{\textit{\underline{0.592}}} & \textbf{\textit{\underline{0.634}}} & \textbf{\textit{\underline{0.566}}} & \textbf{\textit{\underline{0.518}}} & \textbf{\textit{\underline{0.521}}} & \textbf{\textit{\underline{0.516}}} & \textbf{\textit{\underline{0.523}}}  & - & \textbf{\textit{\underline{0.586}}} & \textbf{\textit{\underline{0.536}}} & \textbf{\textit{\underline{0.562}}} & \textbf{\textit{\underline{0.531}}} & \textbf{\textit{\underline{0.504}}} & \textbf{\textit{\underline{0.520}}} & \textbf{\textit{\underline{0.525}}} & \textbf{\textit{\underline{0.538}}} & \textbf{\textit{\underline{0.514}}} & \textbf{\textit{\underline{0.585}}} & \textbf{\textit{\underline{0.539}}} \\ 
BG & 1.105 & 1.030 & 1.060 & 1.007 & 1.065 & 0.987 & 0.904 & 0.911 & 0.902 & 0.913 & 1.873  & - & \textit{0.925} & 0.985 & \textit{0.910} & \textit{0.874} & 0.918 & \textit{0.931} & 0.938 & 0.902 & 1.010 & 0.949 \\ 
RF & 1.196 & 1.114 & 1.140 & 1.088 & 1.154 & 1.065 & 0.977 & 0.983 & 0.974 & 0.986 & 2.013 & 1.083  & - & 1.063 & 0.982 & 0.944 & 0.992 & 1.003 & 1.014 & 0.974 & 1.090 & 1.024 \\ 
GB & 1.135 & 1.057 & 1.065 & 1.033 & 1.098 & 1.007 & \textit{0.922} & \textit{0.928} & \textit{0.919} & 0.930 & 1.887 & 1.029 & 0.948  & - & \textbf{\textit{0.933}} & \textit{0.894} & 0.935 & \textit{0.941} & 0.958 & 0.919 & 1.035 & 0.966 \\ 
NN$_{1}^{1}$ & 1.256 & 1.175 & 1.198 & 1.148 & 1.220 & 1.118 & 1.021 & 1.028 & 1.019 & 1.031 & 2.106 & 1.139 & 1.050 & 1.114  & - & 0.977 & 1.039 & 1.034 & 1.055 & 1.012 & 1.112 & 1.063 \\ 
NN$_{1}^{10}$ & 1.280 & 1.196 & 1.215 & 1.170 & 1.242 & 1.139 & 1.041 & 1.047 & 1.038 & 1.051 & 2.140 & 1.161 & 1.070 & 1.135 & 1.033  & - & 1.058 & 1.059 & 1.077 & 1.033 & 1.147 & 1.086 \\ 
NN$_{2}^{1}$ & 1.236 & 1.153 & 1.163 & 1.128 & 1.199 & 1.095 & 1.000 & 1.007 & 0.997 & 1.010 & 2.038 & 1.119 & 1.032 & 1.092 & 1.013 & 0.971  & - & 1.021 & 1.039 & 0.997 & 1.124 & 1.047 \\ 
NN$_{2}^{10}$ & 1.250 & 1.168 & 1.191 & 1.142 & 1.213 & 1.113 & 1.014 & 1.020 & 1.011 & 1.023 & 2.073 & 1.136 & 1.046 & 1.106 & 1.013 & 0.977 & 1.031  & - & 1.050 & 1.000 & 1.125 & 1.052 \\ 
NN$_{3}^{1}$ & 1.197 & 1.116 & 1.125 & 1.092 & 1.158 & 1.063 & 0.973 & 0.979 & 0.970 & 0.982 & 1.994 & 1.084 & 1.001 & 1.060 & 0.971 & 0.938 & 0.985 & \textbf{\textit{0.990}}  & - & \textit{0.964} & 1.071 & 1.013 \\ 
NN$_{3}^{10}$ & 1.246 & 1.163 & 1.179 & 1.137 & 1.207 & 1.108 & 1.011 & 1.017 & 1.008 & 1.020 & 2.067 & 1.130 & 1.042 & 1.102 & 1.011 & 0.975 & 1.026 & 1.017 & 1.044  & - & 1.119 & 1.051 \\ 
NN$_{4}^{1}$ & 1.145 & 1.066 & 1.083 & 1.043 & 1.108 & 1.015 & \textit{0.929} & 0.935 & \textit{0.927} & 0.937 & 1.901 & 1.035 & 0.953 & 1.011 & \textit{0.910} & \textbf{\textit{0.888}} & \textbf{\textit{0.943}} & \textbf{\textit{0.942}} & 0.952 & \textbf{\textit{0.916}}  & - & 0.961 \\ 
NN$_{4}^{10}$ & 1.189 & 1.109 & 1.120 & 1.084 & 1.152 & 1.055 & 0.963 & 0.969 & 0.961 & 0.972 & 1.965 & 1.077 & 0.992 & 1.050 & 0.962 & \textit{0.928} & 0.977 & \textbf{\textit{0.969}} & 0.995 & \textbf{\textit{\underline{0.952}}} & 1.064  & - \\ 
\hline \hline
\end{tabular}
\smallskip
\begin{scriptsize}
\parbox{0.98\textwidth}{\emph{Note.} We report the out-of-sample realized variance forecast MSE of each model in the selected column relative to the benchmark in the selected row. Each number is a cross-sectional average of such pairwise relative MSEs for each stock. 
The formatting is as follows: \textit{number} (\textbf{\textit{number}}) [\underline{\textbf{\textit{number}}}] denotes whether the Diebold-Mariano test of equal predictive accuracy is rejected more than 50\% of the time at the 10\% (5\%) [1\%] level of significance across individual tests for each asset. 
The hypothesis being tested is $\text{H}_{0}: \text{MSE}_{i} = \text{MSE}_{j}$ against a one-sided alternative $\text{H}_{1}: \text{MSE}_{i} > \text{MSE}_{j}$, 
where model $i$ is the label of the selected row, whereas model $j$ is the label of the selected column.}
\end{scriptsize}
\end{center}
\end{scriptsize}
\end{sidewaystable}

\clearpage

\noindent \underline{\textbf{Training set = 2,000 days}}:

\begin{sidewaystable}
\begin{scriptsize}
\setlength{ \tabcolsep}{0.09cm}
\begin{center}
\caption{One-day-ahead relative MSE and Diebold-Mariano test for dataset $\mathcal{M}_{\text{HAR}}$.
\label{table:har-1d-2000.tex}}
\begin{tabular}{lcccccccccccccccccccccc}
\hline \hline
& HAR & HAR-X & LogHAR & LevHAR & SHAR & HARQ & RR & LA & EN & A-LA & P-LA & BG & RF & GB & NN$_{1}^{1}$ & NN$_{1}^{10}$ & NN$_{2}^{1}$ & NN$_{2}^{10}$ & NN$_{3}^{1}$ & NN$_{3}^{10}$ & NN$_{4}^{1}$ & NN$_{4}^{10}$ \\
\hline
HAR  & - & 1.000 & 1.025 & 1.143 & 1.020 & 1.013 & 0.984 & 1.001 & 0.986 & 1.010 & 1.008 & 1.210 & 1.034 & 1.065 & 1.006 & \textit{0.983} & 1.064 & 1.008 & 1.059 & 1.050 & 1.120 & 1.137 \\ 
HAR-X & 1.000  & - & 1.025 & 1.143 & 1.020 & 1.013 & 0.984 & 1.001 & 0.986 & 1.010 & 1.008 & 1.210 & 1.034 & 1.065 & 1.006 & \textit{0.983} & 1.064 & 1.008 & 1.059 & 1.050 & 1.120 & 1.137 \\ 
LogHAR & 0.984 & 0.984  & - & 1.110 & 1.005 & 0.992 & 0.967 & 0.984 & 0.969 & 0.994 & 0.992 & 1.188 & 1.016 & 1.046 & 0.989 & 0.967 & 1.042 & 0.990 & 1.036 & 1.029 & 1.092 & 1.110 \\ 
LevHAR & \textit{0.904} & \textit{0.904} & \textit{0.916}  & - & 0.919 & 0.906 & \textit{0.888} & 0.903 & 0.891 & 0.912 & 0.911 & 1.079 & 0.930 & 0.960 & 0.907 & \textbf{\textit{0.887}} & 0.950 & 0.908 & 0.948 & 0.941 & 0.994 & 1.011 \\ 
SHAR & 0.985 & 0.985 & 1.011 & 1.124  & - & 0.996 & 0.969 & 0.986 & 0.972 & 0.995 & 0.993 & 1.186 & 1.015 & 1.047 & 0.988 & 0.967 & 1.043 & 0.992 & 1.040 & 1.032 & 1.101 & 1.117 \\ 
HARQ & 0.999 & 0.999 & 1.019 & 1.129 & 1.016  & - & 0.981 & 0.998 & 0.984 & 1.008 & 1.007 & 1.194 & 1.026 & 1.059 & 1.000 & 0.979 & 1.058 & 1.004 & 1.054 & 1.045 & 1.115 & 1.132 \\ 
RR & 1.018 & 1.018 & 1.041 & 1.161 & 1.038 & 1.030  & - & 1.018 & 1.003 & 1.028 & 1.026 & 1.229 & 1.051 & 1.082 & 1.023 & 1.000 & 1.080 & 1.025 & 1.075 & 1.066 & 1.136 & 1.153 \\ 
LA & \textbf{\textit{1.000}} & \textbf{\textit{1.000}} & 1.024 & 1.141 & 1.020 & 1.012 & \textit{0.983}  & - & \textit{0.986} & 1.010 & \textit{1.008} & 1.207 & 1.032 & 1.063 & 1.005 & \textit{0.983} & 1.062 & 1.007 & 1.057 & 1.048 & 1.117 & 1.134 \\ 
EN & 1.015 & 1.015 & 1.039 & 1.159 & 1.035 & 1.028 & 0.997 & 1.015  & - & 1.025 & 1.023 & 1.227 & 1.048 & 1.079 & 1.020 & 0.998 & 1.078 & 1.022 & 1.072 & 1.063 & 1.133 & 1.150 \\ 
A-LA & \textbf{\textit{0.990}} & \textbf{\textit{0.990}} & 1.014 & 1.130 & 1.010 & 1.002 & 0.974 & 0.990 & 0.976  & - & \textit{0.998} & 1.195 & 1.023 & 1.053 & 0.995 & \textbf{\textit{0.973}} & 1.052 & 0.998 & 1.048 & 1.039 & 1.108 & 1.124 \\ 
P-LA & \textbf{\textit{0.992}} & \textbf{\textit{0.992}} & 1.017 & 1.134 & 1.011 & 1.006 & 0.976 & 0.993 & 0.979 & 1.003  & - & 1.201 & 1.026 & 1.057 & 0.998 & \textbf{\textit{0.976}} & 1.056 & 1.000 & 1.051 & 1.042 & 1.111 & 1.128 \\ 
BG & \textbf{\textit{0.853}} & \textbf{\textit{0.853}} & 0.871 & 0.962 & 0.866 & \textit{0.854} & \textbf{\textit{0.837}} & \textit{0.851} & \textbf{\textit{0.839}} & \textit{0.860} & \textit{0.860}  & - & \textbf{\textit{0.870}} & 0.900 & \textit{0.855} & \textbf{\textit{0.836}} & 0.900 & \textit{0.856} & 0.900 & 0.890 & 0.951 & 0.965 \\ 
RF & 0.976 & 0.976 & 0.998 & 1.110 & 0.992 & 0.982 & 0.958 & 0.975 & 0.961 & 0.985 & 0.984 & 1.165  & - & 1.032 & 0.979 & 0.958 & 1.032 & 0.980 & 1.028 & 1.019 & 1.087 & 1.103 \\ 
GB & 0.949 & 0.949 & 0.971 & 1.082 & 0.966 & 0.957 & 0.932 & 0.948 & 0.935 & 0.958 & 0.957 & 1.138 & 0.975  & - & 0.952 & \textit{0.932} & 1.004 & 0.953 & 1.000 & 0.991 & 1.056 & 1.071 \\ 
NN$_{1}^{1}$ & 0.998 & 0.998 & 1.022 & 1.137 & 1.015 & 1.007 & 0.981 & 0.998 & 0.984 & 1.008 & 1.006 & 1.203 & 1.029 & 1.060  & - & \textit{0.979} & 1.057 & 1.003 & 1.052 & 1.044 & 1.114 & 1.130 \\ 
NN$_{1}^{10}$ & 1.019 & 1.019 & 1.043 & 1.161 & 1.037 & 1.029 & 1.002 & 1.019 & 1.004 & 1.029 & 1.027 & 1.228 & 1.051 & 1.083 & 1.022  & - & 1.080 & 1.025 & 1.076 & 1.067 & 1.138 & 1.155 \\ 
NN$_{2}^{1}$ & 0.960 & 0.960 & 0.980 & 1.087 & 0.975 & 0.968 & 0.943 & 0.960 & 0.945 & 0.969 & 0.968 & 1.152 & 0.986 & 1.016 & 0.962 & \textbf{\textit{0.941}}  & - & 0.963 & 1.002 & 0.995 & 1.053 & 1.069 \\ 
NN$_{2}^{10}$ & 0.995 & 0.995 & 1.018 & 1.134 & 1.013 & 1.005 & 0.978 & 0.995 & 0.980 & 1.005 & 1.003 & 1.199 & 1.025 & 1.055 & 0.998 & \textit{0.977} & 1.052  & - & 1.047 & 1.039 & 1.106 & 1.123 \\ 
NN$_{3}^{1}$ & 0.959 & 0.959 & 0.978 & 1.087 & 0.976 & 0.968 & \textit{0.942} & 0.959 & 0.944 & 0.968 & 0.967 & 1.157 & 0.987 & 1.016 & 0.961 & \textbf{\textit{0.941}} & 1.004 & \textit{0.962}  & - & 0.995 & 1.053 & 1.069 \\ 
NN$_{3}^{10}$ & 0.964 & 0.964 & 0.984 & 1.092 & 0.981 & 0.972 & \textbf{\textit{0.946}} & 0.963 & \textbf{\textit{0.949}} & 0.973 & 0.971 & 1.160 & 0.991 & 1.020 & \textbf{\textit{0.966}} & \textbf{\textit{\underline{0.945}}} & 1.010 & \textbf{\textit{\underline{0.966}}} & 1.007  & - & 1.060 & 1.076 \\ 
NN$_{4}^{1}$ & \textbf{\textit{\underline{0.919}}} & \textbf{\textit{\underline{0.919}}} & \textbf{\textit{\underline{0.934}}} & 1.033 & \textbf{\textit{\underline{0.937}}} & \textbf{\textit{0.928}} & \textbf{\textit{\underline{0.902}}} & \textbf{\textit{\underline{0.918}}} & \textbf{\textit{\underline{0.904}}} & \textbf{\textit{0.928}} & \textbf{\textit{\underline{0.926}}} & 1.109 & 0.945 & 0.973 & \textbf{\textit{\underline{0.922}}} & \textbf{\textit{\underline{0.902}}} & \textbf{\textit{0.959}} & \textbf{\textit{\underline{0.921}}} & \textbf{\textit{\underline{0.957}}} & \textbf{\textit{\underline{0.951}}}  & - & 1.017 \\ 
NN$_{4}^{10}$ & \textbf{\textit{\underline{0.904}}} & \textbf{\textit{\underline{0.904}}} & \textbf{\textit{\underline{0.920}}} & 1.019 & \textbf{\textit{\underline{0.921}}} & \textbf{\textit{0.912}} & \textbf{\textit{\underline{0.887}}} & \textbf{\textit{\underline{0.903}}} & \textbf{\textit{\underline{0.890}}} & \textbf{\textit{\underline{0.913}}} & \textbf{\textit{\underline{0.912}}} & 1.091 & \textit{0.930} & 0.956 & \textbf{\textit{\underline{0.907}}} & \textbf{\textit{\underline{0.887}}} & \textbf{\textit{\underline{0.943}}} & \textbf{\textit{\underline{0.906}}} & \textbf{\textit{\underline{0.941}}} & \textbf{\textit{\underline{0.935}}} & \textbf{\textit{0.985}}  & - \\ 
\hline \hline
\end{tabular}
\smallskip
\begin{scriptsize}
\parbox{0.98\textwidth}{\emph{Note.} We report the out-of-sample realized variance forecast MSE of each model in the selected column relative to the benchmark in the selected row. Each number is a cross-sectional average of such pairwise relative MSEs for each stock. 
The formatting is as follows: \textit{number} (\textbf{\textit{number}}) [\underline{\textbf{\textit{number}}}] denotes whether the Diebold-Mariano test of equal predictive accuracy is rejected more than 50\% of the time at the 10\% (5\%) [1\%] level of significance across individual tests for each asset. 
The hypothesis being tested is $\text{H}_{0}: \text{MSE}_{i} = \text{MSE}_{j}$ against a one-sided alternative $\text{H}_{1}: \text{MSE}_{i} > \text{MSE}_{j}$, 
where model $i$ is the label of the selected row, whereas model $j$ is the label of the selected column.}
\end{scriptsize}
\end{center}
\end{scriptsize}
\end{sidewaystable}
\begin{sidewaystable}
\begin{scriptsize}
\setlength{ \tabcolsep}{0.09cm}
\begin{center}
\caption{One-day-ahead relative MSE and Diebold-Mariano test for dataset $\mathcal{M}_{\text{ALL}}$.
\label{table:all-1d-2000.tex}}
\begin{tabular}{lcccccccccccccccccccccc}
\hline \hline
& HAR & HAR-X & LogHAR & LevHAR & SHAR & HARQ & RR & LA & EN & A-LA & P-LA & BG & RF & GB & NN$_{1}^{1}$ & NN$_{1}^{10}$ & NN$_{2}^{1}$ & NN$_{2}^{10}$ & NN$_{3}^{1}$ & NN$_{3}^{10}$ & NN$_{4}^{1}$ & NN$_{4}^{10}$ \\
\hline
HAR  & - & 1.048 & \textbf{\textit{\underline{0.939}}} & 1.115 & 1.049 & 1.467 & \textbf{\textit{0.976}} & \textit{0.984} & \textbf{\textit{0.950}} & 1.010 & 1.041 & 0.996 & \textbf{\textit{\underline{0.918}}} & 0.997 & 1.048 & \textbf{\textit{\underline{0.933}}} & 1.061 & 0.996 & 1.126 & 1.083 & 1.203 & 1.183 \\ 
HAR-X & 0.972  & - & \textbf{\textit{\underline{0.905}}} & 1.056 & 1.001 & 1.323 & \textbf{\textit{\underline{0.937}}} & \textbf{\textit{0.945}} & \textbf{\textit{\underline{0.915}}} & 0.970 & 0.995 & 0.968 & \textbf{\textit{0.893}} & 0.961 & 1.013 & \textbf{\textit{\underline{0.907}}} & 1.019 & 0.958 & 1.077 & 1.031 & 1.134 & 1.118 \\ 
LogHAR & 1.075 & 1.116  & - & 1.180 & 1.118 & 1.507 & 1.040 & 1.051 & 1.015 & 1.080 & 1.110 & 1.071 & 0.988 & 1.068 & 1.123 & 1.003 & 1.133 & 1.064 & 1.199 & 1.151 & 1.276 & 1.258 \\ 
LevHAR & 0.934 & \textit{0.955} & \textbf{\textit{\underline{0.865}}}  & - & 0.955 & 1.240 & \textbf{\textit{0.894}} & \textbf{\textit{0.904}} & \textbf{\textit{0.876}} & 0.928 & 0.951 & 0.927 & \textit{0.856} & 0.920 & 0.970 & \textbf{\textit{\underline{0.871}}} & 0.977 & 0.918 & 1.031 & 0.987 & 1.087 & 1.071 \\ 
SHAR & 0.974 & 1.002 & \textbf{\textit{\underline{0.907}}} & 1.058  & - & 1.326 & \textbf{\textit{0.937}} & 0.946 & \textbf{\textit{0.916}} & 0.971 & 0.996 & 0.968 & \textit{0.893} & 0.961 & 1.013 & \textbf{\textit{\underline{0.908}}} & 1.020 & 0.959 & 1.080 & 1.033 & 1.136 & 1.120 \\ 
HARQ & \textbf{\textit{\underline{0.826}}} & \textbf{\textit{\underline{0.830}}} & \textbf{\textit{\underline{0.758}}} & \textbf{\textit{0.864}} & \textbf{\textit{\underline{0.831}}}  & - & \textbf{\textit{\underline{0.781}}} & \textbf{\textit{\underline{0.790}}} & \textbf{\textit{\underline{0.768}}} & \textbf{\textit{\underline{0.813}}} & \textbf{\textit{\underline{0.827}}} & \textbf{\textit{\underline{0.822}}} & \textbf{\textit{\underline{0.759}}} & \textbf{\textit{\underline{0.808}}} & \textbf{\textit{\underline{0.851}}} & \textbf{\textit{\underline{0.771}}} & \textbf{\textit{\underline{0.856}}} & \textbf{\textit{\underline{0.803}}} & \textbf{\textit{0.898}} & \textbf{\textit{0.853}} & 0.929 & 0.919 \\ 
RR & 1.037 & 1.072 & \textit{0.966} & 1.132 & 1.072 & 1.430  & - & 1.010 & 0.977 & 1.038 & 1.065 & 1.032 & 0.952 & 1.026 & 1.081 & \textbf{\textit{0.966}} & 1.089 & 1.025 & 1.153 & 1.106 & 1.219 & 1.202 \\ 
LA & 1.027 & 1.062 & \textbf{\textit{0.958}} & 1.125 & 1.062 & 1.422 & 0.992  & - & \textbf{\textit{0.968}} & 1.027 & 1.055 & 1.022 & 0.942 & 1.016 & 1.071 & \textbf{\textit{0.957}} & 1.079 & 1.015 & 1.144 & 1.095 & 1.204 & 1.188 \\ 
EN & 1.060 & 1.100 & 0.989 & 1.165 & 1.100 & 1.484 & 1.025 & 1.034  & - & 1.062 & 1.092 & 1.054 & 0.972 & 1.050 & 1.106 & 0.988 & 1.115 & 1.049 & 1.182 & 1.134 & 1.251 & 1.233 \\ 
A-LA & 1.001 & 1.036 & \textbf{\textit{\underline{0.935}}} & 1.097 & 1.036 & 1.398 & \textbf{\textit{0.968}} & \textbf{\textit{\underline{0.975}}} & \textbf{\textit{\underline{0.944}}}  & - & 1.029 & 0.996 & \textbf{\textit{0.918}} & 0.990 & 1.044 & \textbf{\textit{\underline{0.933}}} & 1.053 & 0.990 & 1.118 & 1.070 & 1.175 & 1.160 \\ 
P-LA & 0.978 & 1.007 & \textbf{\textit{\underline{0.911}}} & 1.065 & 1.008 & 1.332 & \textbf{\textit{\underline{0.942}}} & \textbf{\textit{\underline{0.950}}} & \textbf{\textit{\underline{0.920}}} & 0.975  & - & 0.974 & \textbf{\textit{0.897}} & 0.966 & 1.018 & \textbf{\textit{\underline{0.911}}} & 1.024 & 0.964 & 1.084 & 1.038 & 1.137 & 1.124 \\ 
BG & 1.024 & 1.073 & 0.961 & 1.138 & 1.072 & 1.512 & 0.998 & 1.007 & 0.971 & 1.032 & 1.066  & - & \textit{0.932} & 1.017 & 1.071 & 0.954 & 1.081 & 1.020 & 1.148 & 1.111 & 1.235 & 1.215 \\ 
RF & 1.095 & 1.149 & 1.030 & 1.222 & 1.148 & 1.624 & 1.069 & 1.077 & 1.040 & 1.104 & 1.141 & 1.081  & - & 1.090 & 1.147 & 1.020 & 1.160 & 1.092 & 1.233 & 1.190 & 1.320 & 1.299 \\ 
GB & 1.012 & 1.050 & \textit{0.947} & 1.113 & 1.049 & 1.426 & 0.980 & 0.988 & 0.955 & 1.014 & 1.043 & 1.004 & \textbf{\textit{0.927}}  & - & 1.056 & \textit{0.944} & 1.062 & 1.000 & 1.124 & 1.081 & 1.193 & 1.175 \\ 
NN$_{1}^{1}$ & 0.966 & 1.005 & \textbf{\textit{0.904}} & 1.065 & 1.005 & 1.364 & \textit{0.937} & 0.945 & \textbf{\textit{0.913}} & 0.970 & 0.998 & 0.961 & \textbf{\textit{0.886}} & 0.959  & - & \textbf{\textit{\underline{0.898}}} & 1.019 & 0.957 & 1.078 & 1.036 & 1.143 & 1.128 \\ 
NN$_{1}^{10}$ & 1.077 & 1.128 & 1.011 & 1.200 & 1.128 & 1.576 & 1.050 & 1.058 & 1.022 & 1.086 & 1.120 & 1.072 & 0.987 & 1.073 & 1.125  & - & 1.141 & 1.073 & 1.213 & 1.166 & 1.290 & 1.271 \\ 
NN$_{2}^{1}$ & 0.968 & 1.003 & \textbf{\textit{\underline{0.904}}} & 1.064 & 1.004 & 1.344 & \textbf{\textit{0.936}} & \textit{0.944} & \textbf{\textit{0.913}} & 0.970 & 0.995 & 0.960 & \textbf{\textit{0.888}} & 0.956 & 1.010 & \textbf{\textit{\underline{0.902}}}  & - & \textbf{\textit{0.952}} & 1.058 & 1.024 & 1.128 & 1.112 \\ 
NN$_{2}^{10}$ & 1.015 & 1.052 & \textbf{\textit{0.948}} & 1.116 & 1.053 & 1.415 & 0.983 & 0.991 & \textbf{\textit{0.959}} & 1.018 & 1.045 & 1.012 & \textit{0.933} & 1.005 & 1.059 & \textit{0.947} & 1.062  & - & 1.121 & 1.078 & 1.191 & 1.173 \\ 
NN$_{3}^{1}$ & 0.928 & 0.957 & \textbf{\textit{\underline{0.863}}} & 1.015 & 0.960 & 1.266 & \textbf{\textit{\underline{0.896}}} & \textbf{\textit{0.904}} & \textbf{\textit{\underline{0.874}}} & \textit{0.930} & 0.952 & \textbf{\textit{0.923}} & \textbf{\textit{\underline{0.852}}} & \textit{0.916} & \textbf{\textit{0.966}} & \textbf{\textit{\underline{0.866}}} & \textbf{\textit{0.957}} & \textbf{\textit{\underline{0.908}}}  & - & 0.974 & 1.076 & 1.061 \\ 
NN$_{3}^{10}$ & 0.954 & 0.982 & \textbf{\textit{\underline{0.888}}} & 1.041 & 0.984 & 1.285 & \textbf{\textit{\underline{0.919}}} & \textbf{\textit{0.927}} & \textbf{\textit{\underline{0.898}}} & 0.954 & 0.976 & \textbf{\textit{0.952}} & \textbf{\textit{\underline{0.877}}} & 0.941 & 0.993 & \textbf{\textit{\underline{0.890}}} & \textit{0.990} & \textbf{\textit{\underline{0.935}}} & 1.041  & - & 1.101 & 1.086 \\ 
NN$_{4}^{1}$ & \textbf{\textit{\underline{0.884}}} & \textbf{\textit{\underline{0.907}}} & \textbf{\textit{\underline{0.824}}} & \textbf{\textit{0.964}} & \textbf{\textit{\underline{0.909}}} & 1.180 & \textbf{\textit{\underline{0.851}}} & \textbf{\textit{\underline{0.856}}} & \textbf{\textit{\underline{0.831}}} & \textbf{\textit{\underline{0.879}}} & \textbf{\textit{\underline{0.899}}} & \textbf{\textit{\underline{0.881}}} & \textbf{\textit{\underline{0.812}}} & \textbf{\textit{\underline{0.871}}} & \textbf{\textit{\underline{0.918}}} & \textbf{\textit{\underline{0.823}}} & \textbf{\textit{\underline{0.917}}} & \textbf{\textit{\underline{0.867}}} & \textbf{\textit{\underline{0.969}}} & \textbf{\textit{\underline{0.927}}}  & - & 0.995 \\ 
NN$_{4}^{10}$ & \textbf{\textit{\underline{0.890}}} & \textbf{\textit{0.912}} & \textbf{\textit{\underline{0.829}}} & \textbf{\textit{0.968}} & \textbf{\textit{0.914}} & 1.188 & \textbf{\textit{\underline{0.856}}} & \textbf{\textit{\underline{0.862}}} & \textbf{\textit{\underline{0.836}}} & \textbf{\textit{\underline{0.885}}} & \textbf{\textit{\underline{0.906}}} & \textbf{\textit{\underline{0.888}}} & \textbf{\textit{\underline{0.818}}} & \textbf{\textit{\underline{0.876}}} & \textbf{\textit{\underline{0.925}}} & \textbf{\textit{\underline{0.829}}} & \textbf{\textit{\underline{0.922}}} & \textbf{\textit{\underline{0.871}}} & \textbf{\textit{\underline{0.972}}} & \textbf{\textit{\underline{0.931}}} & 1.013  & - \\ 
\hline \hline
\end{tabular}
\smallskip
\begin{scriptsize}
\parbox{0.98\textwidth}{\emph{Note.} We report the out-of-sample realized variance forecast MSE of each model in the selected column relative to the benchmark in the selected row. Each number is a cross-sectional average of such pairwise relative MSEs for each stock. 
The formatting is as follows: \textit{number} (\textbf{\textit{number}}) [\underline{\textbf{\textit{number}}}] denotes whether the Diebold-Mariano test of equal predictive accuracy is rejected more than 50\% of the time at the 10\% (5\%) [1\%] level of significance across individual tests for each asset. 
The hypothesis being tested is $\text{H}_{0}: \text{MSE}_{i} = \text{MSE}_{j}$ against a one-sided alternative $\text{H}_{1}: \text{MSE}_{i} > \text{MSE}_{j}$, 
where model $i$ is the label of the selected row, whereas model $j$ is the label of the selected column.}
\end{scriptsize}
\end{center}
\end{scriptsize}
\end{sidewaystable}

\clearpage

\subsection{Additional VI measures} \label{appendix:variable-importance}

In the figures below, we report the VI measures for those forecasting models that were excluded from Figure \ref{figure:variable-importance} in Section \ref{section:variable-importance} in the main text.

\begin{figure}[ht!]
	\begin{center}
		\caption{VI measure.}
		\begin{tabular}{cc}
			\small{Panel A: LogHAR.} & \small{Panel B: LevHAR.} \\
			\includegraphics[height=5.2cm,width=0.48\textwidth]{{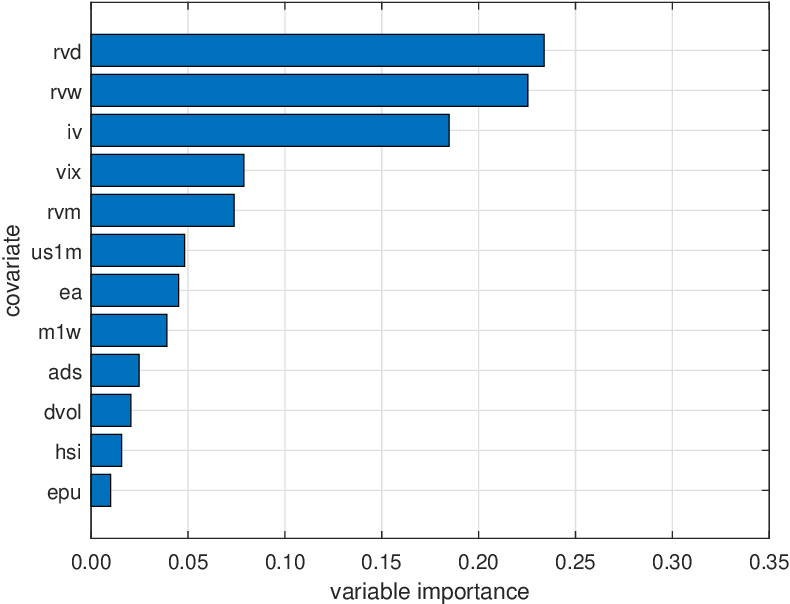}} &
			\includegraphics[height=5.2cm,width=0.48\textwidth]{{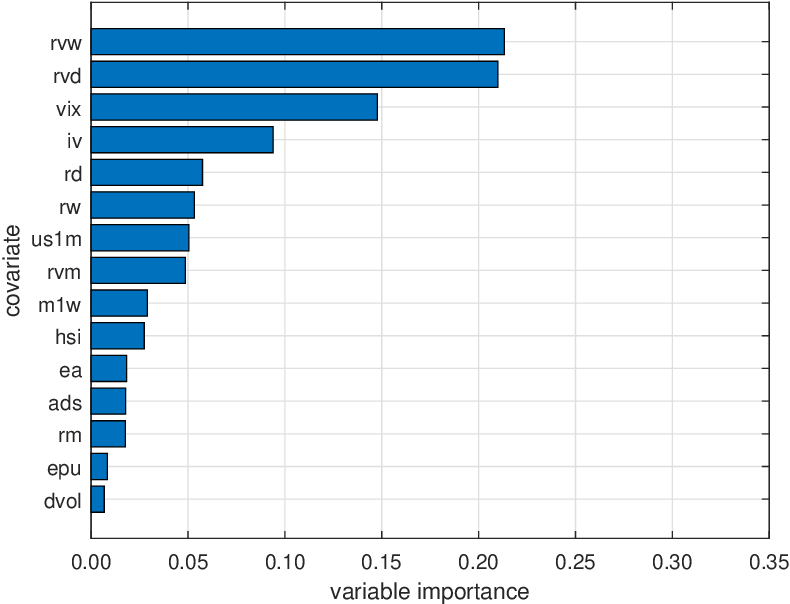}} \\
			\small{Panel C: SHAR.} & \small{Panel D: HARQ.} \\
			\includegraphics[height=5.2cm,width=0.48\textwidth]{{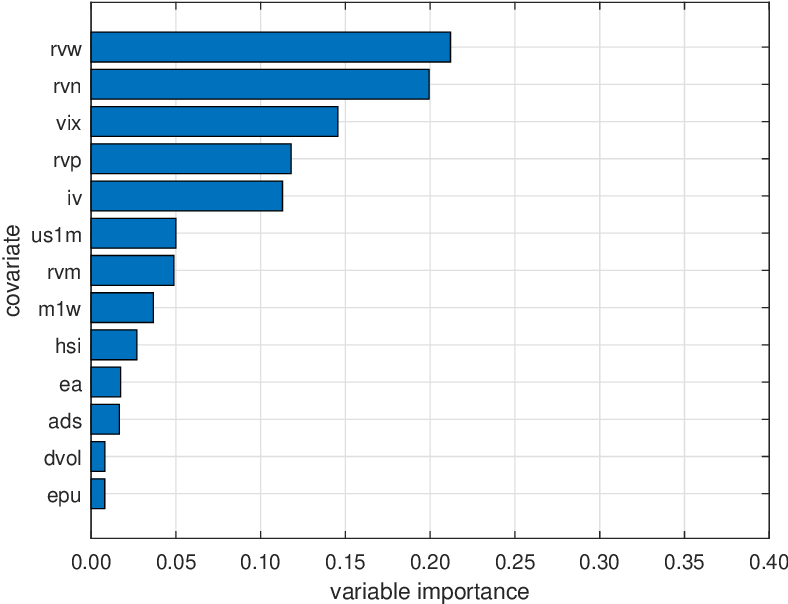}} &
			\includegraphics[height=5.2cm,width=0.48\textwidth]{{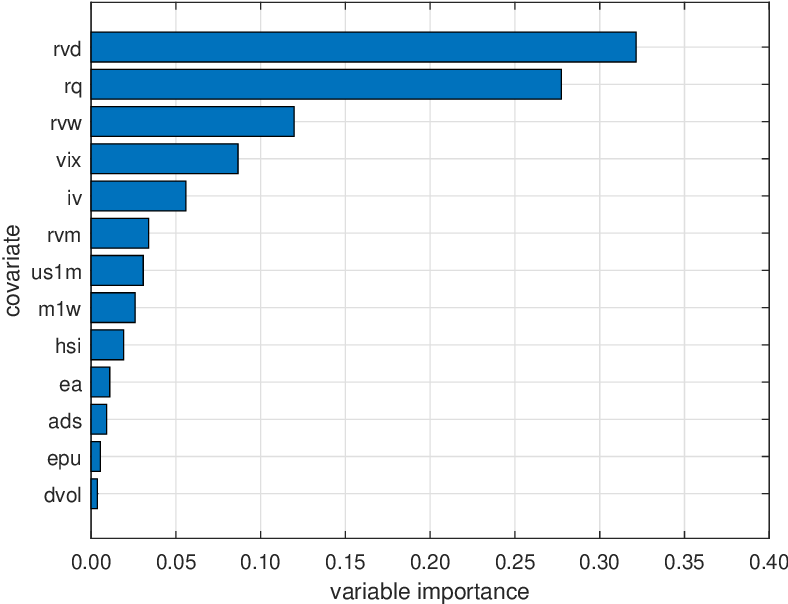}} \\
			\small{Panel E: Ridge.} & \small{Panel F: Lasso.} \\
			\includegraphics[height=5.2cm,width=0.48\textwidth]{{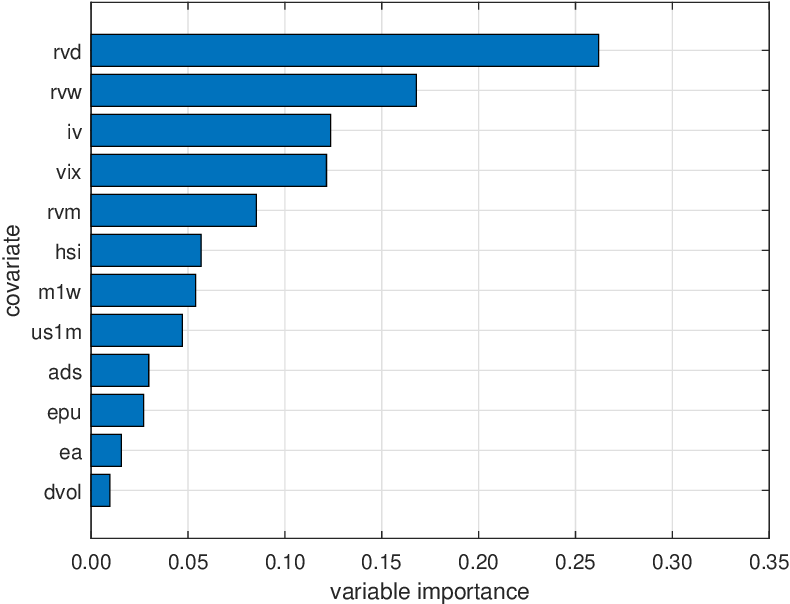}} &
			\includegraphics[height=5.2cm,width=0.48\textwidth]{{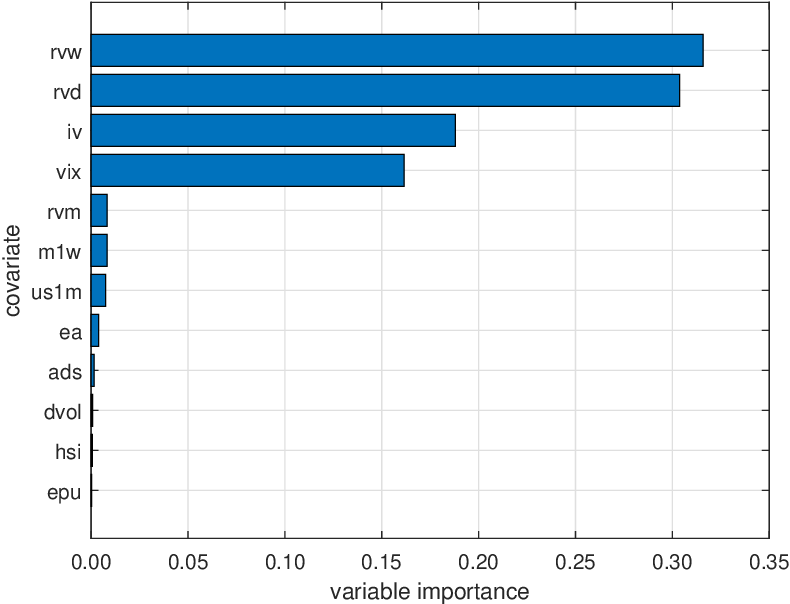}} \\
		\end{tabular}
		\begin{scriptsize}
			\parbox{\textwidth}{\emph{Note.} We report the VI measure for each feature of LogHAR, LevHAR, SHAR, HARQ, ridge, and lasso, sorted according to the numeric value of VI. Please refer to the main text for further details.}
		\end{scriptsize}
	\end{center}
\end{figure}

\begin{figure}[ht!]
	\begin{center}
		\caption{VI measure (continued).}
		\begin{tabular}{cc}
			\small{Panel A: Adaptive lasso.} & \small{Panel B: Post lasso.} \\
			\includegraphics[height=5.2cm,width=0.48\textwidth]{{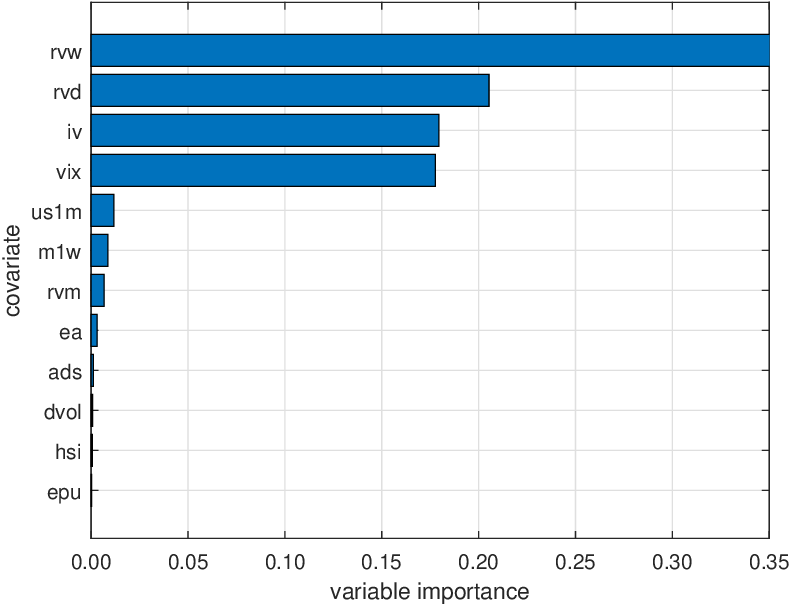}} &
			\includegraphics[height=5.2cm,width=0.48\textwidth]{{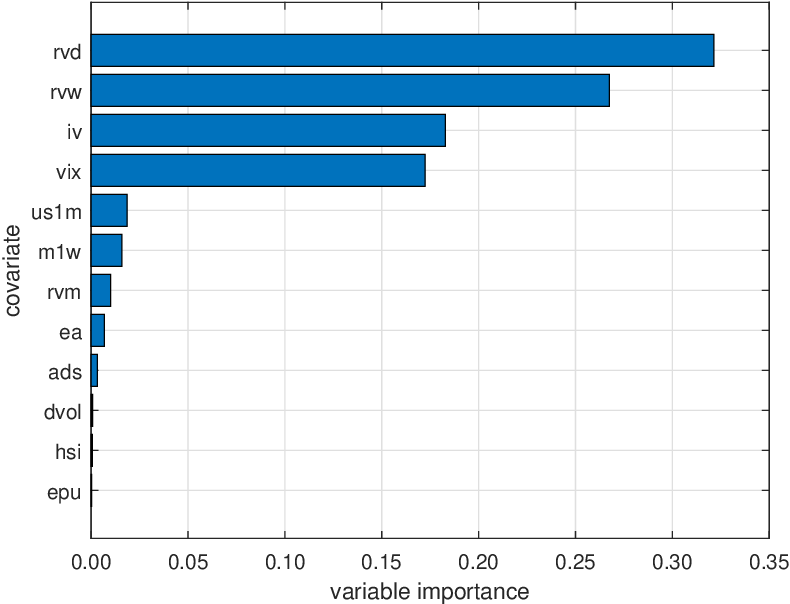}} \\
			\small{Panel C: Bagging.} & \small{Panel D: Gradient boosting.} \\
			\includegraphics[height=5.2cm,width=0.48\textwidth]{{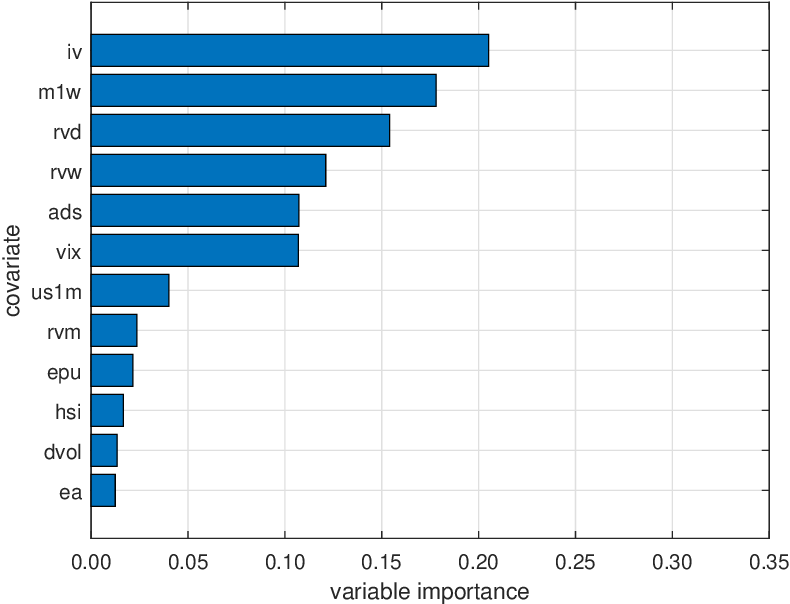}} &
			\includegraphics[height=5.2cm,width=0.48\textwidth]{{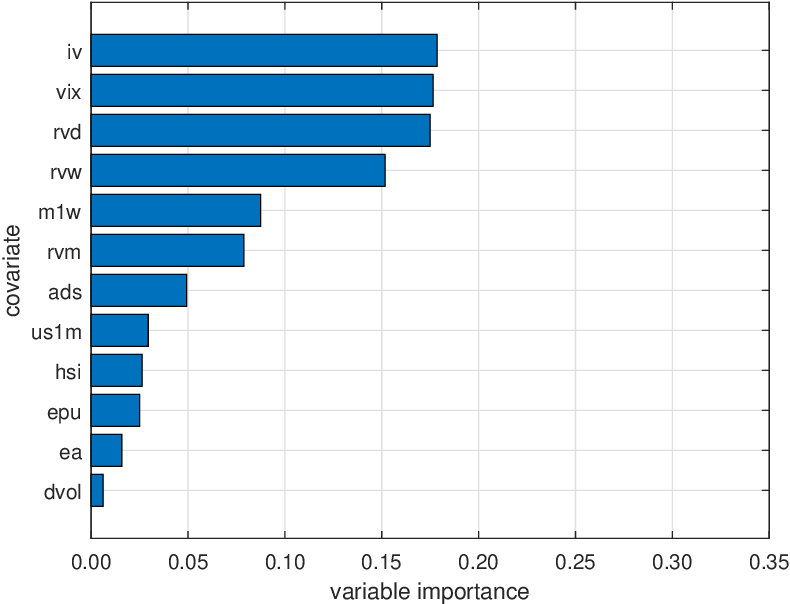}} \\
			\small{Panel E: NN$_{1}^{10}$.} & \small{Panel F: NN$_{4}^{10}$.} \\
			\includegraphics[height=5.2cm,width=0.48\textwidth]{{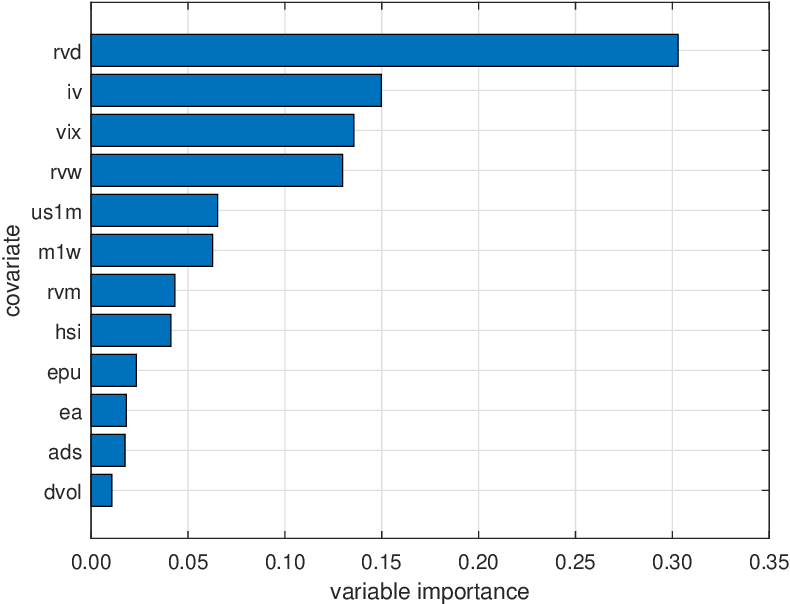}} &
			\includegraphics[height=5.2cm,width=0.48\textwidth]{{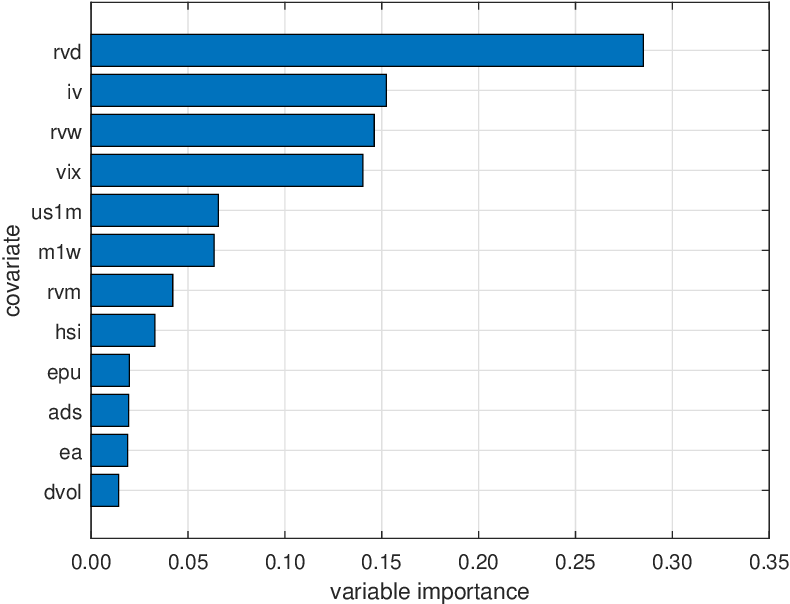}} \\
		\end{tabular}
		\begin{scriptsize}
			\parbox{\textwidth}{\emph{Note.} We report the VI measure for each feature of adaptive lasso, post lasso, bagging, NN$_{1}^{10}$, and NN$_{4}^{10}$ (NN$_{3}^{10}$ is omitted to conserve space), sorted according to the numeric value of VI. Please refer to the main text for further details.}
		\end{scriptsize}
	\end{center}
\end{figure}

\clearpage

\subsection{In-sample parameter estimates of HAR and HAR-X} \label{appendix:har}

\begin{table}[ht!]
\setlength{ \tabcolsep}{2.0cm}
\begin{center}
\caption{Parameter estimates of the HAR and HAR-X model. \label{table:har}}
\begin{tabular}{lcc}
\hline \hline
& HAR &  HAR-X \\
\hline
$\beta_{0} $ & $\underset{(-0.477)}{-0.006}$ & $\underset{(0.378)}{0.005}$ \\
$\beta_{ \text{RVD}}$  & $\underset{( 2.507)}{ 0.185}$ & $\underset{( 1.501 )}{ 0.110}$ \\
$\beta_{ \text{RVW}}$  & $\underset{( 3.782)}{ 0.415 }$ & $\underset{( 3.151 )}{ 0.365}$ \\
$\beta_{ \text{RVM}}$  & $\underset{( 2.635)}{ 0.166 }$  & $\underset{( 1.281)}{ 0.079}$ \\
$\beta_{ \text{IV}}$  & & $\underset{( 6.390)}{ 0.198}$ \\
$\beta_{ \text{EA}}$  &  & $\underset{( 4.656 )}{ 0.043}$ \\
$\beta_{ \text{EPU}}$  & & $\underset{( 1.612 )}{ 0.039}$ \\
$\beta_{ \text{VIX}}$  &    & $\underset{(0.088 )}{0.003}$ \\
$\beta_{ \text{US1M}}$  & & $\underset{(-2.001 )}{-0.075}$ \\
$\beta_{ \text{HSI}}$  &  & $\underset{(-0.508 )}{-0.013}$ \\
$\beta_{ \text{M1W}}$   &  & $\underset{( -3.398)}{-0.067}$ \\
$\beta_{ \text{\$VOL}}$  & & $\underset{( 3.361)}{ 0.046}$ \\
$\beta_{ \text{ADS}}$   & & $\underset{(-0.662)}{-0.022}$ \\
\hline \hline
\end{tabular}
\smallskip
\begin{scriptsize}
\parbox{0.98\textwidth}{\emph{Note.} The table reports parameter estimates of the HAR and HAR-X model in the initial training set for the one-day-ahead forecasts of Apple's realized variance. In parentheses underneath the coefficient estimate, we report the $t$-statistic for testing statistical significance $H_{0}: \beta = 0$, which are based on White's heteroscedasticity-robust standard errors. All variables are standardized.}
\end{scriptsize}
\end{center}
\end{table}

\clearpage

\subsection{Hyperparameter tuning} \label{appendix:hyperparameter}

The objective of the hyperparameters is to control the bias–variance tradeoff. The purpose of the majority of the hyperparameters is also to reduce the complexity of the model. The challenge is to do it in such a way that the model is capable of extracting relevant information from the observations and does not fit noise. There is a shortage of theoretical results in the literature on how to optimally select hyperparameters, so no superior strategy exists. In this brief appendix, we explain how we handle it.

We divide the dataset into a training, validation, and test set\glsadd{Test Set}. In the training set\glsadd{Training Set}, the model is estimated with various sets of hyperparameters. In the validation set\glsadd{Validation Set}, the models are compared based on their forecasts, and the optimal value of each hyperparameter is selected.\footnote{Standard k-fold cross-validation has not been conducted, as it violates the time-series structure in our data.}  The test set serves as the true out-of-sample evaluation, where no estimation or tuning is conducted.

In the paper, we consider both a fixed and rolling estimation window. The less computational heavy is the fixed window. Here, the original data is split at the beginning, and the hyperparameters and weights are estimated once and employed throughout the test set. There is no re-estimation through time. The alternative is the rolling window, which allows for time-varying hyperparameters. Here, the hyperparameters are re-estimated each day with a fixed length of the training and validation set. This approach entails inclusion of more recent observations during estimation, as past data are gradually excluded.

Table \ref{table:hyperparameter} shows which hyperparameters are tuned (marked with an asterisk) and, if so, over which interval, or those that are set equal to the default value suggested by the original author.

The interval for $\lambda$ implies the unregularized regression is a potential solution. The actual grid-searching is here based on partitioning the natural logarithm of $[10^{-5}, 10^{2}]$, as this allows for extra refinement with smaller degrees of regularization. The optimal value of $\lambda$ associated with the minimum validation MSE is practically speaking always an interior point. The lower bound is occasionally binding in the $\mathcal{M}_{ \text{ALL}}$ dataset (about 2.5\% of the times across all estimations), but otherwise not. Typically, only modest regularization is needed. The upper boundary is never hit.

\begin{sidewaystable}[th!]
\setlength{\tabcolsep}{0.40cm}
\caption{Tuning of hyperparameters. \label{table:hyperparameter}}
\smallskip
\begin{tabular}{ccccc}
\hline \hline
Regularization & BG & RF & GB & NN \\ \hline			
$\lambda \in [10^{-5}, 10^{2}]$* & min. node size = 5 & min. node size = 5 & depth $\in \{1, 2 \}$* & learning rate = 0.001\glsadd{Learning Rate} \\
$\alpha \in [0,1]$* & trees = 500 & trees = 500 & trees $\in \{50,100, \ldots, 500 \}$* & epochs = 500\glsadd{Epoch} \\
& & feature split = $J/3$ & learning rate $\in \{ 0.01, 0.1 \}$* & patience = 100 \\
& & & & drop-out = 0.8 \\
& & & & initializer: Glorot normal \\
& & & & Adam = default\glsadd{Adam} \\
& & & & ensemble $\in \{1, 10 \}$ \\
\hline \hline
\end{tabular}
\smallskip
\begin{scriptsize}
\parbox{0.98\textwidth}{\emph{Note.} The table presents the hyperparameters of the various ML algorithms. An asterisk indicates that the hyperparameter is tuned in the validation set\glsadd{Validation Set}. Regularization: We do grid-searching over the interval based on a partition consisting of 1,000 points for $\lambda$ and 10 for $\alpha$. Bagging (BA) and random forest (RF): These are the default parameters based on the original Fortran code from \citet*{breiman-cutler:04a}. Gradient Boosting (GB): These are the default parameters based on the implementation in \citet*{greenwell-boehmke-cunningham:19a}, who extended the AdaBoost algorithm of \citet*{freund-schapire:97a}. Neural network (NN): These are the default parameters based on the Adam optimizer proposed in \citet*{kingma-ba:14a}. The drop-out selection follows \citet*{goodfellow-bengio-courville:16a}.}
\end{scriptsize}
\end{sidewaystable}

\clearpage
	
\subsection{Regularization of the neural network} \label{appendix:regularization}

\textbf{Adaptive learning rate}: To achieve fast convergence and locate a near-minimum function value, an adaptive learning rate is often imposed. The Adam\glsadd{Adam} optimizer, applied in this study, achieves that by shrinking the learning rate\glsadd{Learning Rate} towards zero during training. \\

\noindent \textbf{Drop-out}: The idea behind drop-out is to temporarily remove neurons\glsadd{Nodes Neural network} from an underlying layer, which in practice means that the output of that neuron is multiplied by zero with strictly positive probability \citep*[see, e.g.,][]{srivastava-hinton-krizhevsky-sutskever-salakhutdinov:14a}. The drop-out rate is set to 0.8 following \citet*{goodfellow-bengio-courville:16a}. The drop-out is applied in the training of the network and all connections from the neurons are retained in the validation and test set\glsadd{Test Set}. \\

\noindent \textbf{Early stopping}: In early stopping, the goal is not to reach a local minimum of the training error, but to stop training the network when there is no improvement for some epochs\glsadd{Epoch} in the validation error. To ensure sufficiently high patience, we set the patience to 100. \\

\noindent \textbf{Ensemble}: When conducting an ensemble, multiple neural networks are trained, and hereafter the prediction is constructed as an average across the ensemble. When initializing the weights, the Glorot (also called Xavier) normal initializer is chosen. The initialization draws samples from a truncated normal distribution centered on zero. \\

\noindent \textbf{Initial seed}: To ensure our results are not driven by the random initialization, 100 independent neural networks with different seeds are trained and ranked by validation MSE. This allows to reduce variance induced by stochasticity of the Adam optimizer (in the main text, we report the performance of the best, plus an ensemble of the ten best, network from the validation step). \\

\begin{figure}[H]
\begin{center}
\caption{Robustness check of initial seed for the neural network. \label{figure:nn-initial-value}}
\begin{tabular}{c}
\includegraphics[height=8cm,width=0.6\textwidth]{{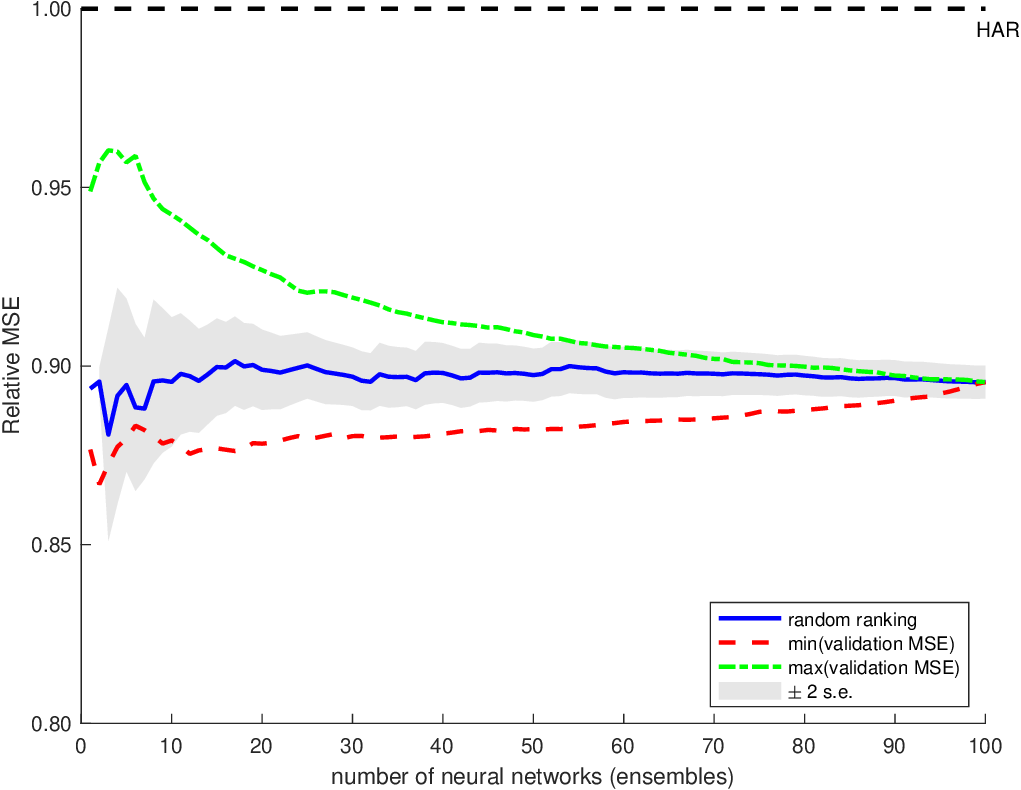}}
\end{tabular}
\begin{scriptsize}
\parbox{\textwidth}{\emph{Note.} On the $y$-axis, the graph shows the average MSE of NN$_{2}$ relative to the HAR model. The $x$-axis denotes the number of neural networks (ensembles) out of a hundred that are averaged, when initial seeds are drawn at random (random ranking). The models are also sorted based on highest or lowest MSE in the validation set. The shaded area is $\pm$ two standard error bands of the random sort. The plot is based on high-frequency data for Apple, but the evolution is representative for other stocks in our sample.}
\end{scriptsize}
\end{center}
\end{figure}

As illustrated in Figure \ref{figure:nn-initial-value}, a fast convergence to the actual performance is seen. Furthermore, changing the initial seed has a negligible impact. To further support this the figure also shows coverages, when ranking by highest and lowest MSE in the validation set. The relative MSE compared to the HAR model is below one for all initial values. The graph is based on high-frequency data from Apple, but the results are representative for other stocks in our sample.

\clearpage

\printnoidxglossary[nonumberlist]

\end{document}